%% file: main.tex
\documentclass[10pt]{article} 
\usepackage[preprint]{tmlr}
\usepackage{colortbl}

\input{preamble}

\usepackage{subfigure}
\usepackage{float}
\usepackage{graphbox}
\usepackage{svg}
\usepackage{nicefrac}
\usepackage{xcolor}

\usepackage{bold-extra}
\usepackage[T1]{fontenc}

\usepackage{array}
\newcolumntype{P}[1]{>{\centering\arraybackslash}p{#1}}
\newcolumntype{M}[1]{>{\centering\arraybackslash}m{#1}}

\definecolor{orange}{rgb}{1,0.5,0}
\definecolor{graynode}{RGB}{20,20,20}
\definecolor{crimsonred}{RGB}{220,20,60}
\definecolor{darkgraynode}{gray}{0.5}
\definecolor{lightgraynode}{gray}{0.8}

\usepackage{rotate}
\usepackage{adjustbox}
\usepackage{array}
\usepackage{capt-of}
\usepackage{tabulary}
\usepackage{setspace}
\usepackage{amssymb}
\usepackage{mathtools}
\usepackage{pifont}

\definecolor{gray}{RGB}{20,20,20}
\definecolor{gray}{RGB}{0.7,0.7,0.7}
\definecolor{greencm}{RGB}{0,153,0}

\newcommand{\hboldline}{\noalign{\hrule height 0.3mm}}
\newcommand{\boldbottomline}{\noalign{\hrule height 0.3mm}}

\definecolor{plotblue}{RGB}	{30,144,255}
\definecolor{plotgreen}{RGB}	{50,205,50}
\definecolor{plotred}{RGB}	{220,20,60}

\definecolor{myyellow}{RGB}{255,255,204}
\definecolor{myred}{RGB}{255,204,204}
\definecolor{myblue}{RGB}{0,200,255}
\definecolor{mygreen}{RGB}{20,130,13}

\newcommand*\hrulefillvar[1][0.4pt]{\leavevmode\leaders\hrule height#1\hfill\kern0pt}

\definecolor{thedarkblue}{RGB}{0,0,120} 
\definecolor{mydarkblue-new}{rgb}{0,0.08,0.45} 

\usepackage{hyperref}
\hypersetup{%
    colorlinks=true,
    linkcolor=mydarkblue-new,
    citecolor=mydarkblue-new,
    filecolor=mydarkblue-new,
    urlcolor=mydarkblue-new}

\definecolor{googleblue}{RGB}{66,133,244}
\definecolor{googlered}{RGB}{219,68,55}
\definecolor{googlegreen}{RGB}{15,157,88}
\definecolor{googlepurple}{RGB}{138,43,226}
\definecolor{lightred}{RGB}{255, 220, 219}
\definecolor{lightblue}{RGB}{204, 243, 255}
\definecolor{lightgreen}{RGB}{200, 247, 200}
\definecolor{lightergreen}{RGB}{230, 255, 230}
\definecolor{lightpurple}{RGB}{230,230,250}
\definecolor{lightyellow}{RGB}{242, 232, 99}
\definecolor{lighterblue}{RGB}{197, 220, 255}
\definecolor{lighterblue-row}{RGB}{217,240,255}
\definecolor{lighterred}{RGB}{253, 249, 205}
\definecolor{lightyellow}{RGB}{207, 161, 13}
\definecolor{darkpurple}{RGB}{218, 210, 250}
\definecolor{darkred}{RGB}{255,198,196}
\definecolor{darkblue}{RGB}{172, 233, 252}
\definecolor{mydarkblue}{rgb}{0,0.08,180}

\DeclareMathAlphabet{\mathbcal}{OMS}{cmsy}{b}{n}
\usepackage{amsmath}
\usepackage{mathrsfs}
\usepackage{comment}

\usepackage{bm}
\usepackage{bbm}
\usepackage{amssymb}
\usepackage{enumitem}
\AtBeginDocument{
  \providecommand\BibTeX{{
    \normalfont B\kern-0.5em{\scshape i\kern-0.25em b}\kern-0.8em\TeX}}}

\usepackage{paralist}

\usepackage{url}

\title{Survey of User Interface Design and Interaction Techniques in Generative AI Applications}

\author{%
      \name Reuben Luera \email raluera@ucsd.edu \\
      \addr University of California -- San Diego
      \AND
      \name Ryan A. Rossi \email ryrossi@adobe.com \\
      \addr Adobe Research
      \AND
      \name Alexa Siu \email asiu@adobe.com \\
      \addr Adobe Research
      \AND
      \name Franck Dernoncourt \email dernonco@adobe.com \\
      \addr Adobe Research
      \AND
      \name Tong Yu \email tyu@adobe.com \\
      \addr Adobe Research
      \AND
      \name Sungchul Kim \email sukim@adobe.com \\
      \addr Adobe Research
      \AND
      \name Ruiyi Zhang \email ruizhang@adobe.com \\
      \addr Adobe Research
      \AND
      \name Xiang Chen \email xiangche@adobe.com \\
      \addr Adobe Research
      \AND
      \name Hanieh Salehy \email deilamsa@adobe.com \\
      \addr Adobe Research
      \AND
      \name Jian Zhao \email jianzhao@uwaterloo.ca \\
      \addr University of Waterloo
      \AND
      \name Samyadeep Basu \email sbasu12@cs.umd.edu \\
      \addr University of Maryland, College Park
      \AND
      \name Puneet Mathur \email puneetm@adobe.com \\
      \addr Adobe Research
      \AND
      \name Nedim Lipka \email lipka@adobe.com \\
      \addr Adobe Research
}

\def\month{MM}  
\def\year{YYYY} 
\def\openreview{\url{https://openreview.net/forum?id=XXXX}} 

\begin{document}

\maketitle

\begin{abstract}
The applications of generative AI have become extremely impressive, and the interplay between users and AI is even more so. Current human-AI interaction literature has taken a broad look at how humans interact with generative AI, but it lacks specificity regarding the user interface designs and patterns used to create these applications. Therefore, we present a survey that comprehensively presents taxonomies of how a human interacts with AI and the user interaction patterns designed to meet the needs of a variety of relevant use cases. We focus primarily on user-guided interactions, surveying interactions that are initiated by the user and do not include any implicit signals given by the user. With this survey, we aim to create a compendium of different user-interaction patterns that can be used as a reference for designers and developers alike. In doing so, we also strive to lower the entry barrier for those attempting to learn more about the design of generative AI applications.
\end{abstract}

\section{Introduction}

The academic and general population have become increasingly enamored with generative artificial intelligence (AI) as it continues to revolutionize just about every field of study it is involved in. So much so, in fact, that the field of Human-Computer Interaction (HCI) has shifted much of its focus to study a sub-field of HCI called Human-AI Interaction \citep{TheInteractionDesignFoundation_2024}. While current Human-AI Interaction literature provides a broad view of the field \citep{shi2023hci}, this survey aims explicitly to capture the current state of user interfaces and the respective user interactions being utilized within generative AI applications. Specifically, this survey takes a snapshot of current trends and design techniques that involve user-guided interactions (Sec. \ref{sec:userguideddef}). 

In doing so, we aim to create a design compendium that generative AI designers, researchers, and developers can reference to understand the current state of the user experience (UX) and user interface (UI) designs of generative AI. The overall goal is to lower the barrier to entry for those interested in the UX and UI of generative AI by giving them a foundation upon which to build. For designers and developers specifically, this paper can serve as a design library to inspire their designs for generative AI applications. This paper prevents designers and developers from needing to partake in large competitive analyses and allows them to learn from design patterns currently utilized by other generative systems. For researchers, this survey can guide further explorations of human-AI interaction. This paper lists dozens of different ways that humans and generative AI applications are currently interacting and can serve as a foundation for researchers to dive into a specific area of human-AI interaction or to dive into the area in a general sense.

\subsection{Summary of Main Contributions}

This survey contains several major contributions to the human-AI interaction field based on our survey of more than a hundred relevant generative AI articles. The key contributions include key definitions and disambiguation, relevant taxonomies, and research-based design principles. Specifically, the key contributions of this work are as follows:

\begin{compactenum}

    \item \textbf{A formalization of the key notions and definitions and a disambiguation and expansion of key terms relating to UI \& interactions for generative AI applications
    (Section~\ref{sec:preliminaries}). }
   We formalize vital definitions that are relevant to understanding how a user interacts with a generative artificial intelligence system. We also disambiguate user interactions by proposing a novel concept of user-guided interactions, which are interactions that a user engages in willingly and deliberately. These do not include implicit interactions or interactions that the generative system detects without the user's knowledge. Additionally, we outline the different modality types that users can utilize to interact with generative systems. In doing so, we aim to enhance the general understanding of user interaction terms and their pertinence to generative AI.

    \item \textbf{A survey and taxonomy of UI interaction techniques for generative AI systems (Section \ref{sec:userguided}).}
    We highlight and categorize common user-guided interaction design patterns and the context in which they are used. We focus on user-guided interactions, as these are examples of how users deliberately and intentionally communicate with the generative systems. We organize these interactions into a taxonomy highlighting prompting, selecting, system and parameter manipulation, and object manipulation. We aim to create a compendium of user-guided interaction techniques that designers can refer to as they plan the designs for their own generative AI applications.

    \item \textbf{A survey and taxonomy of user interface layouts for generative AI systems (Section \ref{uilayouts}).} 
    We survey key user interface design patterns utilized in various generative applications. In doing so, we present a taxonomy that categorizes and highlights common user interface structures we found through the survey. By generalizing standard UI layouts, we prevent designers from starting from scratch and encourage using these common generative AI design patterns.

    \item \textbf{A survey and taxonomy of human-AI engagement levels for generative AI systems (Section \ref{sec:interaction-levels}).}
    We survey human-AI engagement levels, which consist of the intensity of deliberate interaction and collaboration between a system and the user. We aim to expand on existing literature that has characterized human-AI engagement levels by presenting more recent generative AI examples and by including additional human-AI engagement levels. In doing so, we aim to present a holistic survey of the current state of human-AI interaction levels.

    \item \textbf{A survey and taxonomy of AI applications and use cases for generative AI systems (Section \ref{applications})}.
    We highlight and categorize the different applications of generative AI systems to survey how they are used in various domains. We aim to survey these different use case areas to discover which user interfaces and interactions are most appropriate for use in various situations. With this knowledge, designers can discern which interaction patterns and user-guided techniques are best used in their respective domains.

    \item \textbf{An overview of key open problems and challenges that future work should address (Section \ref{sec:open-problems-challenges}).} We outline problems and open issues that future research can focus on to address generative AI accessibility, growth, and ethics. In doing so, we can continue to ensure that designers continue to create designs that incorporate solutions to larger issues.
\end{compactenum}

\subsection{Scope of this Article}
This survey focuses on user-guided interaction techniques for generative AI applications. 
Such user-guided interactions are sometimes referred to as controllability techniques.
One simple example of such techniques is when a user prompts a generative model via text and/or images; similarly, another example is when a user selects text or a specific part of an image to control the generation.
We do not attempt to survey interaction or controllability techniques that are not user-guided, nor do we survey techniques that leverage user/system feedback and the like. Given this, the purpose of this paper is to provide a recommendation on which interaction techniques are most effective when applied to specific use cases.

\section{Background \& Preliminaries}\label{sec:preliminaries}

We begin with basic definitions and notations to formalize the terms that will be used in subsequent sections as we discuss user-guided interactions and connected concepts. 
We formally define and discuss the notion of user-guided interactions and explain concepts such as the interplay and differences between prompting and inputs.
Then, finally, we discuss different modalities that users can utilize to interact with the generative AI systems.

\subsection{User-Guided Interactions} \label{sec:userguideddef}

We begin by defining user-guided interactions in order to distinguish them as a specific type of user interaction. While user interactions are \textit{any} interactions, whether explicit or implicit, made by a user to affect a system, user-guided interactions focus solely on interactions that are explicitly made by users to affect a system in a pre-desired way. Given this, this paper will only focus on user-guided interactions and how they are used in the context of generative AI.

\begin{Definition}[\sc User Guided Interaction]
User-guided interactions are defined as explicit user-initiated actions that a user deliberately makes that affect the respective computer system. 
\end{Definition}
In terms of generative AI, user-guided interactions consist of any actions that a user explicitly takes to affect the generative system. This can be anything from prompting the system to complete a certain task, to selecting and manipulating objects within a system, to adjusting a system's parameters to create a specific output. For example, a user might write a prompt that generates an image. Then they might adjust sliders or select specific parts of the image to manipulate it further. All of these are examples of user-directed interactions. This definition does not, however, include implicit user interactions, which are implicit or indirect actions made by the user that the system acts upon without being explicitly tasked to do so. Implicit user interactions include implicit behavioral interactions that the system uses to create a user profile or implicit feedback where the system infers user satisfaction based on word cues or interaction delays from the user. For instance, a system might alter its answers based on a user's chat history or surface different news stories based on a user's implied political beliefs. We acknowledge that these interactions are a crucial part of the generative process, but these interactions fall outside the scope of user-guided interactions.

All in all, we focus solely on user-guided interactions in order to provide a holistic, but focused survey of these types of interactions. Doing so affords us the ability to delve deeper into the complexities of user-guided interactions and offer a wide palette of interaction solutions. In addition, we constrain our scope solely to user-guided interactions as these are the interactions, between the user and system, that are explicitly visible. For example, one can visibly see the interaction of a user selecting a UI element or prompting a system, but they cannot, however, see a system collect implicit feedback or user data. Given that user-guided interactions are those with visible interactions, they are also the interactions that can actually be visibly designed. Given this, this survey's goal is to be especially helpful to those product, visual, user experience (UX), and any other type of designer that has been tasked with designing a generative AI user interface. To conclude, we deliberately highlight and focus on user-guided interactions as their implementation can greatly enhance the overall generative AI user experience. Utilizing this scope ensures that this survey is focused and understandable for designers of all levels.

\subsection{Prompts and Inputs}

\begin{Definition}[\sc Inputs]
An input is a piece of data, information, or content that the user uploads to the system. An input, if available, is what the prompt acts upon.
\end{Definition}

\begin{Definition}[\sc Prompts]
A prompt is a type of user-guided interaction in which the user asks the generative system to complete a certain job. 
\end{Definition}

Prompting a generative system is often the most commonly thought of user-guided interaction associated with generative AI. As mentioned, it consists of a user asking a generative system to complete a specific task. While text-prompting is the most commonly thought of prompting modality, other prompting modalities include visual, audio, and multi-modal prompting. In essence, prompting is an essential way that users interact with and guide the systems that they are working with. Meanwhile, inputs are data, information, or content that is uploaded to the generative system. Like prompts, inputs can be text-based, visual, or audio. In tandem, or sometimes on their own, these are two important aspects of the generative AI user flow.

\begin{figure}[t] \label{promptsvinputs}
\centering
\includegraphics[width=.60\linewidth]{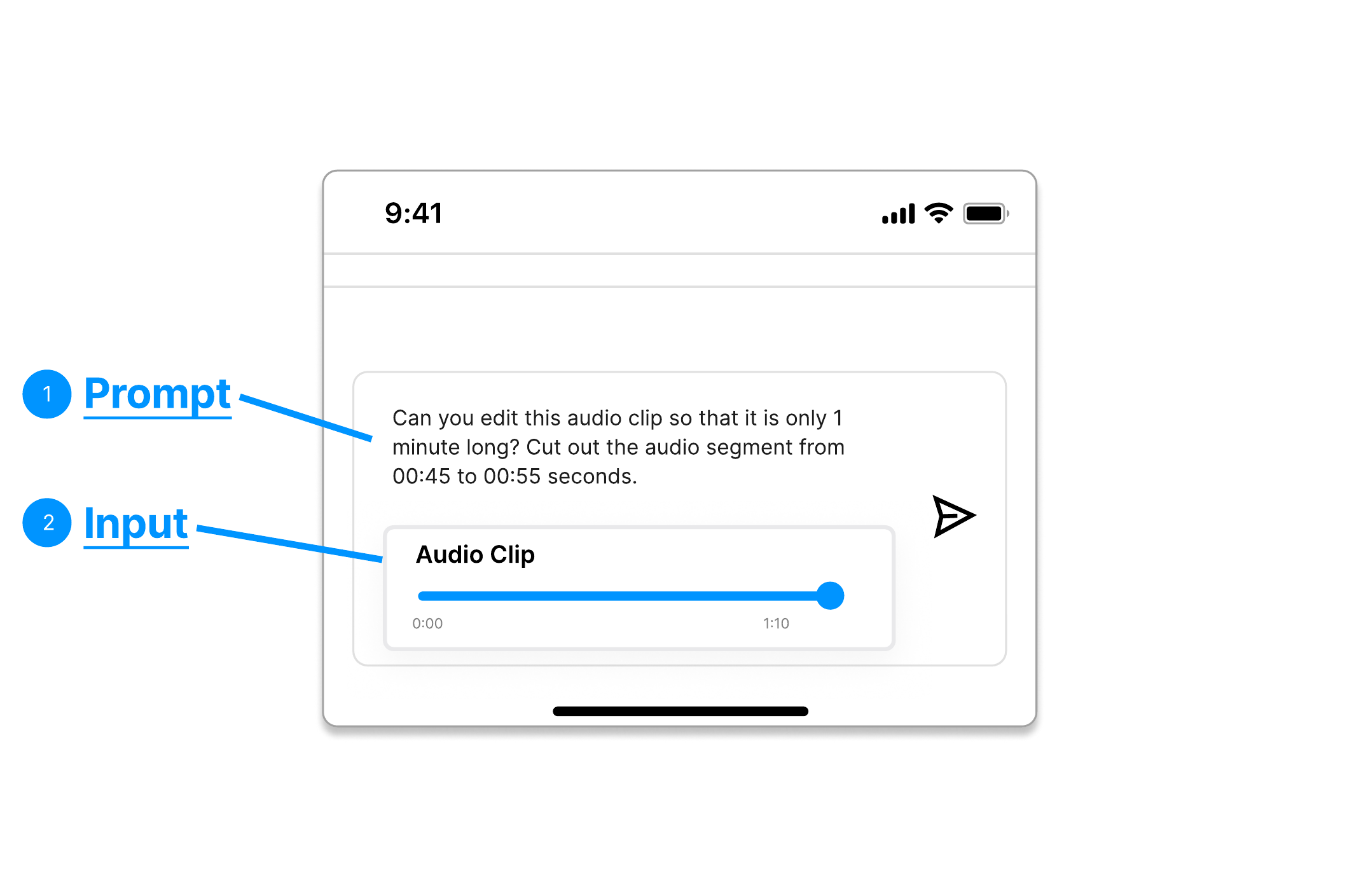}
\caption{%
\textbf{Prompt vs Inputs (Sec. \ref{sec:input-modalities}): A visual summary of the distinction between prompts and inputs. A prompt is a user-guided interaction where the user asks the system to complete a task. Whereas the input is the piece of data, information, or content that the prompt is acting upon.}}
\label{fig:promptvinp}
\end{figure}

An important distinction is that this section does not consider "prompts" as an input. Instead, we view user prompts and inputs as often two separate entities, where a prompt is used to query the system, and the input is what is being acted upon. As seen in Fig. \ref{fig:promptvinp}, if there is an audio editing generative system, then the \textit{input} would be an audio file, but the user \textit{prompt} can be text such as "Can you edit this audio clip so that it is only one minute long?". Thus, the input and prompting are distinct since the prompt is a user-guided interaction that acts on the input video. 

We separate "prompting" and "inputting" as two different terms because, from a user interaction point of view, they are distinctly different. Whereas an input interaction is essentially some data that the generative system is acting upon, prompting is the user-guided interaction that consists of actually instructing the system to complete a specific task. Furthermore, distinguishing between the two components simplifies and focuses the upcoming discussion sections: Input Modalities (Sec. \ref{sec:input-modalities}) and Prompting (Sec. \ref{sec:prompting}).

\subsection{Input Modalities} \label{sec:input-modalities}

This subsection will focus solely on the input modalities that a user can use when interacting with generative AI systems. As they pertain to generative AI user interactions, we define inputs as data or information that the generative AI system is capable of analyzing or processing to generate a different output. After surveying the modalities of different generative AI systems (Fig. \ref{fig:inout}), we present the different input modalities currently in use: text-based inputs (Sec. \ref{sec:mtext}), visual inputs (Fig. (Sec. \ref{sec:mvisual})), and sound inputs (Sec. \ref{sec:msound}).

\subsubsection{Text-based} \label{sec:mtext}

\textbf{Natural Language: } Text-based natural language is a modality commonly used to interact with generative AI systems \citep{achiam2023gpt, padiyath2021desainer, suh2023structured, wang2024weaver,petridis2024constitutionmaker,zhao2023chatbridge,kim2023metaphorian, cho2024sora, wang2024aesopagent}. As noted, the text inputs in questions are \textit{not} the same as prompts. Text inputs can be anything from PDF files, structured texts, and unstructured texts, but they are unique as they do not explicitly ask the system to perform an action on its own. It is the prompts that do that, as illustrated in figure \ref{promptsvinputs}. Given this point, \cite{wang2024aesopagent}'s AesopAgent is a prime example of inputting natural language text into a generative system to create something novel and unique. With AesopAgent, a user can input a short text story and separately prompt the system to create a full script for the given short text story. The system obliges and outputs a full script with storyboard images and can even create a matching video. In doing so, the system essentially can create a short animated video from just an inputted text story.

\textbf{Data:}
Generative AI systems are often used as a tool to synthesize, clean, or gain insights from inputted data \citep{achiam2023gpt, 10.1145/2984511.2984588,singhfigura11y, schneider2019wav2vec,swanson2024generative, deepscope2019}. In relation to generative AI, inputted data can be thought of as either raw or structured information or data that can be in the form of text files, structured data files, system log files, etc. Datasets, in general, can be large, untenable obstacles when it comes to completing both academic and industrial goals. Having generative AI help synthesize and digest data can make existing tasks easier and make new tasks possible. Take Synthemol \citep{swanson2024generative} for example. This system takes 
in chemical data to synthesize and design new chemical compounds that are novel and synthesizable. In essence, using data in the generative process, whether structured or unstructured, can help users complete tasks, like chemical synthesis, that would be extremely difficult to do without generative AI.

\textbf{Code:} Code is a common type of text that is both inputted and outputted from generative systems \citep{ross2023programmer,achiam2023gpt, barke2023grounded, chen2021evaluating, finnie2022robots, yen2023coladder, li2023motcoder, okuda2024askit}. When talking about code as an input or output, this can take the form of programming languages, structured scripts, markup languages, query languages, etc., and users could input it into a system to get help completing it \citep{achiam2023gpt}, debug it \citep{achiam2023gpt}, or ask questions to understand it \citep{ross2023programmer}.  For example, in \cite{ross2023programmer}'s Programmer's Assistant, users can input raw code and ask the generative system natural language questions about either the inputted code or how to create new code. By inputting already established code into the system, the system can interact with the input and present new potential interactions that the user can partake in.

\begin{figure}[t]
\centering
\includegraphics[width=.99\linewidth]{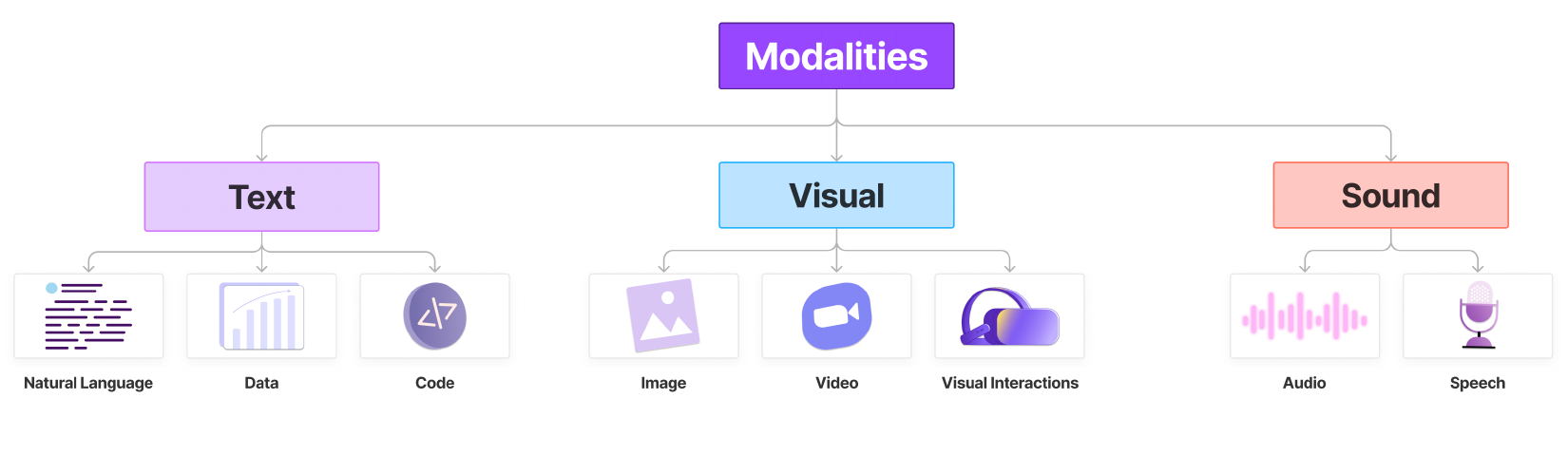}
\vspace{-5mm}
\caption{%
\textbf{Modalities: A high-level visual summary of the different modalities that generative AIs use (Sec.~\ref{sec:input-modalities}).}}

\label{fig:Modality}
\end{figure}

\subsubsection{Visual} \label{sec:mvisual}

\textbf{Images:}
Using images as an input to interact with generative systems has become common, as users attempt to do everything from generating new images \citep{padiyath2021desainer,jeon2021fashionq,betker2023improving}, to captioning images \citep{singhfigura11y,alayrac2022flamingo}, to creating infographics \citep {10.1145/2984511.2984588}. In essence, the image modality consists of the user interacting with images of any type, whether they are infographics, photographs, illustrations, etc., and many systems have differing and unique uses for image inputs. For example, one aspect of \cite{singhfigura11y}'s FigurA11y is creating alt text for figures in research papers. A user inputs figures and the system will create captions and alt text to increase accessibility or even help the reader get a deeper understanding of the paper. Overall, image inputs have a wide range of uses and are an extremely versatile modality as it pertains to generative AI systems.

\textbf{Videos:} Videos are also a common input and output modality for generative systems and are especially common in multi-modal LLMs (MMLLMs) \citep{liu2023interngpt, cho2024sora, gao2023assistgpt, wu2023nextgpt, goyal2023minotaur}. Users can use these systems to do anything from generating or finishing videos \citep{cho2024sora,liu2023interngpt} to highlighting or annotating them \citep{liu2023interngpt}. For example, in InternGPT \cite{liu2023interngpt}, a user can input a video and prompt the system to complete a task pertaining to that video. So in the system, a user could prompt it to edit an inputted video so it matched a TikTok format. Generative AI interactions with video modalities, such as this, have a wide range of uses for both professional and amateur users.

\textbf{Visual Interactions:} Visual interactions are an input modality that consists of any visual interaction or gestural movement that the system records as an input. This includes visual movements in virtual or augmented reality spaces \citep{giunchi2024dreamcodevr, konenkov2024vr, deepscope2019} or directly manipulating UI elements in a way the system takes as an input \citep{10.1145/3613904.3641920, 10.1145/3586183.3606737, suh2023structured, masson2023directgpt, kim2023lmcanvas}. For example, \cite{konenkov2024vr}'s VR GPT allows users to gesture at certain items to ask the system to interact with it. So if a user is trying to learn how to correctly pack a medical bag, they may gesture at a first-aid item and ask, "What is that?" In doing so, their pointing gesture is recorded in the system as an input and is used in tandem with the spoken prompt to interact with the system in a unique way.

\subsubsection{Sound}\label{sec:msound}

\textbf{Speech:} Speech is a growing medium that more and more generative systems can interact with. In most use cases, speech is often used or generated by the system to help the user complete a speech-related task. Take \cite{borsos2023audiolm}'s AudioLM, which is capable of taking a recorded spoken input and generating an "end" to the recording. So if a user generates the first half of a speech, AudioLM can generate the rest of the speech. Interactions that use speech as a modality, such as this one, often create novel and relevant use cases for their respective users.

\textbf{Audio:}
Similar to the video modality, audio inputs and outputs have become a versatile way that the user can interact with generative AIs \citep{borsos2023audiolm, wang2024aesopagent, zhao2023chatbridge, gao2023assistgpt, wu2023nextgpt, copet2023musicgen, agostinelli2023musiclm}. In \cite{wu2023nextgpt}'s NExT-GPT, a user can input a sound recording of something like a plane taking off. From there, the user can prompt the system to do something along the lines of "create a video that this sound would come from." The system will then use the inputed audio recording to create an accompanying video of a plane taking off. Inputting audio recordings offers another dimension that can be utilized to prompt the system to complete a unique set of tasks.

\begin{figure}[t]
\centering
\includegraphics[width=.70\linewidth]{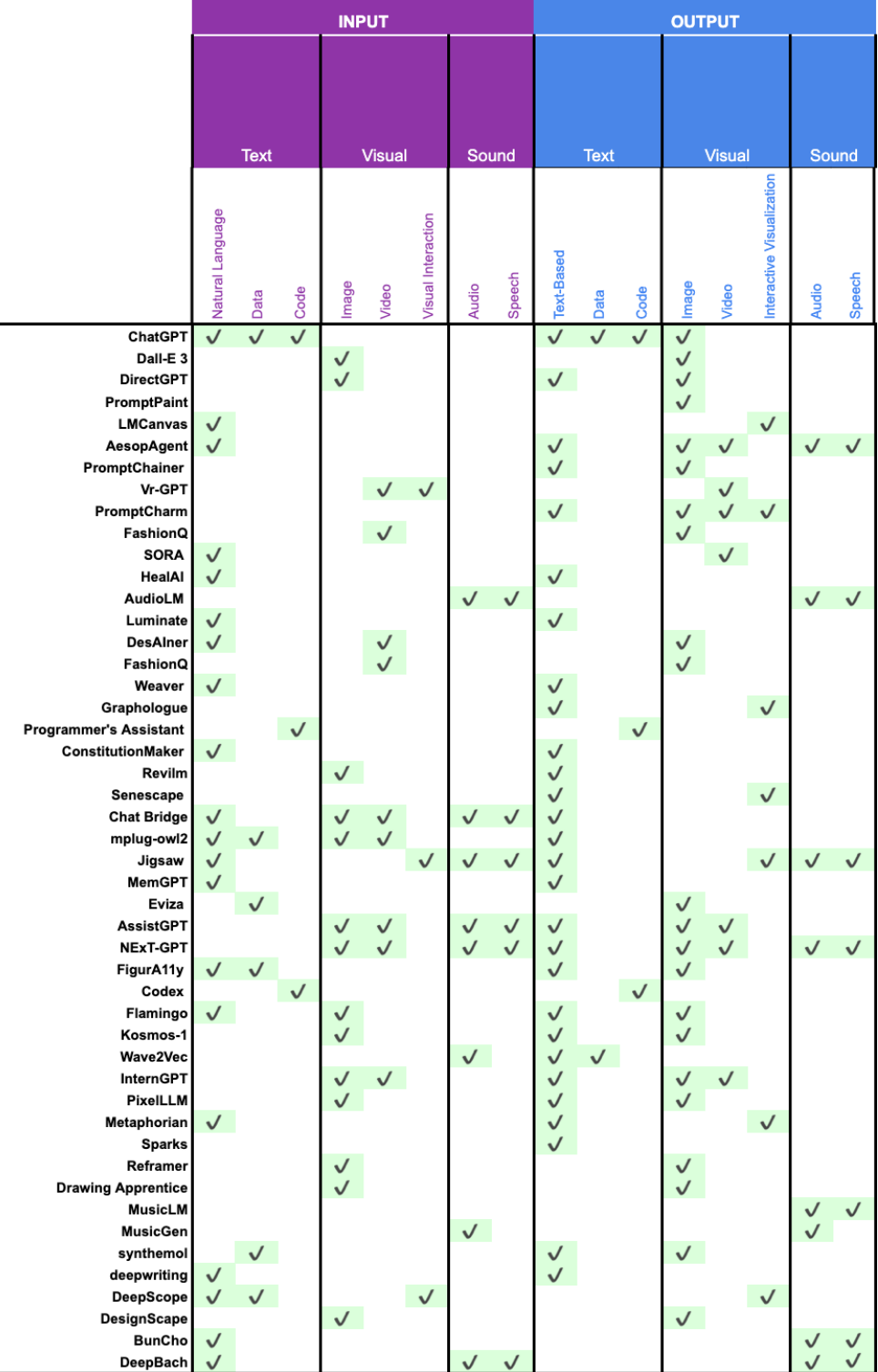}
\caption{%
\textbf{Taxonomy of works by their input/output modalities.} \label{fig:inout}
}
\label{fig:UIChart}
\vspace{-2mm}
\end{figure}

\section{User-Guided Interactions} \label{sec:userguided}

\textbf{Definitions and Scope:} In the case of generative AI, user-guided interactions are defined as explicit and deliberate engagements between the user and the system that the user initiates or guides in the generative process. Therefore, the scope of this section will focus primarily on the actual user interactions explicitly performed by the user, and not implicit interactions such as how a user’s interactions can implicitly inform an AI system over time about their preferences. Many of these interactions are performed by interacting with UI elements, which, in this case, are defined as the visual components that users interact with to manipulate the generative process.

In the context of generative AI, we propose the following taxonomy that categorizes user-guided interactions into the following categories:
prompting (Section~\ref{sec:prompting}),  selection techniques (Section~\ref{sec:selecting}),  system and parameter manipulation (Section~\ref{sec:manipulating}), and object manipulation and transformation (Section~\ref{sec:Object}).

\begin{compactenum}
   \item 
   \textbf{Prompting (Sec.~\ref{sec:prompting})}: 
   Prompting is a user-guided interaction in which a user asks or "prompts" the system to complete a certain task or job. Such user-guided prompting interactions include text-based prompts (Sec.~\ref{sec:Ptext}), audio prompts (Sec.~\ref{sec:Paudio}), visual prompts (Sec.~\ref{sec:Pvisual}), and multi-modal prompts (Sec.~\ref{sec:Pmultimodal}).

   \item \textbf{Selection Techniques (Sec.~\ref{sec:selecting})}: Selection techniques use various tools and methods to highlight or choose a specific UI element (\eg, prompting blocks, image or text previews, parts of an image, etc.) within the generative AI system to be further interacted with during the generative process. Such selection interactions include single selection (Sec.~\ref{sec:Ssingle}), multi-selection  (Sec.~\ref{sec:Smulti}), lasso and brush selection (Sec.~\ref{sec:Spaint}), and multi-modal selection (Sec.~\ref{sec:Smulti}).

   \item \textbf{System and Parameter Manipulation (Sec.~\ref{sec:manipulating})}: System and parameter manipulation consist of user interaction techniques that allow the user to adjust the parameters, settings, or functions of an overall generative AI system. These interactions are often used to personalize generated outputs to meet user needs. Such user-guided system and parameter manipulation interactions include menus (Sec.~\ref{sec:MParameter}), sliders (Sec.~\ref{sec:MSliders}), and explicit feedback(Sec.~\ref{sec:MExplicitFeedback}).
   
   \item \textbf{Object Manipulation and Transformation (Sec.~\ref{sec:Object})}: Object manipulation and transformation interactions occur in situations where the user directly modifies, adjusts, and/or transforms a specific UI element, like a building block, puzzle piece, or similar entity. Doing so gives the user deeper control over the system and allows them to interact with the UI elements in a unique and novel way. Such user-guided object manipulation and transformation interactions include drag and drop interactions (Sec.~\ref{sec:ODragnDrop}), connecting (Sec.~\ref{sec:OCombining}) and resizing (Sec.~\ref{sec:OResizing}).   
 \end{compactenum}

\begin{figure}[t]
\centering
\vspace{-10mm}
\includegraphics[width=.70\linewidth]{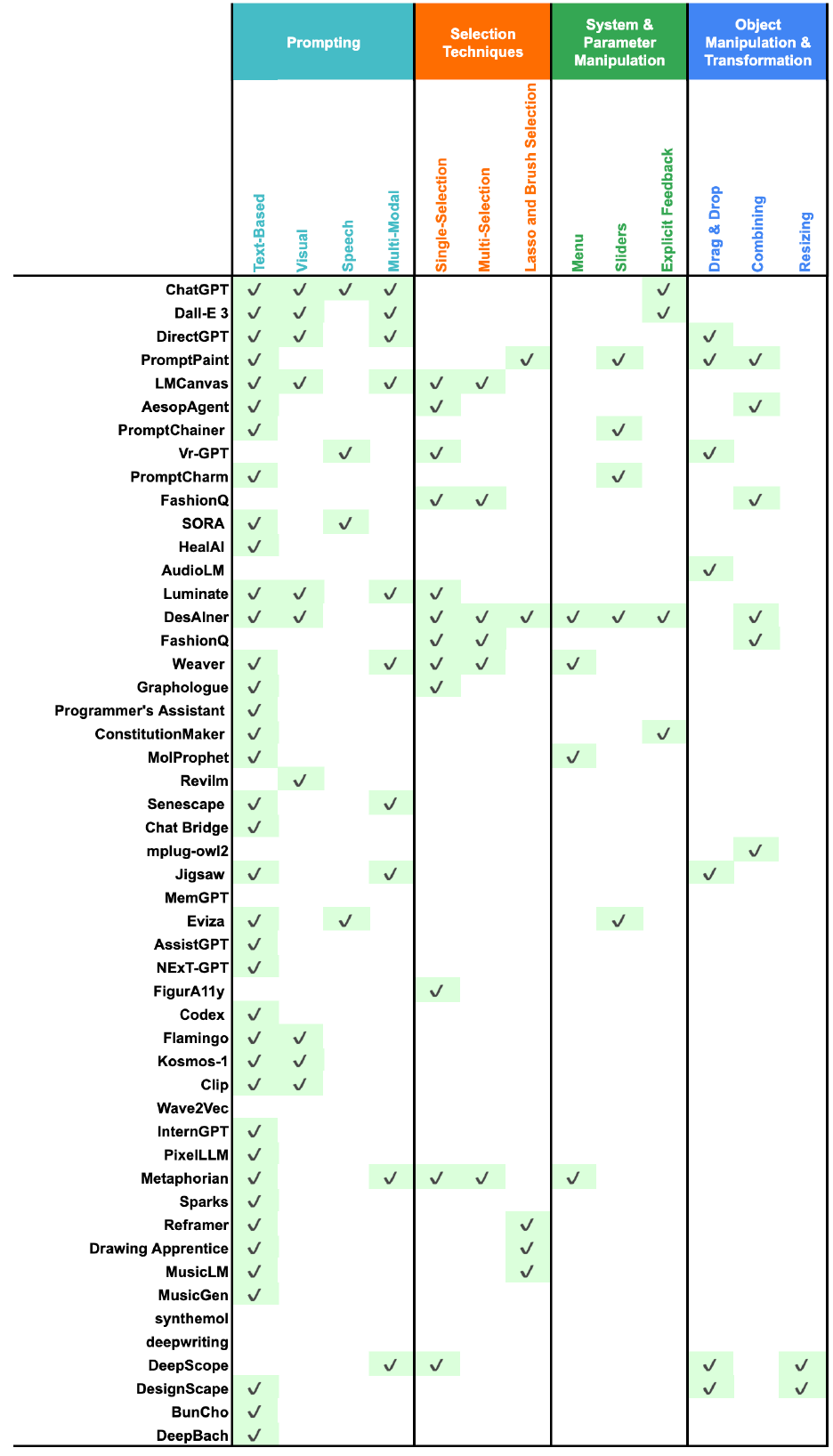}
\vspace{-2mm}
\caption{%
\textbf{User-Guided Interaction Taxonomy.}
Generative AI systems and tools are summarized using the proposed user-guided interaction taxonomy (Sec.~\ref{sec:userguided}).}
\label{fig:user-guided-interaction-taxonomy}
\vspace{-2mm}
\end{figure}

\textbf{Motivation:} There is likely no generative AI system that utilizes all of the techniques that will be outlined, nor should there be. The goal of these user-guided techniques is to ensure a seamless and user-friendly navigation of the respective systems to improve the generative process. Moreover, surveying such a wide array of user-guided techniques in generative AI will help the reader understand which generative AI techniques are most appropriate in a number of different situations. The goal is also to expose several novel user interaction techniques that are not widely known and to empower designers and developers to leverage them to create powerful and accessible generative AI systems.

\textbf{Variability:} As such, we survey works that have used any of the UI interactions (\eg, text, visual prompts, drag-and-drop, sliders, selection, in-painting, and so on), and categorize the systems that use each, and how they use them, and the different ways each are used to guide the generation process. 
This is useful to understand how each UI interaction (\eg, sliders) was used by existing systems (\eg, in the case of sliders, some work used sliders to adjust the generative model hyperparameters, while others used it to adjust the attention weights of specific user-guided selections such as text that was generated).

\subsection{Prompting}\label{sec:prompting}

\textbf{Definition \& Scope:} Prompting is a user-guided interaction in which a user asks or "prompts" the system to complete a certain task or job. Prompting, specifically text-based prompting, is often thought of as the primary interaction method utilized to interact with generative AI systems. Prompting is different than just inputting content like images or videos, as the user is explicitly asking the system to perform a task.
An important distinction is that in terms of generative AI user interactions, a prompt is not the same as an input. For example, if there is an AI system like video editing, the input could be a video, but the user-guided prompt can be text, such as "edit the video so it is no longer than 10 seconds." Thus, the inputs and prompts are distinct, where the prompt is a user-guided interaction that acts on the input video.

\begin{figure}[t]
\centering
\subfigure[Text-based \!Prompt (\S.\!\ref{sec:Ptext})]{
\includegraphics[width=0.23\linewidth]{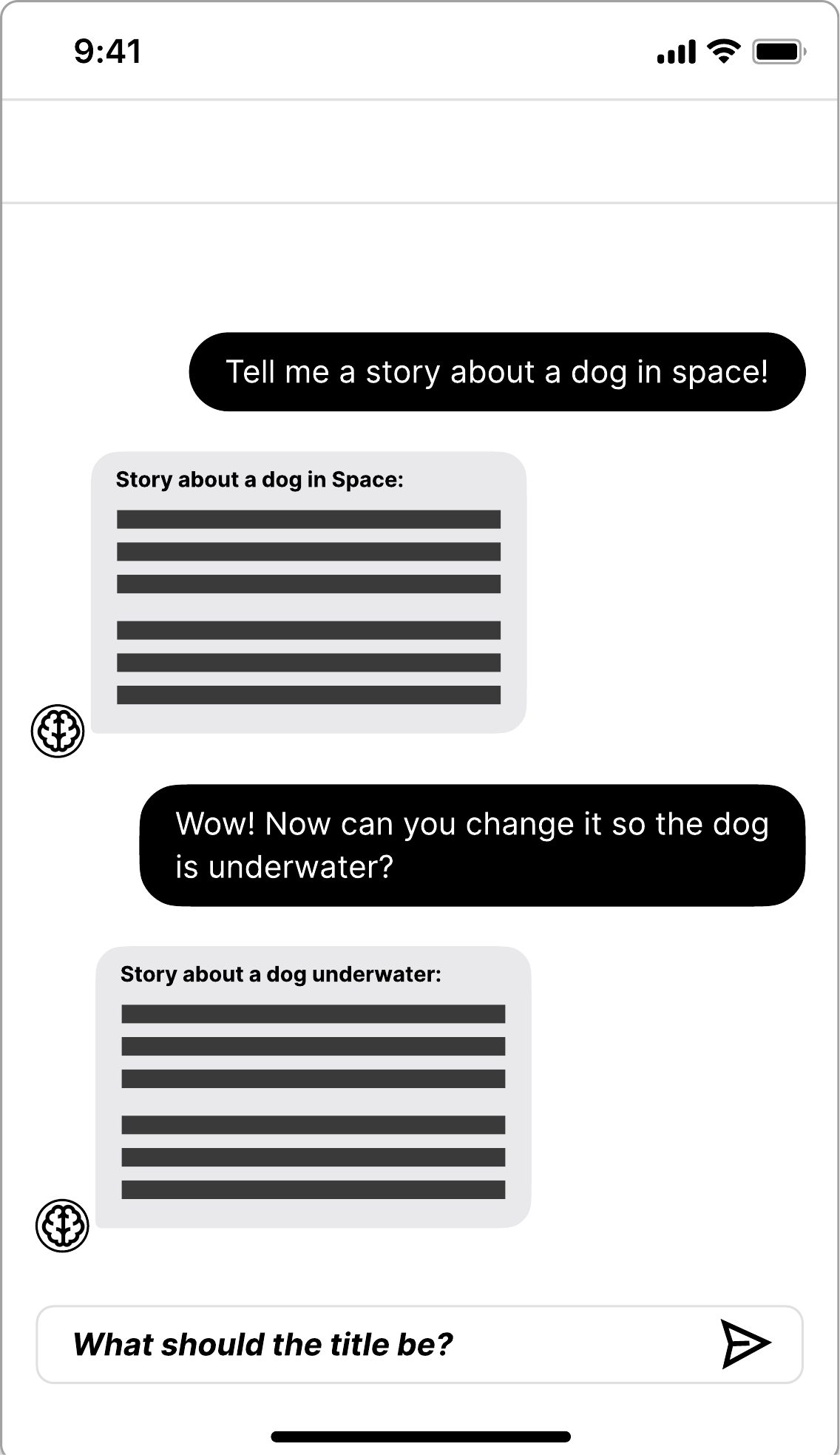}
\label{figure1a}
}
\subfigure[Visual Prompts (\S.\ref{sec:Pvisual})]{
\includegraphics[width=0.23\linewidth]{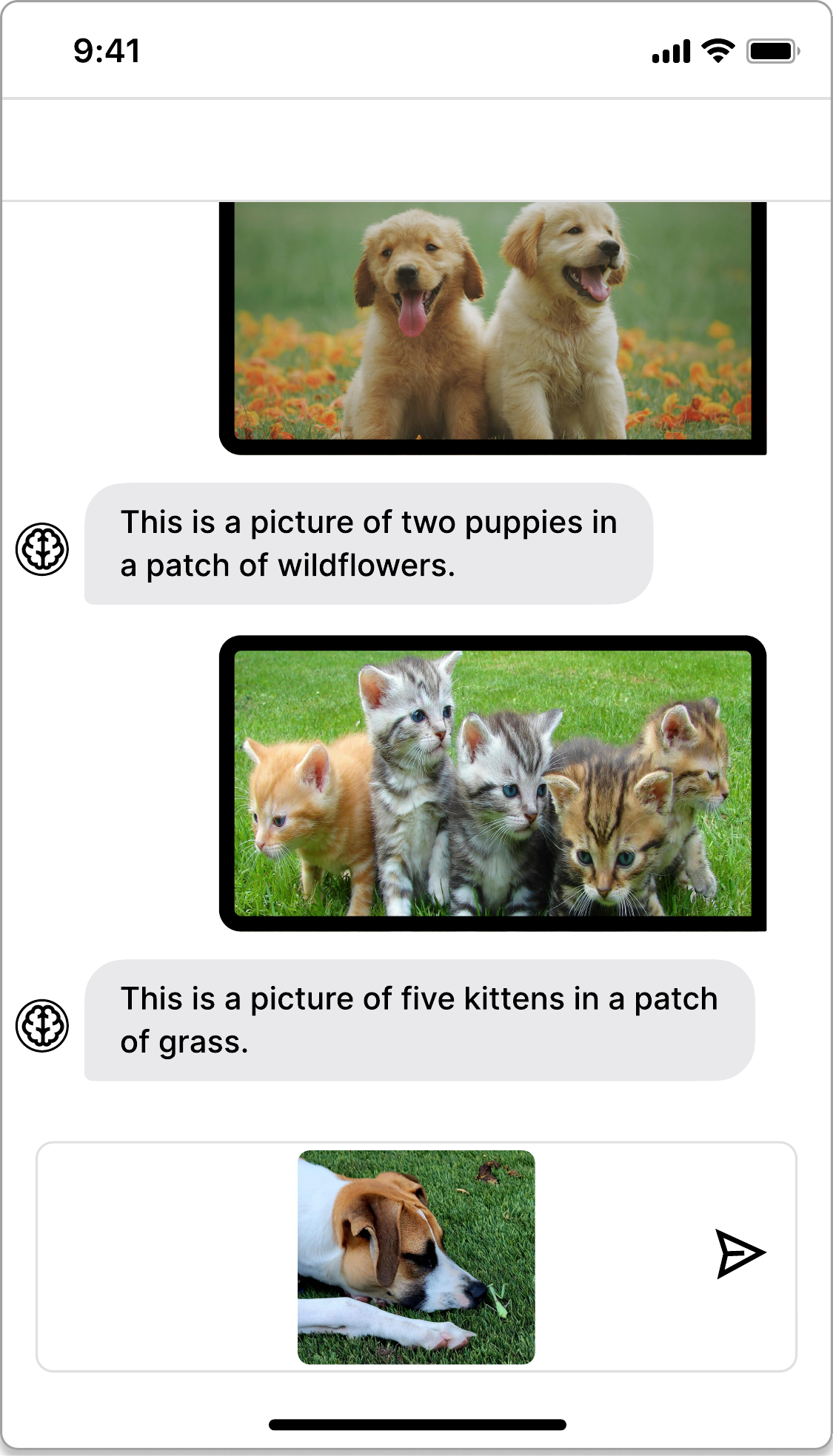}
\label{figure1b}
}
\subfigure[Audio Prompts (\S.\ref{sec:Paudio})]{
\includegraphics[width=0.23\linewidth]{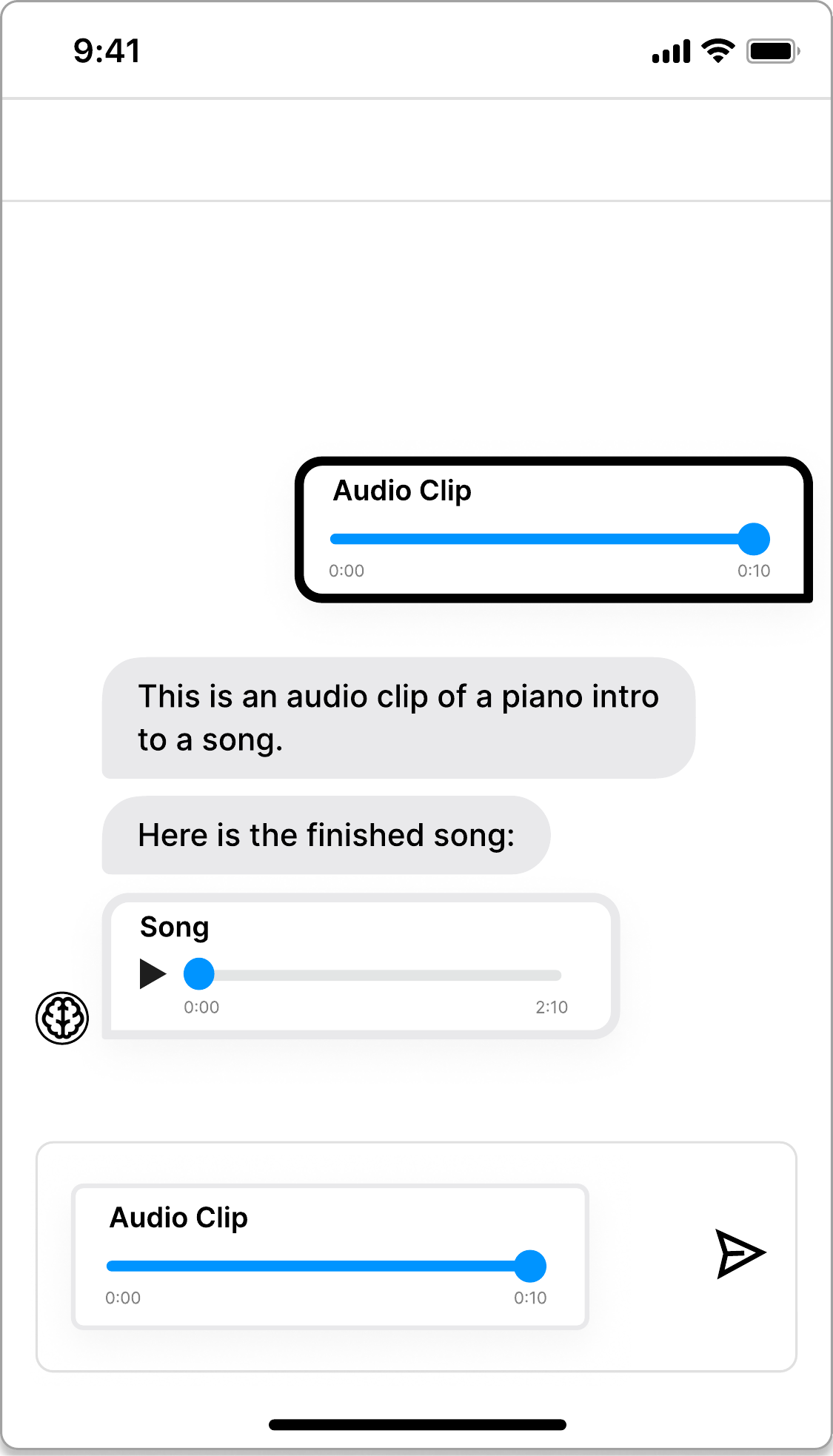}
\label{figure1c}

}
\subfigure[Multi-Modal \!Prompts (\S.\!\ref{sec:Pmultimodal})]{
\includegraphics[width=0.23\linewidth]{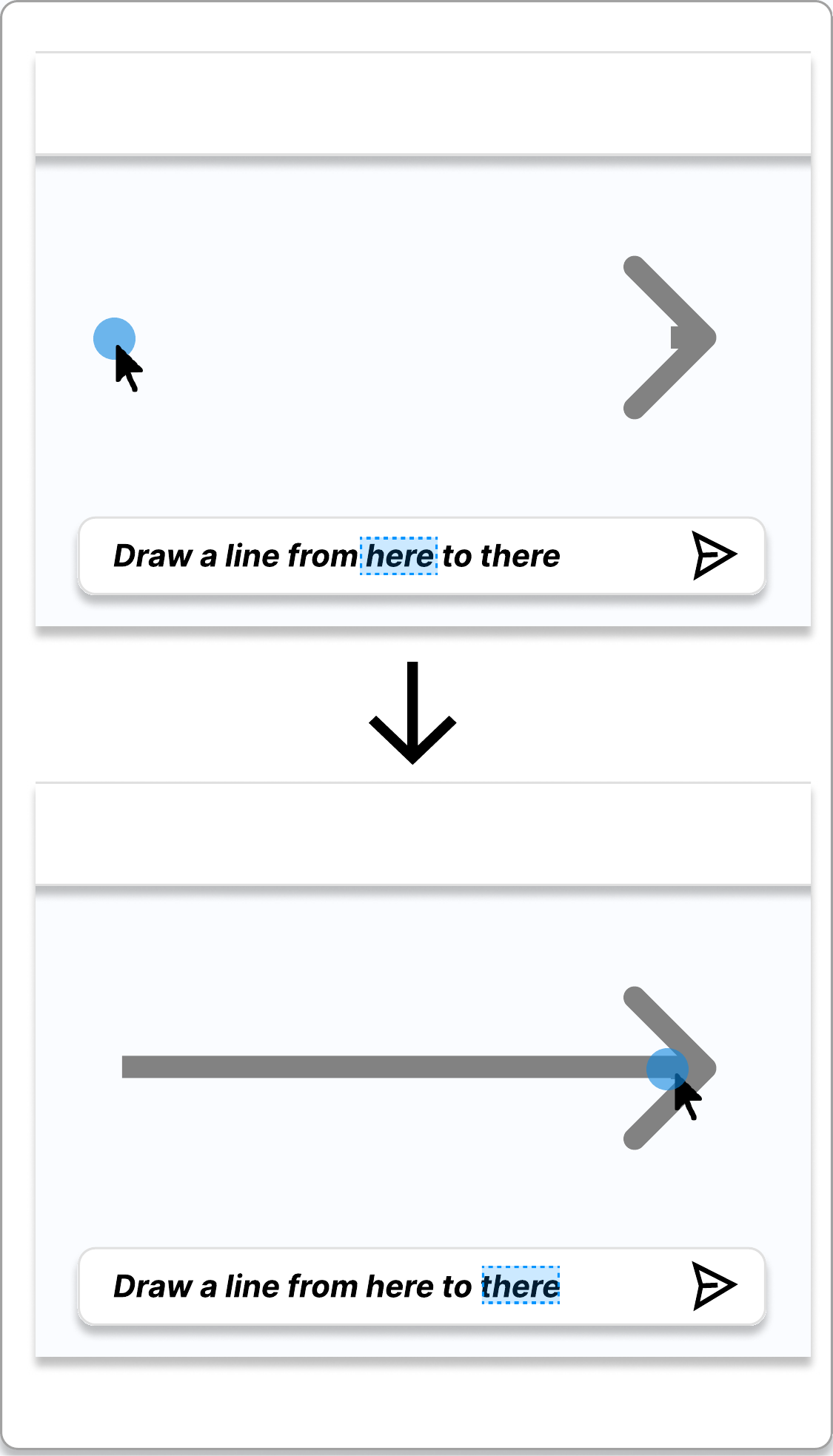}
\label{figure1d}
}
\caption{%
\textbf{Prompting Visual Summary (Sec. \ref{sec:prompting}): An overview of the four main prompting subcategories. Prompting is a user-guided interaction where a user asks or "prompts" the generative AI system to complete a certain task.}
}
\label{fig:Prompting}
\vspace{-2mm}
\end{figure}

\begin{compactitem}

    \item \textbf{Text-based Prompts (Sec.~\ref{sec:Ptext}; Fig. \ref{figure1a})}: Text-based prompts consist of using written text, often in the form of natural language, to prompt the system to complete a certain task.
    
    \item \textbf{Visual Prompts  (Sec.~\ref{sec:Pvisual}; Fig. \ref{figure1b})}:  Visual prompting consists of using visual communication, like gestures, to prompt the system to complete a certain task.
    
    \item \textbf{Audio Prompts  (Sec.~\ref{sec:Paudio}; Fig. \ref{figure1c})}: Audio prompting consists of using speech or any other type of audio to prompt the system to complete a certain task.

    \item \textbf{Multimodal Prompts (Sec.~\ref{sec:Pmultimodal}; Fig.  \ref{figure1d})}: Multimodal prompting consists of using a mix of the previous methods to prompt the system to complete a certain task.
\end{compactitem}

\subsubsection{Text-based Prompts} \label{sec:Ptext}

\textbf{Text-to-text prompting:}
Text-based prompting is the most common interaction medium a user uses when interacting with Generative AI systems and has a wide range of applications ~\citep{goyal2024healai,achiam2023gpt, betker2023improving,gero2022sparks}. While there are several prompt mediums, systems like ChatGPT \citep{achiam2023gpt} and Dall-E 3 \citep{betker2023improving} heavily rely on text-based inputs. In ChatGPT \citep{achiam2023gpt}, for example, users primarily input prompts like "explain the Krebbs cycle to me as if I were a child," and the system will output an explanation of the Krebbs Cycle in plain English (Fig. \ref{figure1a}). Meanwhile, in systems like \cite{chen2021evaluating}, users can use text prompts to ask generative models to create or continue a section of code for them.

\textbf{Text-to-multimodal prompting:}
Meanwhile, multi-modal systems can also utilize text-based prompts and produce outputs of a different medium like visuals \citep{alayrac2022flamingo,liu2023interngpt} or audio \citep{copet2023musicgen,agostinelli2023musiclm,hadjeres2017deepbach}. Dall-E 3, for example, incorporates a text-to-image generation system that takes text-based prompts and outputs an image. Another example is text-to-video, where generative AI systems, like AESOP's agent \citep{wang2024aesopagent}, take text-based prompts and create AI videos. Text prompts are common in many multimodal generative AI system interactions, with some examples consisting of, but not limited to: text-to-image, text-to-video, text-to-audio, and so on. 
Similarly, some generative AI applications \citep{zhao2023chatbridge,ren2023pixellm} allow text-based prompts to ask questions about or manipulate multi-modal inputs like an image or video. All in all, text prompts are often the backbone of most user interactions because of their versatility and the large amounts of different actions that can be performed.

\subsubsection{Visual Prompts} \label{sec:Pvisual}

\textbf{Visual manipulation: } Visual prompting consists of using visual communication, like object manipulation, to prompt the system to complete a certain task (Fig. \ref{figure1b}). For example, Jigsaw ~\citep{10.1145/3613904.3641920} created a set of puzzle pieces that each have a corresponding instruction that the system can complete. So visually, the user can manipulate, drag and drop, and connect these puzzle pieces (each puzzle piece has a partial prompt) to create a new, larger prompt. In doing so, they are visually manipulating UI elements to prompt the system uniquely. 

\textbf{Image-Prompts: } As mentioned, there is a distinct difference between prompts and inputs. A prompt is used to ask the system to complete a certain task, and some inputs can be used to do the same. In turn, this makes it possible for some inputs, like visuals, to explicitly prompt the system to perform a certain task. This is most common in few-shot learning examples where a model can generate content based on a limited number of training inputs \citep{alayrac2022flamingo,kosmos2023,radford2021learning}. In \cite{alayrac2022flamingo}, an image of a solved math equation can be inputted (i.e. 1+1=2, 1+2=3, etc.). Then, using another image of an unsolved math equation (i.e., 1+3=?) would prompt the system to generate a fill to that last equation. Whereas the original picture of solved math equations would purely be seen as inputs, the image of the unsolved math equations would be considered a prompt, as it is \textit{prompting} the system to provide an answer to the unsolved equation.

\subsubsection{Audio Prompts} \label{sec:audiop} \label{sec:Paudio}
\textbf{Speech: } When a user is embedded in a virtual environment and do not have access to traditional text-based inputs, audio and speech become one of the strongest ways to interact with AI models. In ~\cite{konenkov2024vr}, the user can interact with the visual language model (VLM) by speaking into the VR headset's microphone. As it is a VR training application, users can prompt the system by verbally asking it questions like "what is my next step." In situations like this, a user relies on speech as the primary way to prompt the VLM for a response. Just as LLMs such as ChatGPT rely on text-based prompts, VLMs embedded in VR environments often rely on spoken prompts in the same way.

\textbf{Audio Files: } Some models \citep{borsos2023audiolm,schneider2019wav2vec}, can be prompted with just an audio clip. AudioLM is a novel system that uses two audio clips, one being the input clip and the other being the prompt clip, to extend the prompted audio clip. So for example, if a user has a ten-second speech (inputted clip) they can then prompt the system with the first three seconds of the clip and then the system will generate a new and unique continuation from those original three seconds. This type of prompting is interesting because there is no natural language involved. Because the system's only function is to generate a continuation of the inputted audio, the only prompting needed is the truncation of the original clip. This is significant because few systems do not require either a spoken or written prompt. This system is only able to use this novel system of prompting because it only has a single function with essentially no extra parameters, so it does not require extra written or spoken directions.

\subsubsection{Multi-Modal Prompts} \label{sec:Pmultimodal}
\textbf{Dual-modality: }DirectGPT ~\citep{masson2023directgpt} illustrates that it is common for selection techniques to exemplify different modalities. For example, ~\citep{masson2023directgpt} explains how the DirectGPT system incorporates a text-based and selection-based input simultaneously. In this instance, users can highlight multiple words simultaneously and then type a prompt that encourages the system to replace those words with a synonym. Using a hybrid of several interaction techniques is especially useful when a single-dimensional interaction does not address the customer's need on its own. Furthermore, hybrid interactions can significantly reduce the amount of time it takes for a user to complete a task, as seen in ~\cite{masson2023directgpt} where users completed tasks 50 percent faster than those who utilized single dimensional interactions like in ChatGPT.

\textbf{Multi-modal LLMs: }Meanwhile, multi-modal LLMs (MM-LMMs) are becoming more common, as they are almost universal generative tools that work in a number of different situations \citep{alayrac2022flamingo, achiam2023gpt,wu2023nextgpt} NExT GPT \citep{wu2023nextgpt} is an empowered MM-LLM, meaning that it takes inputs from any modality and can output in any modality. It relies on a mix of searching for keywords and detecting the input modalities to decide what the output modality should be.

\subsection{Selection Techniques}\label{sec:selecting}

\textbf{Definition \& Scope:} Selecting, in terms of generative AI systems, consists of choosing or highlighting a specific UI element to further interact with it. Selecting UI elements and utilizing selection tools has become a major user interaction method utilized in many generative AI systems. Selectable UI elements and buttons, selectable inputs and outputs, and dropdown menus all offer a way to directly interact with the system. Meanwhile, selection tools such as selection boxes, lassos, or marquee tools enable users to choose precise areas or objects generated by the AI. By incorporating these user-interaction techniques, users are able to select content with more control and accuracy.

\begin{figure}[t]
\centering
\subfigure[Single Selection]{
\includegraphics[width=0.23\linewidth]{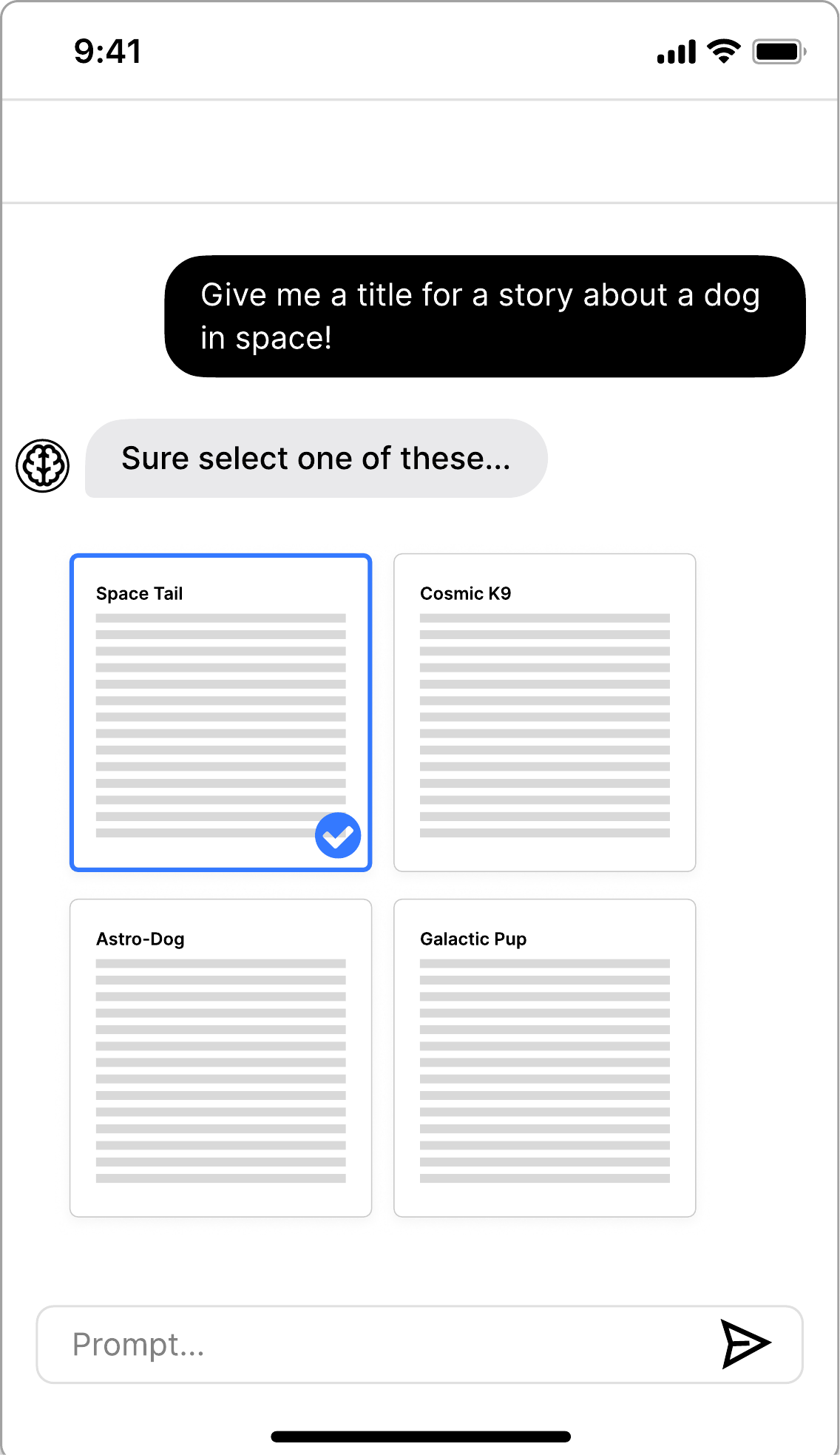}
\label{figure2a}
}
\subfigure[Multi-Selection]{
\includegraphics[width=0.23\linewidth]{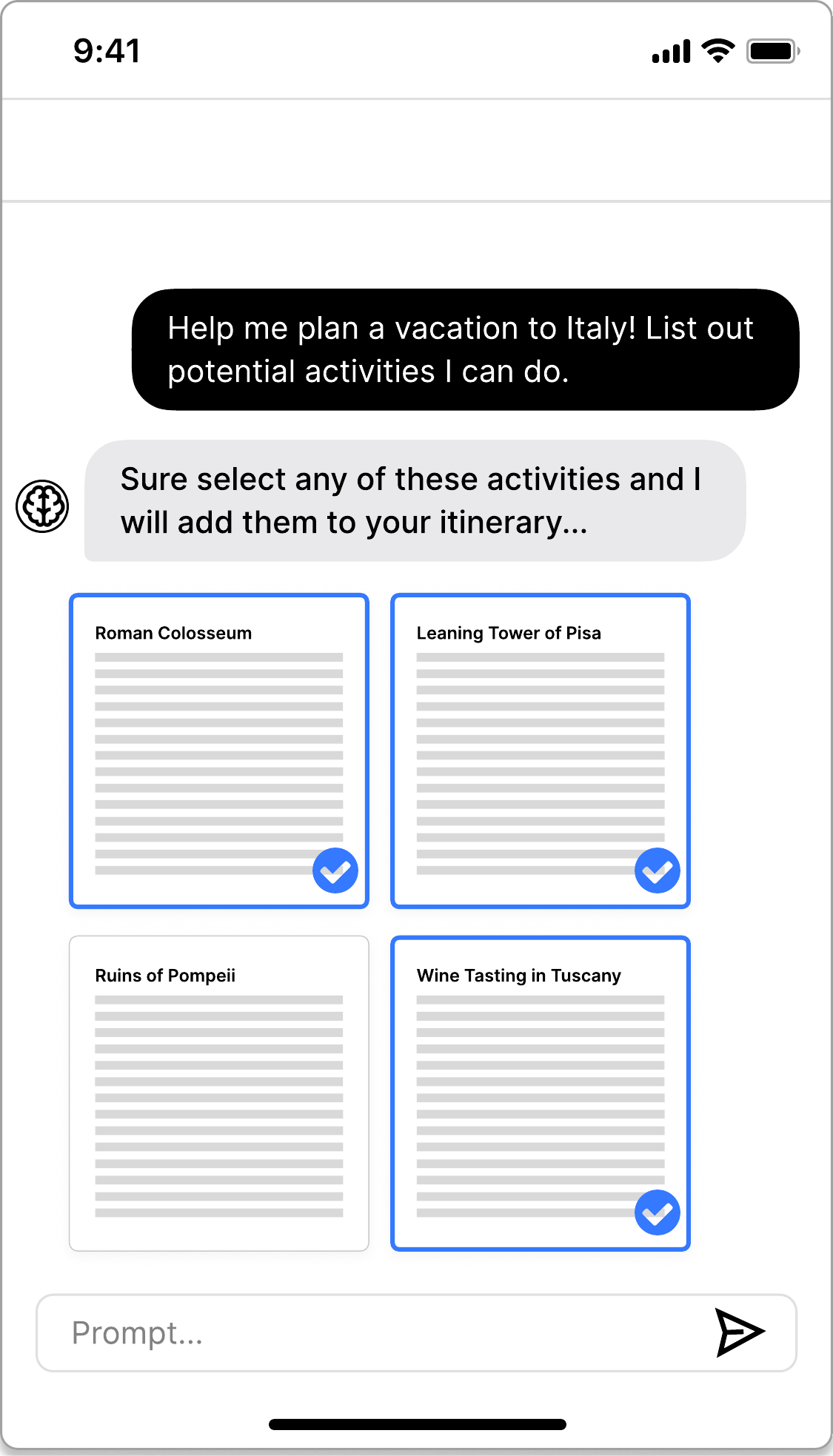}
\label{figure2b}
}
\subfigure[Lasso and Brush Selection]{
\includegraphics[width=0.23\linewidth]{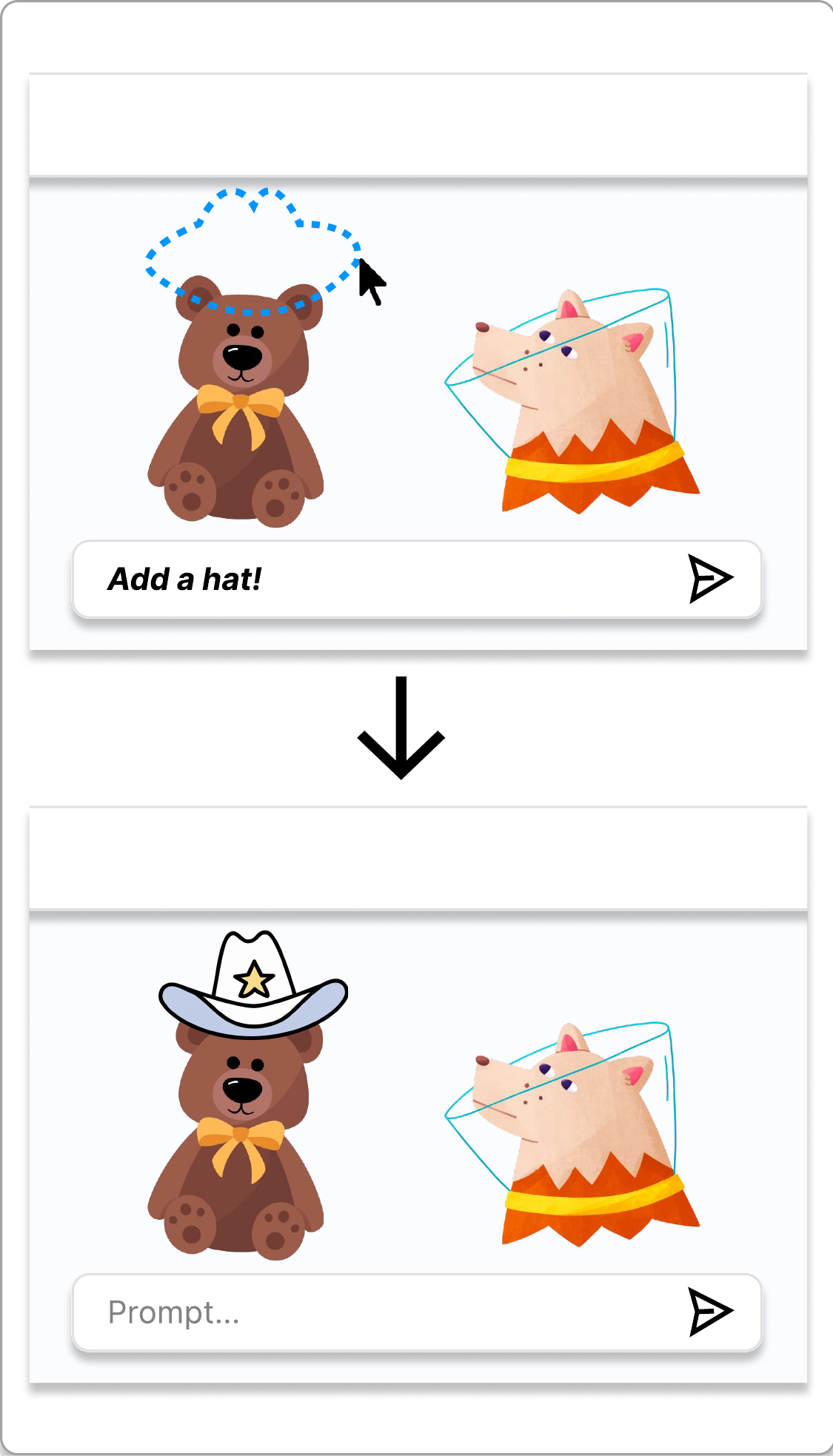}
\label{figure2c}

}

\caption{%
\textbf{Selection Techniques (Sec. \ref{sec:selecting}): Selecting, in terms of generative AI systems, consists of choosing or highlighting a specific UI element in order to further interact with it.}
}
\label{fig:selectionTechniques}
\vspace{-2mm}
\end{figure}

\begin{compactitem}

    \item \textbf{Single Selection  (Sec.~\ref{sec:Ssingle}; Fig. \ref{figure2a})}: A single-selection interaction consists of clicking or choosing a single GUI element that will be interacted with further. An example would be a user choosing one of 3 outputs that they wish to iterate on further.

    \item \textbf{Multi-Selection (Sec.~\ref{sec:Smulti}; Fig. \ref{figure2b})}: A multi-selection interaction consists of clicking or choosing multiple UI elements that will be interacted with further.

    \item \textbf{Lasso and Brush Selection  (Sec.~\ref{sec:Spaint}; Fig. \ref{figure2c})}:  Lasso and brush selections are selection techniques where a lasso or brush is used to create a bounding box that controls the region where a specific prompt is applied (Figure~\ref{figure2c}). 

\end{compactitem}

\subsubsection{Single-Selection} \label{sec:Ssingle}
A single-selection interaction consists of clicking or choosing a single UI element (\eg, selecting one of several outputs, choosing part of an image, etc) that will be interacted with further \citep{lee2023bogen, suh2023structured}. Luminate ~\citep{suh2023structured} is an LLM that intakes a prompt from an author writing a story and outputs several different story options based on the original prompt. Utilizing single-selection, the user clicks on and chooses which of the story options they want to further iterate on. So if a user prompts the system to write a short story about an astronaut, the system will output several stories. From there, the user can click and select the story that best fits their needs. Selection interactions, as opposed to prompting, are often utilized because of their ease of use and efficiency.

\subsubsection{Multi-Selection} \label{sec:Smulti}

\textbf{Multi-Select to Alter: } Multi-selection consists of interactions where users can select multiple elements simultaneously with the goal of interacting with multiple elements at once. For example, DirectGPT ~\citep{masson2023directgpt} allows the user to make one or more selections within an inputted text and create a tool that affects all of the selected words. For example, say a user is writing a story, but they feel that their story is a bit bland, and they want to make the vocabulary more descriptive. They can use multi-select the words that they want to make more interesting,  such as ``run'', "eat," and "said," and then they can type "replace with synonyms" in the prompt module. The system will take all of the words that they selected and replace them with more interesting words like "dashed," "devoured," and "exclaimed." In using the multi-select interaction, users can find and select multiple elements or words that they want to interact with further.

\textbf{Multi-Select to connect: } Furthermore, multi-selection is often used to select multiple outputs so that the user can connect them together \citep{padiyath2021desainer,jeon2021fashionq}. For example, \cite{padiyath2021desainer} is a generative fashion AI system that creates several dress outputs for a designer. The user can then use multi-select to select multiple aspects from different dresses and add them together to create a new dress. So, the fashion designer could select the sleeves they like from one dress, the corset they like from another, and the straps from another output and use all these parts of a dress to prompt the system to create new designs. In cases like this one, multi-select is used to select and add together multiple elements to prompt the generative AI system with the sum of those elements.

\subsubsection{Lasso and Brush Selection} \label{sec:Spaint}

\textbf{Inpainting: }Lasso and brush selection, also known as inpainting, allows users to use a tool to select very specific parts of a larger element. Whereas other selection methods make you choose an entire element, inpainting allows users to fine-tune and edit their selection. PromptPaint~\citep{chung2023promptpaint} focuses on enhancing the generation of images by brushing over the image to better control the generation process where multiple prompts can be mixed. Different prompts can be applied to different regions of the images. The paint-like interactions can be seen as a type of visual selection via lasso, where the lasso bounding box controls the region that a specific prompt is applied. Another work called PromptCharm~\citep{wang2024promptcharm} allows the user to create a mask $M$ over a generated image $I$ by brushing over the area of interest, and then a new image $I^{\prime}$ is generated by modifying only the pixels of $I$ that lie within the mask $M$. The user can guide the inpainting process by typing a text prompt $\vx_M$ that is applied only to the masked region of $I$.

\begin{figure}[t]
\centering
\subfigure[Menus]{
\includegraphics[width=0.23\linewidth]{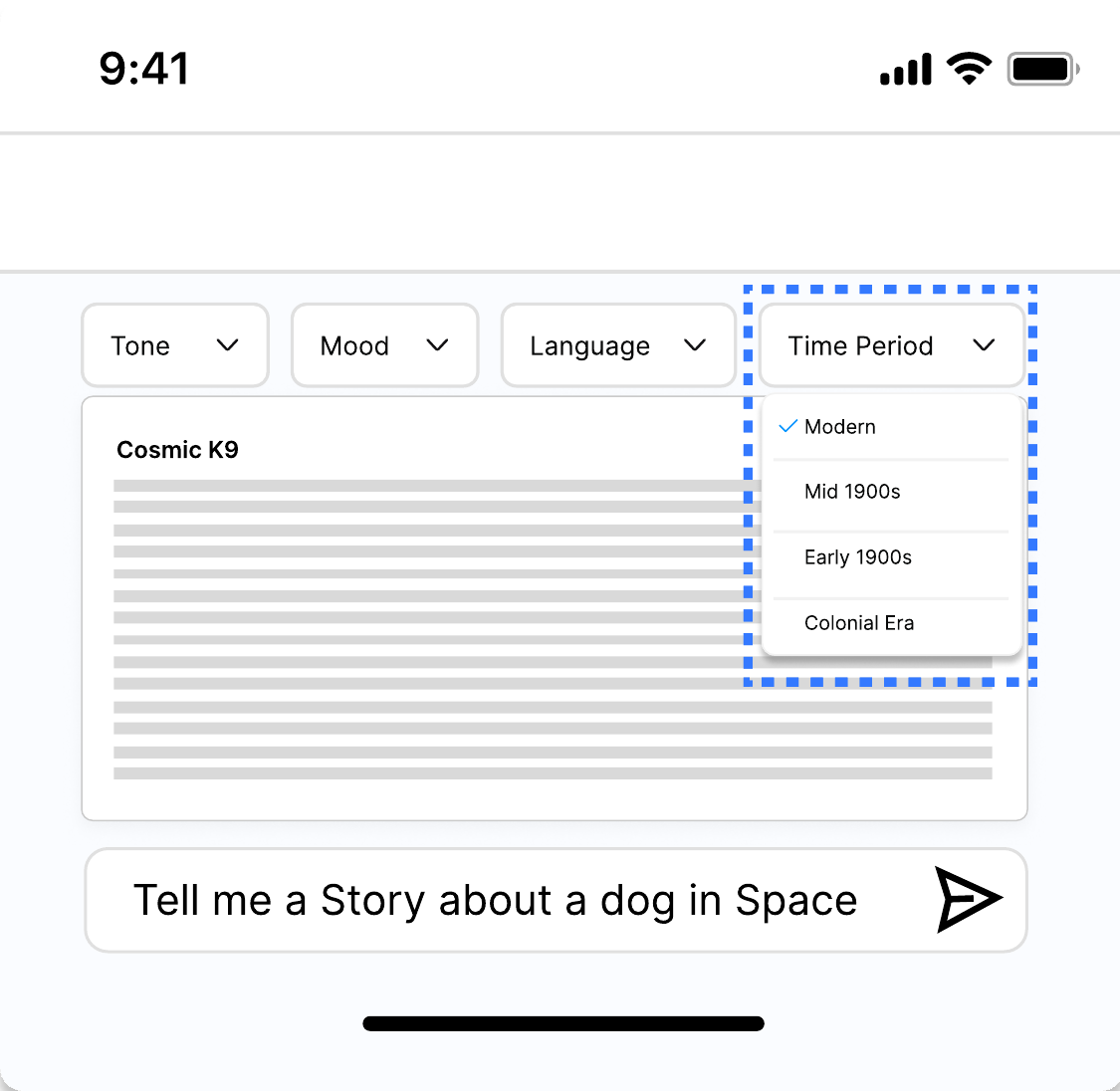}
\label{figure3a}
}
\subfigure[Sliders]{
\includegraphics[width=0.46\linewidth]{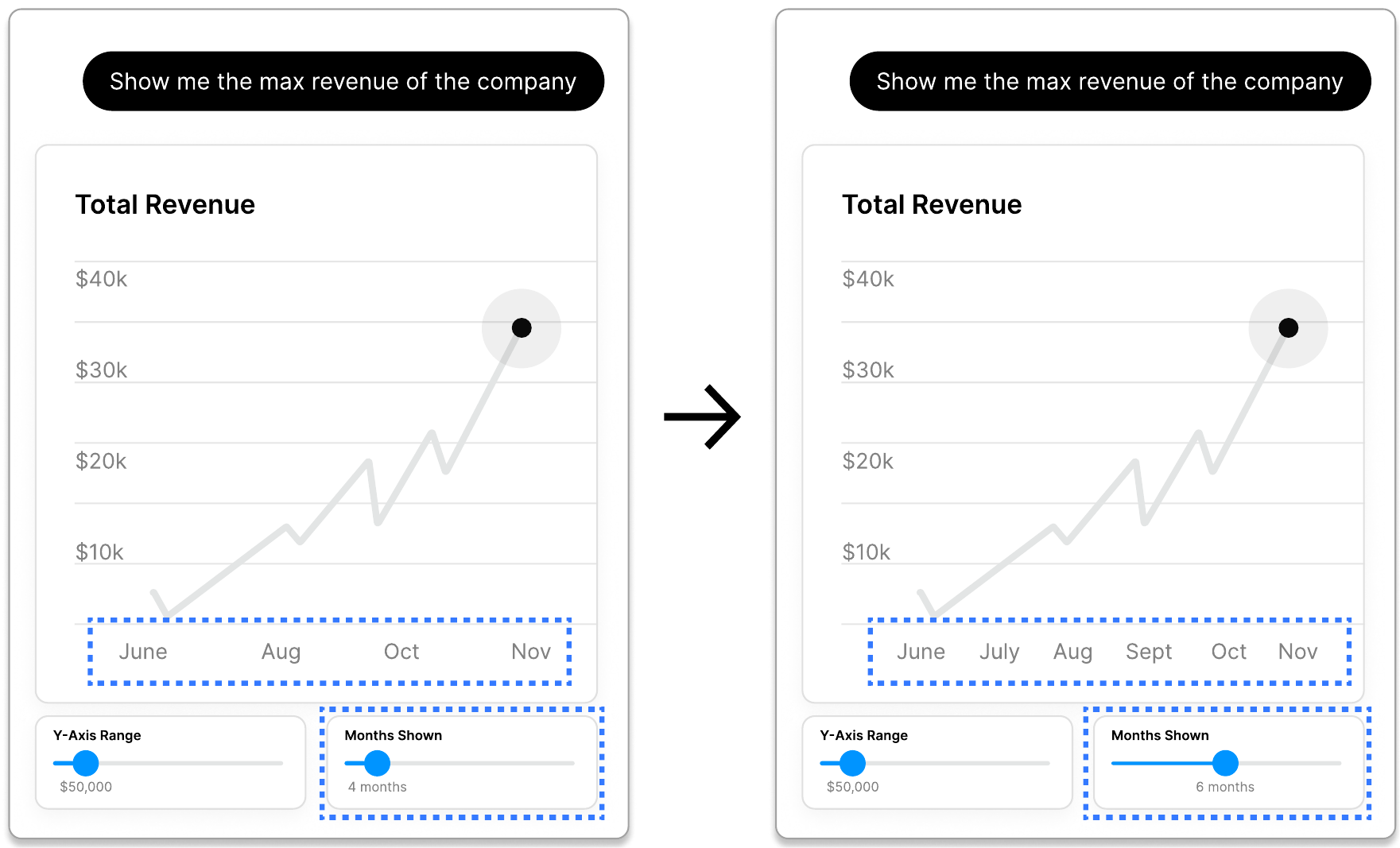}
\label{figure3b}
}
\subfigure[Explicit Feedback]{
\includegraphics[width=0.23\linewidth]{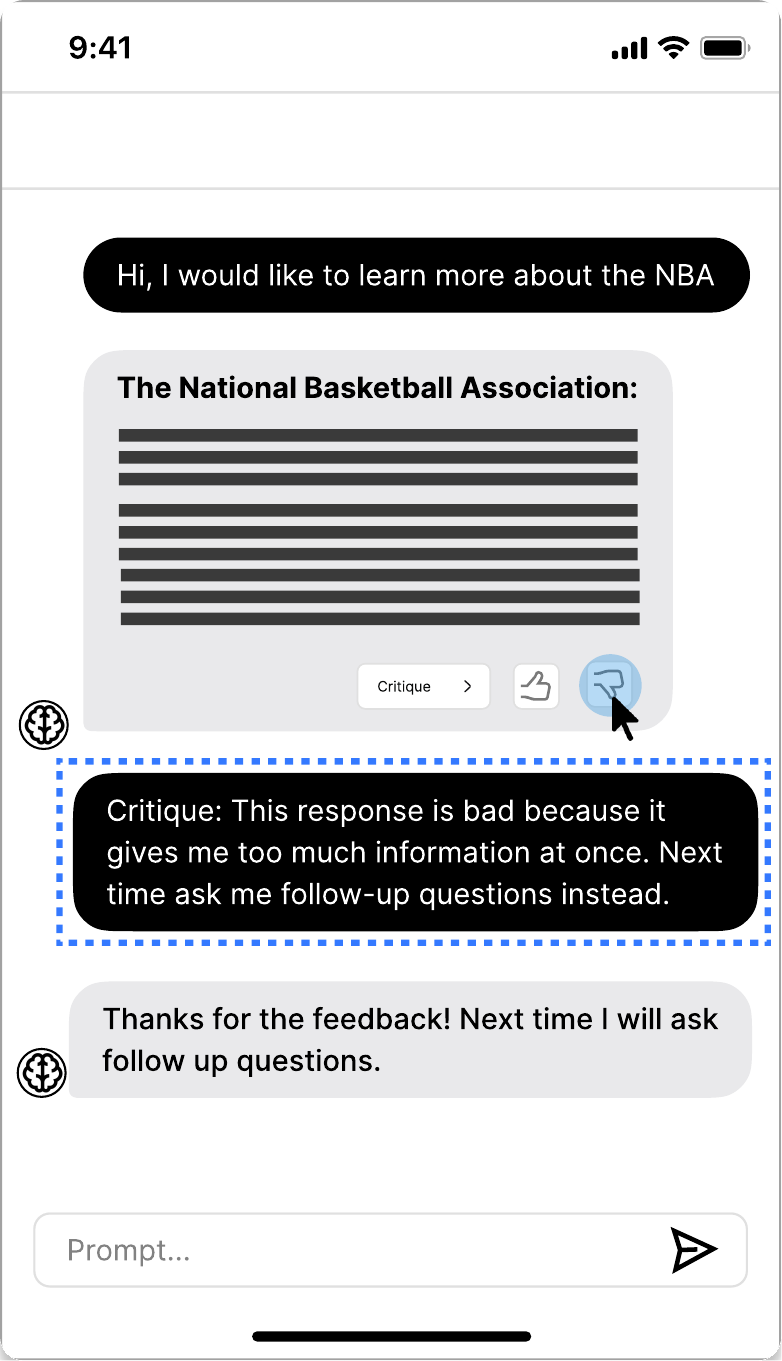}
\label{figure3c}
}
\caption{%
\textbf{System and Parameter Manipulation (Sec. \ref{sec:manipulating}): User interaction techniques that allow the user to adjust the parameters, settings, or functions of an overall generative AI system.}
}
\label{fig:sysPara}
\vspace{-2mm}
\end{figure}

\subsection{System and Parameter Manipulation}\label{sec:manipulating}

\textbf{Definition \& Scope:} The System and parameter manipulation category consists of user interaction techniques that allow the user to adjust the parameters, settings, or functions of an overall generative AI system. These interactions are often used to personalize generated outputs to meet the needs of a user. Some examples of this type of user-guided interaction consist of Menus (Section~\ref{sec:MParameter}), Sliders  (Section~\ref{sec:MSliders}), and  Explicit Feedback (Section~\ref{sec:MExplicitFeedback}).

\begin{compactitem} 
    \item \textbf{Menus (Sec.~\ref{sec:MParameter}; Fig. \ref{figure3a})}: when a user either inputs their own parameters or chooses from preset options to change the parameters of the generative process. An example would be a user using dropdown menu options to alter the output parameters.

    \item \textbf{Sliders (Sec.~\ref{sec:MSliders}; Fig. \ref{figure3b})}: A UI element that can be "slid" to adjust the parameters of the generative AI system.

    \item \textbf{Explicit Feedback (Sec.~\ref{sec:MExplicitFeedback}; Fig. \ref{figure3c})}:  Explicit feedback (i.e., thumbs up/down, written critiques, etc.) is a user interaction that is used to expressly personalize the system to the user's preferences.

\end{compactitem}

\subsubsection{Menus} \label{sec:MParameter}
\textbf{Preset Option Menus:} Menus are user-guided interaction features that allow users to either input their own parameters or choose from preset options ~\citep{wang2024weaver,suh2023structured, 10.1145/3586183.3606737,kim2023metaphorian}. A common example of a parameter-selection menu is a drop-down menu that allows users to adjust the parameters of a system with preset parameters. An example of this can be seen in ~\cite{suh2023structured}, a system that helps users write stories or poems. In this system's UI, there are drop-down menus that allow users to impact the output's tone, mood, and overall structure. In menus like these, the user can choose between pre-set parameters to adjust the overall output. Being able to utilize UI features to quickly adjust output parameters drastically speeds up the user experience and lowers the user's cognitive load.

\textbf{Input Menus:} While drop-down menus are a common interaction used to adjust parameters, some systems have menus that solely rely on manual user input. These menus allow the user to input text parameters that work in tandem with the prompt to create a concise output. For example, in ~\cite{10.1145/2984511.2984588}, users can create a data visualization given a data file input and can adjust the generated data visualization by changing typing in different parameters. So if a user prompts the system to "map out all earthquakes in California," there is a parameter menu next to visualization where they can manually type parameters like time range, location range, etc. that will affect the generated visualization. Similarly, PromptCharm~\citep{wang2024promptcharm} is an image generation platform that allows users to generate images with text prompts and menus. In terms of menus, for example, PromptCharm has an "image style" menu where a user can input text separate from the original prompt that allows them to specifically impact the image style. Overall, menus allow users to add another dimension to their already existing prompts and guide them to manipulate specific parameters within the generative process.

\subsubsection{Sliders}\label{sec:MSliders}

\textbf{Range Adjustments:} 
Sliders are visual UI elements that are able to manipulate the parameters of the generative AI system \citep{chung2023promptpaint, 10.1145/2984511.2984588,wu2022promptchainer} For example, in Eviza \citep{10.1145/2984511.2984588}, sliders are used to control the parameters of the type of data that is being shown. Eviza is a data visualization software that overlays data onto a map and the user can interact with the sliders to affect the types of data that is being shown. So if the system generates a historical map of earthquakes in California, the user can manipulate the output by using the sliders to ensure that only earthquakes with magnitudes greater than 5.0 are shown. Essentially, in cases like this one, sliders are used to manipulate data being and work in tandem with the original prompt.

\textbf{Attention Weights:} During the generation process, PromptCharm~\citep{wang2024promptcharm} allows users to select specific tokens in the prompt to then adjust their attention weights via a slider that is dynamically shown for the token selected. This is another user-guided interaction to better control the generated image and how the model attends to the specific tokens in the prompt, which can be viewed as importance/influence scores. Hence, the user can ensure that the tokens most important to the user are also consistent with the importance in the model, giving another axis of fine-grained controllability during the generation process. This can also be viewed as an explainability feature that can help the user refine the prompt directly as well (\eg, tokens in the current prompt can be highlighted via a colormap that corresponds with the attention weights of each, and then the user can modify the prompt by adding, removing, or rephrasing the prompt accordingly).

\subsubsection{Explicit Feedback}\label{sec:MExplicitFeedback}

\textbf{Binary Feedback: }A user's explicit feedback is a user interaction type that can change a system's parameters and, therefore, the generative system as a whole~\citep{petridis2024constitutionmaker,achiam2023gpt}. Giving feedback to the system can be anything from giving an answer a thumbs up or thumbs down \citep{achiam2023gpt}, to explicitly instructing the system to respond a certain way, i.e. "Critique: when I talk about my mom, ask how she is doing" \citep{petridis2024constitutionmaker}. For example, systems like \cite{achiam2023gpt} offer binary feedback on whether the answer that the generative model gave was good or bad. More specifically, the system provides a thumbs up or down option to allow the user to give explicit feedback as to whether or not they were satisfied with the outputted answer. 

\textbf{Comprehensive Feedback: }Meanwhile, systems like \cite{petridis2024constitutionmaker} allow users to give more specific feedback than just a binary "good" or "bad" response. Constitutionmaker allows users to rate an output by giving it kudos, critiquing it, or rewriting it. As seen in Figure ~\ref{figure3c}, a user can write out a specific critique, such as "Critique: This response is bad because it gives me too much information at once. Next time ask me follow-up questions instead." The system will then take this feedback and add it to its own constitution, and all of the subsequent outputs will follow the rules set forth in the constitution. Essentially, feedback techniques allow a user to adjust the parameters of a system to their liking.

\subsection{Object Manipulation and Transformation}\label{sec:Object}

\textbf{Definition \& Scope: }Object manipulation and transformation interactions involve the user directly modifying, adjusting, or transforming a specific UI element. Doing so gives the user a deeper control over the system and allows them to interact with the UI elements in a unique and novel way. Some examples of this type of user-guided interaction consist of Dragging and Dropping UI elements (Section~\ref{sec:ODragnDrop}) and Resizing a UI element within the system, and (Section~\ref{sec:OResizing}).

\begin{figure}[t]
\centering
\subfigure[Drag and Drop]{
\includegraphics[width=0.23\linewidth]{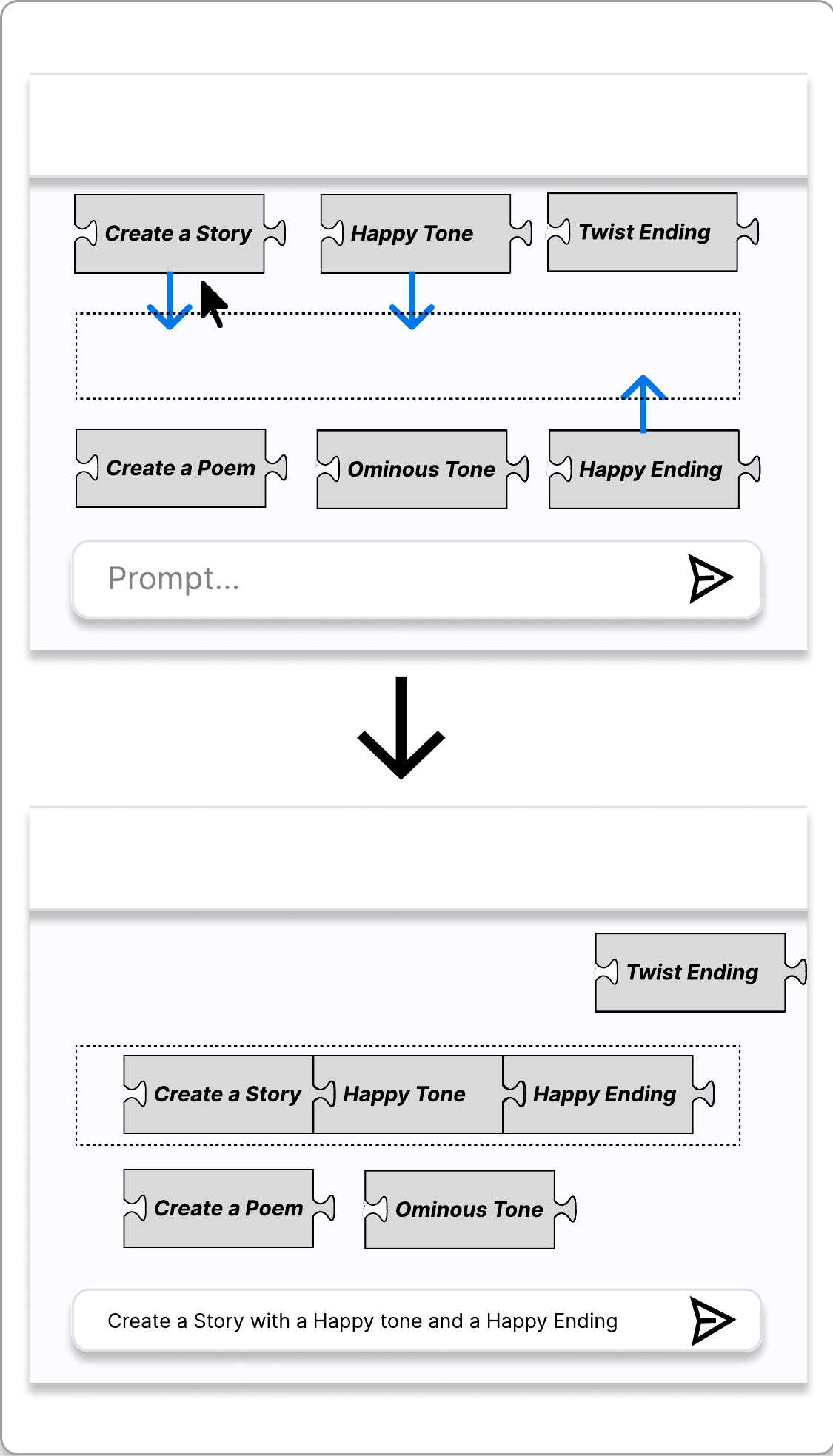}
\label{figure4a}
}
\subfigure[Connecting]{
\includegraphics[width=0.23\linewidth]{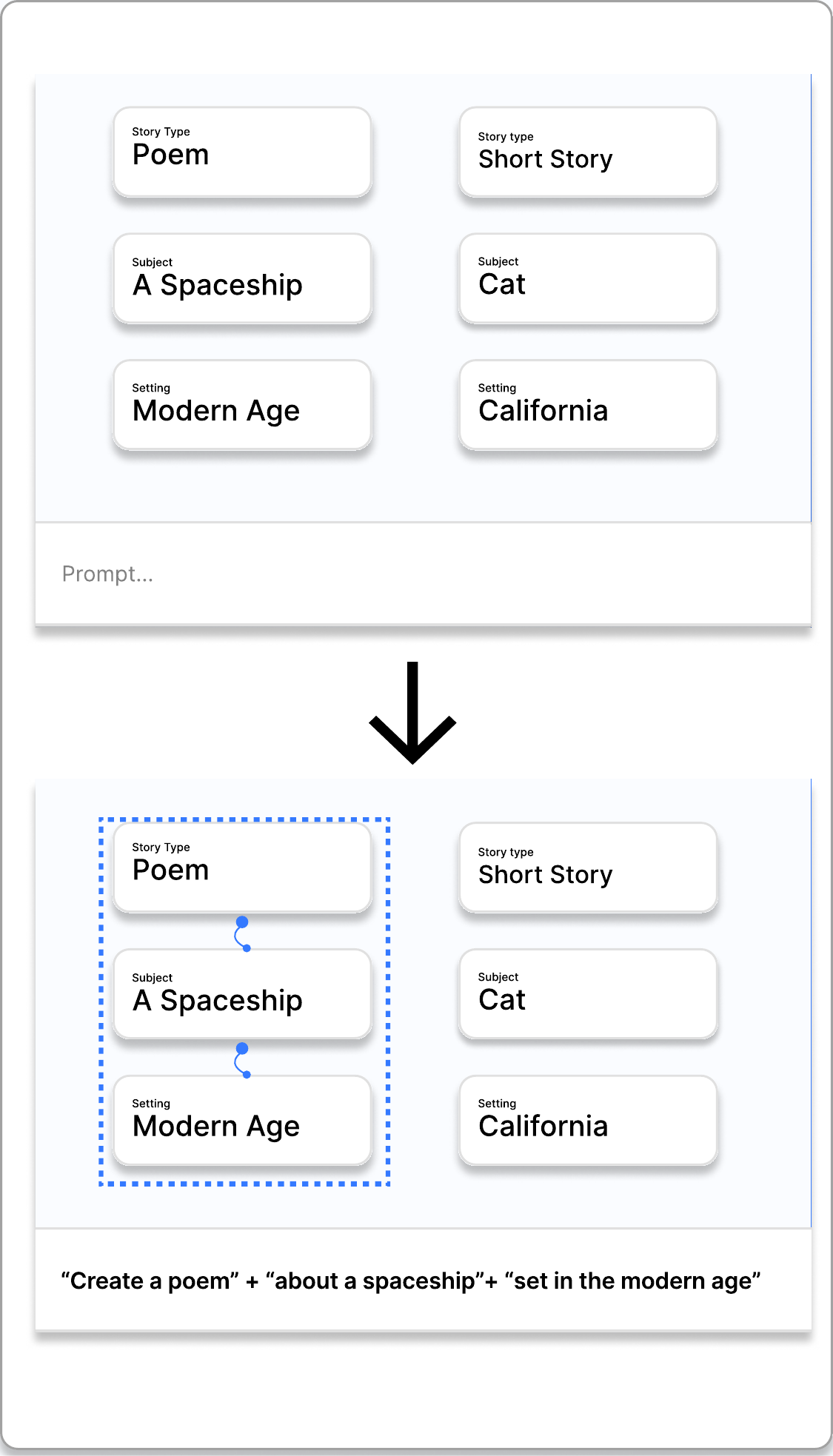}
\label{figure4b}
}
\subfigure[Resizing]{
\includegraphics[width=0.23\linewidth]{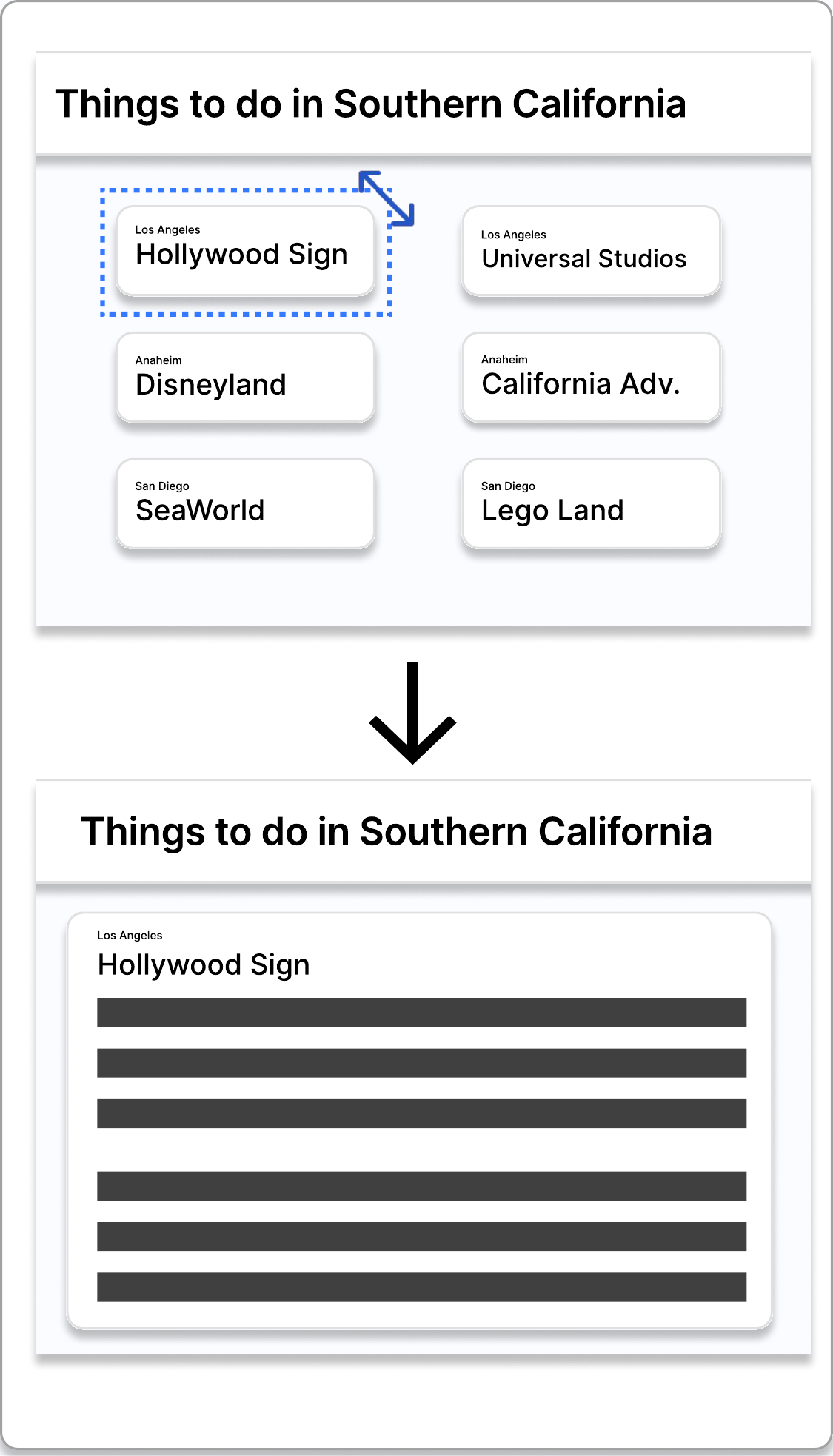}
\label{figure4c}
}
\caption{%
\textbf{ Object Manipulation and Transformation (Sec. \ref{sec:Object}): User interaction techniques that modify, adjust, and/or transform a specific UI element, like a building block, puzzle piece, or similar entity. }
}
\label{fig:ObjManTrans}
\vspace{-2mm}
\end{figure}

\begin{compactitem}
    \item \textbf{Drag and Drop  (Sec.~\ref{sec:ODragnDrop}; Fig. \ref{figure4a})}: Moving an element to a specific location or in a way that manipulates the generative system.

    \item \textbf{Connecting (Sec.~\ref{sec:OCombining}; Fig. \ref{figure4b}}) In connecting
    interactions, a user stacks and connects two UI elements in a way that affects the overall generative process.

     \item \textbf{Resizing  (Sec.~\ref{sec:OResizing}; Fig. \ref{figure4c})}:  Altering the size of a UI element in a way that changes its function in the generative process.

\end{compactitem}

\subsubsection{Drag and Drop}\label{sec:ODragnDrop}

\textbf{For creation:} Drag and drop interactions consist of moving an element to a specific location or in a way that manipulates the generative system. One work by~\cite{masson2023directgpt} leverages drag-and-drop interactions to help in modifying a vector graphic, \eg, they type ``add a line from'', then use drag-and-drop on the vector graphic to refer to the specific locations, which are then included in the prompt being created via direct manipulations.

\textbf{To Connect} Dragging and dropping, as an interaction, often goes hand in hand with the subsequent user interaction technique, connecting (Sec. \ref{sec:OCombining}) \citep{kim2023lmcanvas,10.1145/3613904.3641920}. Oftentimes, users will drag and drop UI elements toward each other with the end goal of connecting them. Take Jigsaw ~\citep{10.1145/3613904.3641920} for example, a system that created a set of puzzle pieces that each have a corresponding system instruction that the system can complete. The user can then drag and drop the puzzle pieces to combine them, which simulates connecting system instructions within a prompt. For example, a user might Connect a puzzle piece that has the prompt "Upload an image" with two other puzzle pieces that say "remove people" and "increase resolution." By dragging and dropping these puzzle pieces together, the user is essentially creating a prompt without having to type it out. Dragging and dropping puzzle pieces with different system instructions on them helps spur novel prompt creations and helps inform the user what the application is capable of.

\textbf{To Disconnect} While the drag and drop interaction is often used to connect two UI components, often to prompt a system in a unique way \citep{10.1145/3613904.3641920, kim2023lmcanvas}, dragging and dropping can be used to disconnect components. LMCanvas \citep{kim2023lmcanvas} is a system that uses sentence blocks as starting points that users can interact with and generate content. In LMCanvas \citep{kim2023lmcanvas}, users can write a sentence in a sentence block,  use multi-select to select half of the sentence, and then drag that half of the sentence away from the other half of the sentence. In doing so, the user has split the original sentence block, creating two sentence blocks that they can further interact with. Essentially, drag and drop to split is another way that users can interact with prompts in the generative system. All in all, object manipulation techniques, such as this one, may be more appealing to users who are more hands-on or are not as familiar with the capabilities of generative AI.

\subsubsection{Connecting}\label{sec:OCombining}
\textbf{Prompt-to-input:} As mentioned, connecting goes hand in hand with drag-and-drop interactions. In connecting interactions, a user stacks and connects two UI elements in a way that affects the overall generative process. For example, \cite{kim2023lmcanvas} aims to facilitate an easier writing process by allowing users to create prompts and input "blocks" that can be connected to simulate connecting a prompt to an input. For instance, there may be a prompt/model block that says, "translate this story from English to French." Then, a user can drag an input block with an English story onto that system instruction block, and the system will recognize this as the user prompting the system to complete the task or translating the inputted story to French. Interactions like these are beneficial for visual or kinesthetic learners who may not be familiar with the basic functionalities of traditional generative AI models like ChatGPT.

\textbf{Prompt-to-prompt:} A similar application that has already been talked about, Promptpaint \citep{chung2023promptpaint}, allows users to combine existing prompts together in a novel way. As mentioned, Promptpaint is a generative AI system that allows users to utilize multiple prompts to create an image. A user can pick and choose which prompts on a prompt list they want to apply to a generated image. An interesting feature that Promptpaint has is \textit{prompt mixing,} where a user can add prompts to a painter's palette where each prompt is assigned a sort of color blob, just as a normal painter's palette would have. From there, the user can "connect" the prompts together, which essentially connects the contents of the prompt to create an image. This visual representation of connecting prompts helps users visualize the connecting of two often dissimilar prompts.

\subsubsection{Resizing}\label{sec:OResizing}
 \textbf{General resizing}In terms of generative UI interactions, resizing an object can have drastically different outcomes depending on the system. \cite{10.1145/3586183.3606756}'s proposed system, Senescape, is a generative AI system that utilizes a resizeable canvas to make digesting information much easier for the user. In Senescape, the user can resize the view port of the canvas to adjust the amount of information that is generated and subsequently shown to them. By essentially zooming in on an element on the canvas, the system will generate more detailed information about the respective element. Whereas if a user zooms out, they will see only high-level information. In essence, using a resizing user-interaction technique allows users to control the amount of information that is shown to them. Doing so significantly reduces cognitive load and effectively gives the user control over the system.

\section{User Interface Layouts for Generative AI} \label{uilayouts}

\textbf{Definitions and Scope:}  In this section, we propose a taxonomy based on different high-level user interface (UI) categories, looking at the overall structures of generative AI user interfaces. Generally, a user interface (UI) is defined as a means by which a human and a computer communicate with one another \citep{norman1988psychology, chignell1990taxonomy}.
In the context of generative AI, we organize UIs into the following categories:
conversational user interface (Section~\ref{sec:UIConversational}),  canvas user interface (Section~\ref{sec:UICanvas}),  modular user interface (Section~\ref{sec:UImodular}), and simulated user interfaces (Section~\ref{sec:UISimulated}).

\begin{figure}[t]

\centering
\includegraphics[width=.99\linewidth]{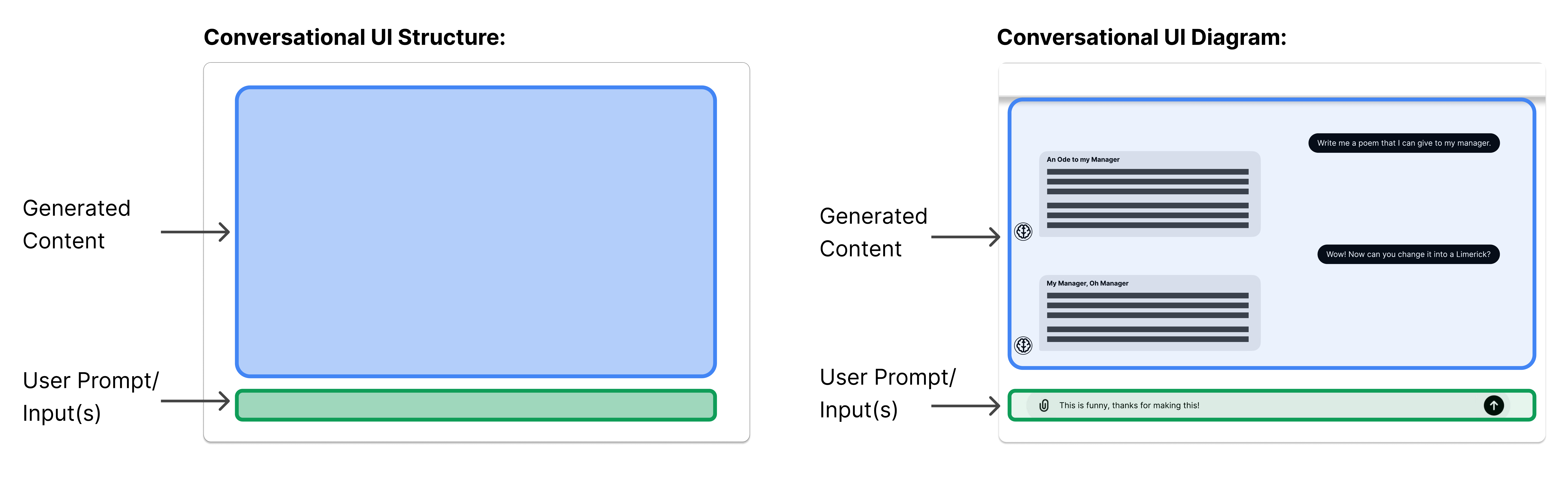}
\caption{%
\textbf{Conversational UI: A conversational UI is structured so that a user interacts with the user prompt/input box. From there, their output(s) and output history exist in a larger space within the UI (Sec.~\ref{sec:UIConversational}).}}

\label{fig:ConversationalUI}
\end{figure}

\begin{figure}[t]

\centering
\includegraphics[width=.99\linewidth]{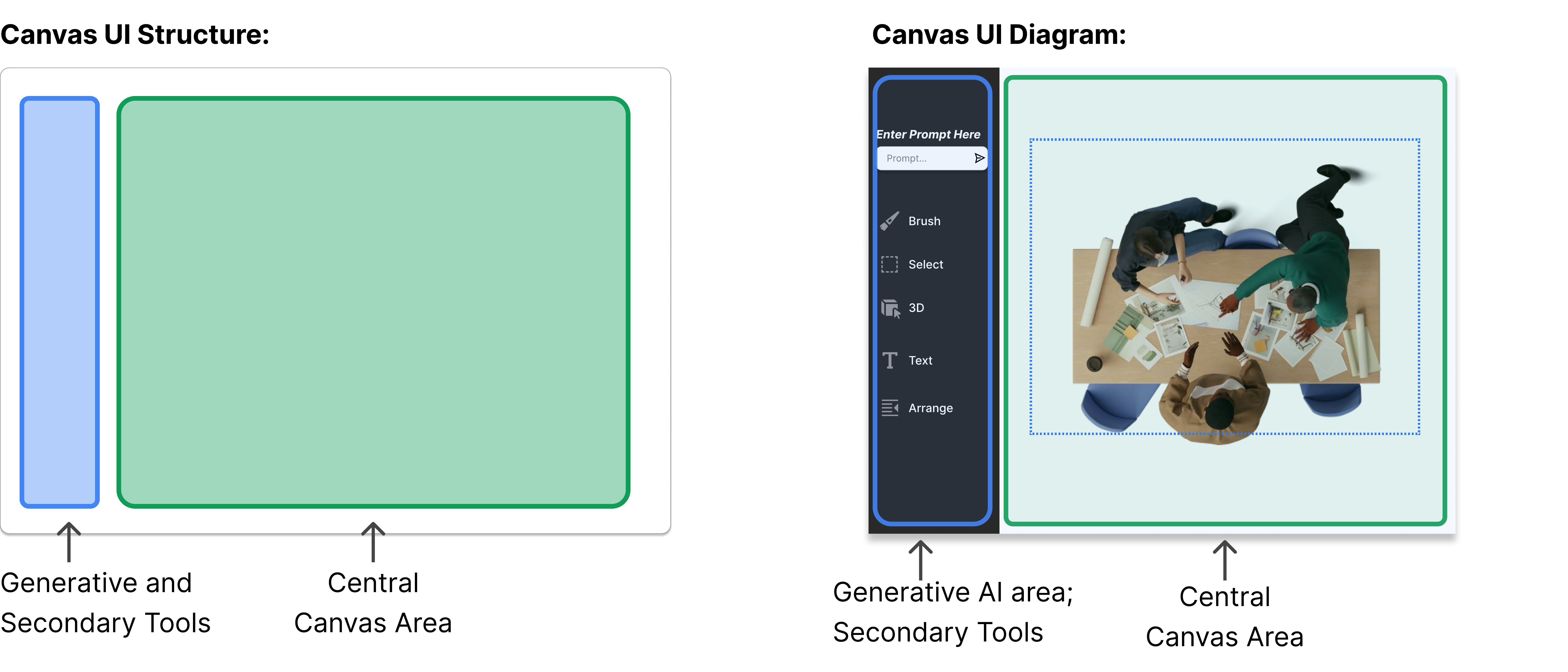}
\caption{%
\textbf{Canvas User Interface: A UI structure with a central canvas area that houses the primary content. The generative and other tools are often in the periphery or off to the side. (Sec.~\ref{sec:UICanvas}).}
}
\label{fig:CanvasUI}
\end{figure}

\begin{compactenum}
   \item \textbf{Conversational User Interface (Sec.~\ref{sec:UIConversational}, Fig. \ref{fig:ConversationalUI})}: A conversational interface is an input-output based interface where users interact with the AI on a turn-based cadence. This structure mimics human conversation, allowing users to ask questions, input media, and receive responses in a sequential format. Examples include AI assistants like chatbots, zero-shot agents, and other turn-based AI agents.

   \item \textbf{Canvas User Interface (Sec.~\ref{sec:UICanvas}), Fig. \ref{fig:CanvasUI})}: A canvas interface often revolves around generating and interacting with a central piece of content. In this interface, the content in question is often in the center of the UI, and generative tools are often on the peripheral. This interface category includes generative systems that edit and generate content like images, videos, word documents, code, data visualizations, and audio.

   \item \textbf{Contextual User Interface (Sec.~\ref{sec:UIContextual}, Fig. \ref{fig:ContextualUI})}: Contextual UIs consist of a user interface where the generative element is in a smaller UI element that exists, structurally, in line with a particular part of the larger UI. Oftentimes, this contextual generative UI element exists near where the user is interacting with the larger application.
   
   \item \textbf{Modular User Interface (Sec.~\ref{sec:UImodular}, Fig. \ref{fig:ModularUI})}: Modular UIs consist of user interfaces that are broken up into multiple main interaction areas, with each of these interaction areas having a different function in the generative process. Modular user interfaces are often used in systems with multiple levels of generation.

   \item \textbf{Simulated Environment (Sec.~\ref{sec:UISimulated})}: A simulated environment UI replicates real-world or hypothetical scenarios in a virtual space, allowing users to interact with and navigate through these scenarios as if they were real. This type of UI is commonly used in virtual reality (VR), augmented reality (AR), and training simulations, where users can immerse themselves in the environment and interact with objects and elements within it. The goal is to provide a realistic and interactive experience that can be used for training, entertainment, or exploration.

 \end{compactenum}

\textbf{Motivation:} We include this taxonomy as a compendium that designers can use when designing a new generative system. The UIs that follow all have specific use cases where they serve the user best, and using the correct UI in the appropriate scenarios will create a better user experience overall. The goal of this section is to explain how and when each UI structure can be used and to give real-world examples of how these UI structures are being used today.

 \subsection{Conversational User Interface} \label{sec:UIConversational}
\begin{figure}[t]
\centering
\includegraphics[width=.99\linewidth]{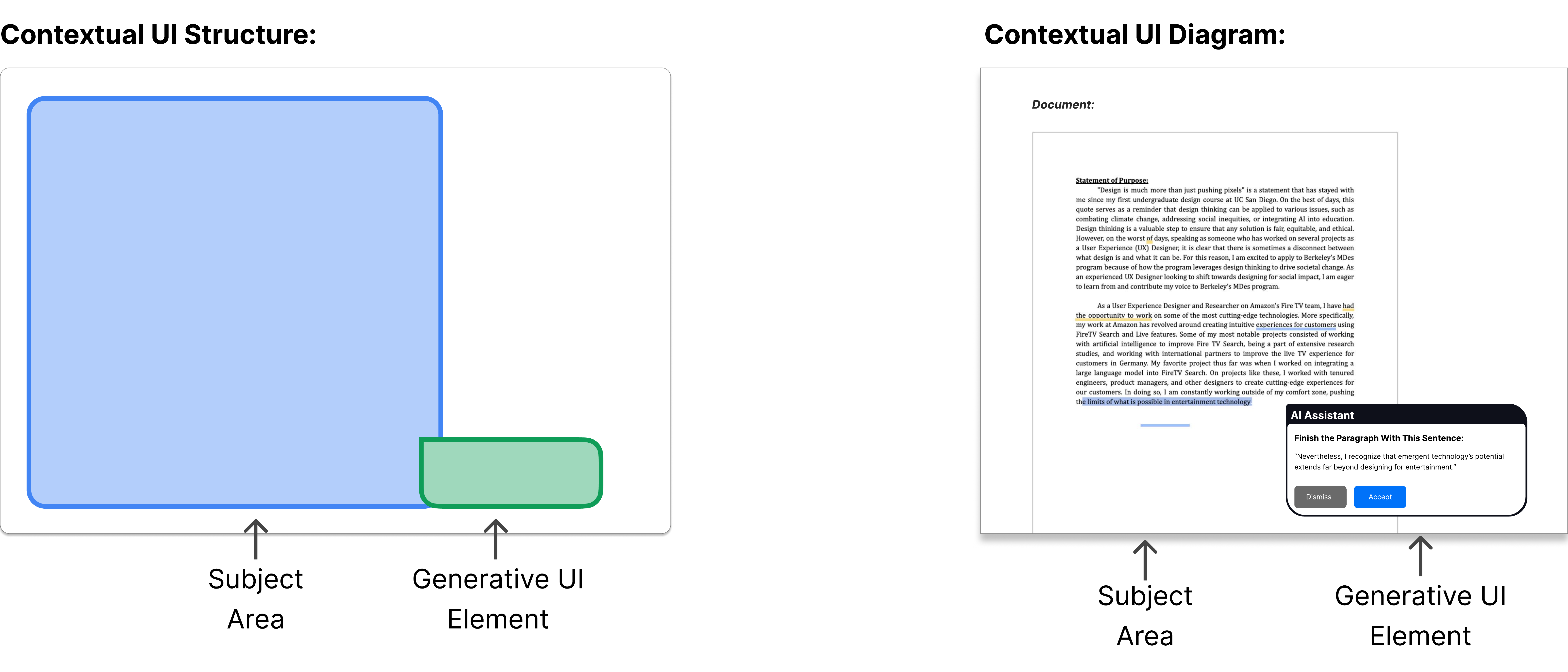}
\caption{%
\textbf{Contextual User Interface: UI Structures where the generative UI element appears inline in context to the primary user interaction space (Sec.~\ref{sec:UIContextual}).}}
\label{fig:ContextualUI}
\end{figure}

 A conversational UI is characterized as a user interface that is visually structured in a way that mimics a conversation \citep{lister2020accessible,achiam2023gpt,fu2023challenger,miller2017parlai}, with there being an exchange between a user and the AI system following a turn-based cadence. This interaction space often consists of an input or prompt box along with a section of the UI for interaction history (Fig. \ref{fig:ContextualUI}). Furthermore, this category will focus on conversational user interfaces within GUIs, and not VUIs like Amazon Alexa and Apple's Siri. 

In terms of generative AIs, visual conversational user interfaces primarily have the same or similar user interface structures. Generally, much of the focus is on an input or prompting box where a user can prompt the system to complete a certain task or ask it a question \citep{achiam2023gpt,fu2023challenger}. In conversational UIs a majority of user-guided interactions will occur in this prompt or input box, as it is the primary interaction space. Given this, the secondary interaction space in conversational user interfaces is the chat history or output section of the UI (Fig. \ref{fig:ConversationalUI}). Given a prompt, the output or chat history section houses the output and/or a chat history of past interactions. This section can store anything from past input and act as a chat history \citep{fu2023challenger,achiam2023gpt} or even hold a single or a gallery of user outputs that a user can select from \citep{betker2023improving, lee2023bogen, wang2024aesopagent}.

Although conversational UIs are mostly constrained by a turn-based cadence and an input-and-output model, they are capable of a variety of use cases. For example, systems like \cite{fu2023challenger}'s Gemini can be used to generate text and have general conversations, while systems like \cite{lee2023bogen}'s BOgen can generate a gallery of 3D models. Both of these scenarios also encapsulate one of the major strengths of conversational UI. Conversational UIs excel at asking the system to make continuous incremental changes as the generative process goes on \citep{achiam2023gpt, betker2023improving, fu2023challenger, liu2023interngpt}. Similarly, conversational UIs excel at recalling information from earlier in the conversation to inform more recent outputs. For example, systems like \cite{alayrac2022flamingo}'s Flamingo learn from earlier conversations and inputs to perform one-shot tasks where they learn how to identify images and patterns using past chat history. This is a strength of visual conversational UIs, as both the system and the user can reference earlier conversations. All in all, conversational UIs are versatile in that they can be used to perform a variety of different tasks.

 \subsection{Canvas User Interface} \label{sec:UICanvas}
 As it pertains to the user interfaces of generative AI, we define canvas-focused user interfaces as those that are structured with a focus on a central canvas area within the UI. Structurally, a majority of the user interactions occur in this central canvas, with the generative tools existing in the periphery (Fig. \ref{fig:CanvasUI}). Canvas-focused user interfaces are broken into two subcategories: content canvases and information visualization canvases.

 \textbf{Content Canvases}: Content Canvas user interfaces are a subcategory of canvas UIs Where the primary canvas area is occupied by a piece of content such as an image or drawing \citep{chung2023promptpaint, liu20233dalleintegratingtexttoimageai, lawton2023tool, du2024deepthink, padiyath2021desainer} a set of text \citep{shi2022effidit}, code \citep{ross2023programmer, prather2023s, barke2023grounded}, or data visualizations \citep{10.1145/2984511.2984588}. An example of a content canvas is \cite{du2024deepthink}'s DeepThInk. DeepthInk \citep{du2024deepthink} is an AI art therapy tool that enables the average person to draw in art therapy sessions. Its UI is structured, so there is a central canvas, and the generative and other art tools are on the periphery. So the user essentially primarily interacts with the art in the middle of the screen on the central canvas and can elicit help from the generative tools on the periphery whenever it is appropriate. By using this structure, most of the interaction is funneled to the central canvas, whereas secondary interactions, like AI generation, occur on the periphery (Fig. \ref{fig:CanvasUI}).

\textbf{Information Visualization Canvases}: Information visualization canvases focus more on visualizing input and output interactions on a central canvas. While content canvases focus on a central piece of content, information visualization canvases are essentially sandboxes where users can directly manipulate components that alter the system. These systems are used to help users visualize the interplay between either the inputs \citep{10.1145/3613904.3641920,masson2023directgpt, kim2023lmcanvas} or the outputs \citep{suh2023structured, kim2023metaphorian, 10.1145/3586183.3606737, 10.1145/3586183.3606756}. Take \cite{10.1145/3586183.3606737}'s Graphologue, for example. Graphologue \citep{10.1145/3586183.3606737} is a system that takes common prompts like "plan a vacation for me in San Diego" and outputs a treemap that represents an itinerary for said trip. Whereas conversational UIs (Section \ref{sec:UIConversational}) primarily output a block of text, visualization canvases UIs output the same information in an easy-to-digest chart, graph, or other visualization. Doing so lowers the cognitive load needed to digest the outputted information and allows the user to interact with specific parts of the output as needed. In the case of Graphologue \citep{10.1145/3586183.3606737}, the system generates a canvas that visualizes the output as a hierarchical treemap in which every "branch" represents a part of the itinerary. Each branch can be expanded to reveal a list of restaurants, a list of hikes, etc.. Again, this type of UI can be utilized to break large pieces of information into easy-to-digest segments, lowering the cognitive load of the user

\subsection{Contextual User Interface} \label{sec:UIContextual}
 As it pertains to generative AI, a contextual user interface is a type of UI where the generative interaction occurs in line with a specific aspect of the larger subject area (Fig. \ref{fig:ContextualUI}). Unlike canvas UIs, contextual UIs's generative action does not occur in the periphery and instead occurs in line where the user is most likely looking. Oftentimes, the generative UI elements within contextual UIs appear unprompted, but instead, as a result of something the user does within the subject area \citep{Jakesch_2023, chang2023citesee, fitria2021grammarly, packer2024memgpt}. Other times, this UI structure is used to put prompting or other toolbar interactions contextually within the subject area, as seen in \cite{masson2023directgpt}. All in all, this type of UI structure is an effective strategy for lowering the cognitive load of the user as it displays relevant interactions in context to the user.

 Take, for example, \cite{packer2024memgpt}'s MemGPT, a generative AI system that contextually gives reminders to the user about the person that they are talking to. This system works contextually within a messaging application and "remembers" relevant facts about the person the user is speaking with. Then, at appropriate times, a pop-up window appears that reminds the user of relevant information about the person they are chatting with. So, for example, if a user was talking to their friend on their birthday, the system may contextually remind them to say happy birthday and can go so far as to tell them what birthday plans this person might enjoy doing. All of these interactions happen contextually within the system and occur in what is essentially another chat bubble. Contextual user interfaces are especially useful in situations such as this one, where the generated output is relevant in context to a specific unprompted input.

 \begin{figure}[t]

\centering
\includegraphics[width=.99\linewidth]{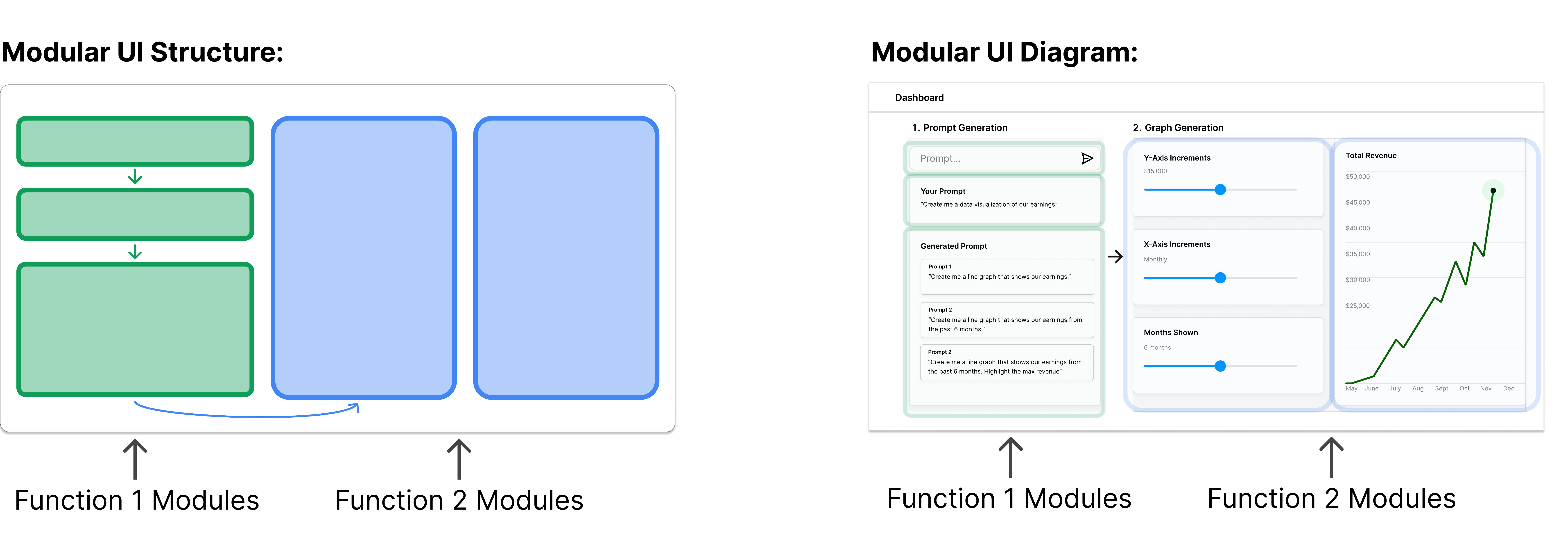}
\caption{%
\textbf{Modular User Interface: A user interface layout that is broken up into several interaction spaces, each with a different function (Sec.~\ref{sec:UImodular}).}}

\label{fig:ModularUI}
\end{figure}

\subsection{Modular User Interface} \label{sec:UImodular}

As it pertains to generative AI, modular UIs consist of user interfaces that are broken up into multiple main interaction areas, with each of these interaction areas having a different function in the generative process (Fig. \ref{fig:ModularUI}). Modular user interfaces are often used in systems with multiple levels of generation. These systems necessitate a modular design as each "module" is handling a different level of interaction \citep{yan2023xcreation, wang2024promptcharm,singhfigura11y, petridis2024constitutionmaker}. Take, for example, \cite{wang2024promptcharm}'s PromptCharm. PromptCharm \citep{wang2024promptcharm} is a text-to-image generative system with a modular user interface. It is unique in that it has an interstitial step where the system helps refine the user prompt to allow the text-to-image part of the system to better capture the user's original intent. This system necessitates a modular design because it has two generative functions: refining the user's original prompt and generating an image based on said prompt. For this reason, PromptCharm has three "modules" dedicated to the generation and refinement of the original prompt and two modules dedicated to the generation and refinement of the image. Given this, modular UIs are useful in that they are versatile and can handle several levels of interaction. Designers should consider utilizing them to design interfaces for multi-level LLMs and other multi-functional generative systems.

\subsection{Simulated User Interface}
  \label{sec:UISimulated}
 Simulated User Interfaces allow users to interact with generative systems in either virtual reality or some form of augmented reality. Simulated UIs are often necessary when traditional GUIs are incapable of helping a user complete a specific task. These interfaces allow users to interact with the generative system in a way that can teach them how to perform a certain task in a simulated environment \citep{konenkov2024vr,giunchi2024dreamcodevr} or interact with data in a tangible way \citep{deepscope2019}. Take, for example, \cite{deepscope2019}'s DeepScope, a tangible user interface created in augmented reality that simulates the urban planning of a particular city. Deepscope essentially helps users create a 3 dimensionsional blueprint of a city that can be used to discuss and visualize urban planning concepts. While they are currently more difficult to create than GUIs, simulated interfaces, such as DeepScope, are especially useful as they offer a tangible interaction that GUIs lack. In turn, they are able to better train users how to complete certain tasks or help them interact with "tangible" data in real-time

\begin{table}[b!]
\centering
\caption{%
\textbf{Taxonomy of Human-GenAI Engagement.}
We summarize the main categories of human-GenAI engagement and provide intuitive definitions and examples of each.
}
\vspace{2.5mm}
\label{table:engagement-taxonomy}
\renewcommand{\arraystretch}{1.18} 
\footnotesize
\setlength{\tabcolsep}{5.5pt} 
\begin{tabularx}{1.0\linewidth}{p{4.8cm} p{5cm}  p{5.6cm} }
\toprule
\textbf{Engagement}
& \textbf{Definition}
& \textbf{Examples}
\\

\hboldline

\rowcolor{darkblue}
\textsc{\textcolor{googleblue}{Passive Engagement} (\S\ref{sec:interaction-passive})} 
& \cellcolor{lightblue}
No direct user interaction during the generation process leverages only user profile and preferences
& \cellcolor{lightblue}
- immersive news writing~\citep{oh2020understanding}\newline
- personalized curated sports articles~\citep{kim2019designing}\newline
- AI-generated user engagement metrics ~\citep{gatti2014large}
\\

\hline

\rowcolor{darkred} 
\textsc{\textcolor{googlered}{Deterministic Engagement} (\S\ref{sec:interaction-deterministic})} 
& \cellcolor{lightred}
Similar to passive, though user provides basic instructions to the genAI model to start or stop the generative process.
 
& \cellcolor{lightred}
- AI generated hierarchical tutorials~\citep{10.1145/3411764.3445721}\newline
- automated newsgathering ~\citep{nishal2024envisioning}\newline
- chemical synthesis~\citep{10.1145/3411764.3445721}
\\

\hline

\cellcolor{lightgreen}
\textsc{\textcolor{googlegreen}{Assistive Engagement} (\S\ref{sec:interaction-assistive})} 
& \cellcolor{lightergreen}
Offers indirect assistance to users such as making suggestions.
Systems using assistive engagement must understand the user intentions and high-level goals.

& \cellcolor{lightergreen} 
- follow-up question generation~\citep{valencia2023less} \newline
- autocompletion \citep{Jakesch_2023} \newline
- writing suggestions~\citep{fitria2021grammarly}
\\

\hline

\cellcolor{darkpurple}
\textsc{\textcolor{googlepurple}{Turn-based Collaborative\newline Engagement} (\S\ref{sec:interaction-collab-sequential})} 
& \cellcolor{lightpurple}
The generative process between the user and generative model occurs in a sequential fashion (\ie, turn-based)
& \cellcolor{lightpurple}

Turn-based conversational interfaces where the user makes a request, then AI generates content, and the process repeats in a turn-based fashion.
\\

\hline

\rowcolor{lighterblue}
\textsc{\textcolor{mydarkblue}{Simultaneous Collaborative\newline Engagement} (\S\ref{sec:interaction-collab-simultaneous})} 
& \cellcolor{lighterblue-row}
User and GenAI work together in parallel to generate the final content
& \cellcolor{lighterblue-row}
A drawing system where user and generative AI draw concurrently in real-time~\citep{lawton2023tool}
\\

\boldbottomline

\end{tabularx}
\end{table}

\section{Human-AI \emph{Engagement} Taxonomy: From Passive to Collaborative}\label{sec:interaction-levels}

In this section, we propose a taxonomy based on the different levels of engagement in human-GenAI interaction, going from least user interaction to increasingly more.
More formally, engagement is defined as the process by which interactors start, maintain, and end their perceived connections with each other during an interaction~\citep{oertel2020engagement,sidner2003engagement,salam2023automatic}.

\begin{figure}[t] 
\centering
\includegraphics[width=.99\linewidth]{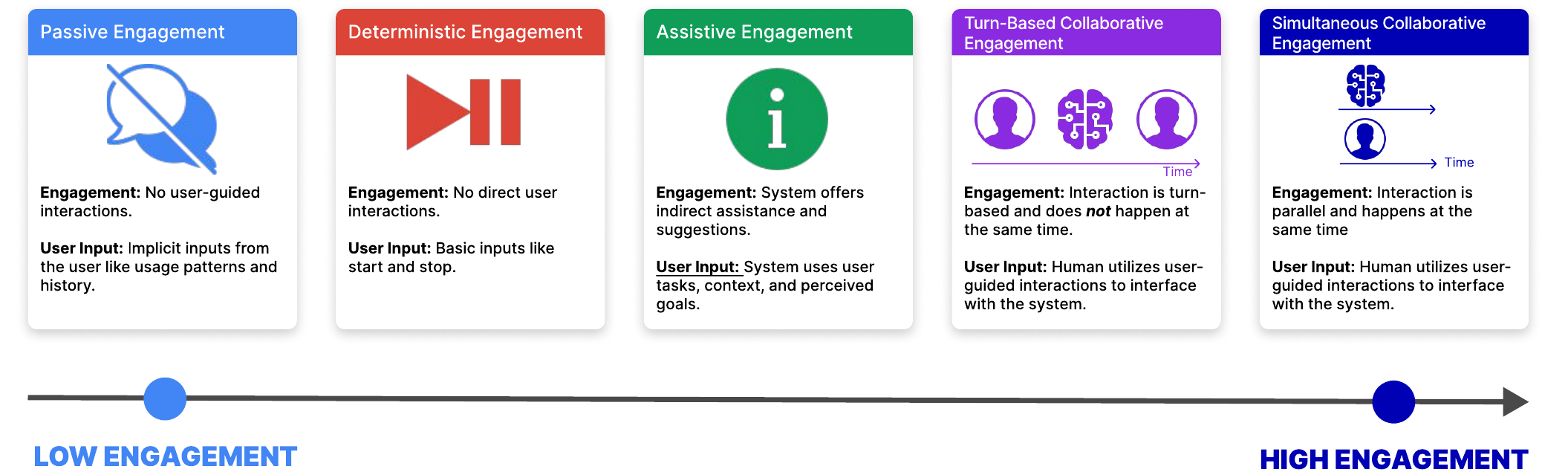}
\caption{%
\textbf{Human-AI Engagement Taxonomy: A high-level visual summary of the different Human-AI Engagement Levels (Sec.~\ref{sec:interaction-levels}).}}
\label{fig:engagement}
\end{figure}

Specifically, we propose a spectrum of engagement levels going from passive to fully collaborative.
In particular, the main engagement types identified include passive engagement (Sec.~\ref{sec:interaction-passive}), 
deterministic engagement (Sec.~\ref{sec:interaction-deterministic}), 
assistive engagement (Sec.~\ref{sec:interaction-assistive}), and
sequential collaborative engagement (Sec.~\ref{sec:interaction-collab-sequential}), and
simultaneous collaborative engagement (Sec.~\ref{sec:interaction-collab-simultaneous}).
We provide an intuitive summary of the engagement level taxonomy in Table~\ref{table:engagement-taxonomy} and provide intuitive examples of each in Figure~\ref{fig:engagement}.
The engagement level dictates the application scenarios supported by the generative AI system and the interaction techniques.

\subsection{Passive Engagement}\label{sec:interaction-passive}
Passive engagement is defined as a system that generates content based solely on implicit information gained by the user. The implicit inputs can be anything from usage patterns, user preferences, or user search history. Passive engagement consists of using these implicit inputs to generate content, and in these scenarios, there is no active interaction by the user. The systems generate content independent of user-guided interactions and are often agent-initiated systems. Examples of passive engagements in generative AI consist of social media engagement systems that measure and generate user engagement metrics \citep{gatti2014large,8352646}, predictive AI medical models \citep{dogheim2023patient,farrokhi2024artificial, jeddi2020remote}, systems that curate personalized news based on user preferences \citep{kim2019designing,oh2020understanding}, and systems that recommend personalized design assets \citep{cai2022personalized, kadner2021adaptifont}.

An example of a system that leverages passive engagement is PINGS (Personalized and Interactive News Generation System) \citep{kim2019designing}. This system automatically generates personalized sports news stories based on a user's preferences. Systems that utilize passive engagement are often used to create outputs that rely heavily on user preferences or other implicit interactions. The best passive systems successfully integrate themselves into user's lives and require little interaction from the user. Furthermore, the success of these systems is often measured by how well they can interpret a user's passive inputs.

\subsection{Deterministic Engagement}\label{sec:interaction-deterministic}

Deterministic engagement consists of a user-AI interaction cadence where the AI system works almost entirely, void of user input or interaction. In these scenarios, the system is provided preset parameters that it uses to complete predetermined task(s). Often times the extent of user interaction at this level of engagement consists of a user inputting preset parameters and/or telling the system to START or STOP. Deterministic generative AIs often only have a single use that can range from anything from news gathering \citep{nishal2024envisioning}, chemical synthesis \citep{yang2024molprophet}, and educational content generation \citep{10.1145/3411764.3445721, 10.1145/3491102.3501839},

A novel example of this is a generative system that is responsible for a "newsgathering" task at a newspaper outlet \citep{nishal2024envisioning}. Essentially, this generative system is responsible for gathering leads, data, and potentially newsworthy leads within a given news cycle. This process is automated, and the generative system determines what is newsworthy based on preset parameters. When it is finished, it presents the topics that it feels to be newsworthy. This is a wholly automated process, with the only human engagement happening when the system starts and when it stops.

\subsection{Assistive Engagement}\label{sec:interaction-assistive}

The assistive engagement level is characterized by interactions that offer indirect assistance to the users such as helping them complete a question \citep{fitria2021grammarly,Jakesch_2023}, auto-completing incomplete code \citep{10.1145/3586030,10.1145/3581641.3584037, prather2023s}, and helping users keep their focus \citep{arakawa2023catalyst}. This level of engagement enhances a user's output without taking creative control. Generative AI that engages at an assistive level is incapable of functioning as an independent agent and is not prompted to act by the user. Assistive engagement is often ongoing and can occur several times while a user is working on a project

Assistive engagement is especially valuable in creative or academic use cases where users need to generate unique or personal content. In these scenarios, users can benefit from generative AI without being flagged for plagiarism or feeling disconnected from the creative process. For instance, a person designing art software might receive suggestions about color palettes to use, or a computer science student might be offered fixes for compiling errors. Essentially, assistive-level engagement is utilized in designing for use cases where users desire autonomy over the generative process while still benefiting from generative AI.

\subsection{Turn-based Collaborative Engagement}\label{sec:interaction-collab-sequential}
\label{sec:interaction-collab}

Turn-based collaborative engagement consists of a user working in tandem with generative systems to complete a task or create a final product. Commonly, the UI structure of this collaboration method follows an almost sequential, turn-based cadence with the user and the system taking turns inputting and outputting information \citep{achiam2023gpt,masson2023directgpt,petridis2024constitutionmaker,ross2023programmer, betker2023improving}. Scenarios such as these involve information exchanges and are typically conversational in style. Moreover, the human and AI are working towards a single goal such as creating and editing a data visualization \citep{10.1145/2984511.2984588}, editing a piece of visual content \citep{lawton2023tool, wang2024aesopagent, davis2015drawing, yan2023xcreation, bangerl2024explorations}, or writing a story or other piece of creative content \citep{suh2023structured, wang2024weaver, 10.1145/3586183.3606737,kim2023metaphorian}. Furthermore, this interaction dynamic requires the user to have some level of subject matter experience to guide the system. In order for the user and the system to work collaboratively with one another, there may be instances where the user may have to act as a guide.

While common turn-based collaborative models include the likes of ChatGPT \citep{achiam2023gpt} and Dall-E \citep{betker2023improving}, one novel turn-based collaborative system can be found in ~\cite{davis2015drawing}'s Drawing Assistant. This system is a drawing model that generates art asynchronously with its user. Essentially, the human and the AI take turns drawing, with the AI basing its drawing on the user input. The system operates on a turn-based cadence, with the human and AI not working simultaneously. While collaborative interactions do not need to occur in real-time, this section is underscored by a back-and-forth interaction between the user and the system and is the most involved of any interaction level.

\subsection{Simultaneous Collaborative Engagement}\label{sec:interaction-collab-simultaneous}
In this section, we discuss simultaneous collaborative engagement that occurs at the same time. More formally, the human and AI are collaborating concurrently on a given task \citep{deshpande2020towards, lawton2023tool}. As an example, a user and AI may be working simultaneously on an image editing task where the user and AI are making changes to different parts of the image in parallel. This is in contrast to turn-based collaborative engagement discussed previously in Section~\ref{sec:interaction-collab-sequential} where the user and AI take turns.

One novel engagement related to simultaneous engagement can be found in Reframer \citep{lawton2023tool},
a human-AI drawing system that enables the user and generative model to draw concurrently in real-time,
where the user interacts with the system through drawing, and these interactions, in turn impact what the AI
draws, and vice-versa. The system also includes a prompt, drawing models, such as draw, focus, and explore,
and sliders to control more advanced drawing features, such as a slider to allow the AI more freedom, a
recency slider that indicates to the AI to either use more history when drawing concurrently or the immediate
past and so on.

More recently, there has been work on multi-agent approaches ~\citep{10.1145/3611643.3616271}. 
In terms of simultaneous collaborative engagement, there is also the situation where multiple AIs or agents (multi-agent)can work together simultaneously towards the same goal. Take Repilot ~\citep{10.1145/3611643.3616271} for example, a multi-agent approach that works side by side with \cite{barke2023grounded}'s Co-pilot, refining its outputs. Essentially, Repilot was created as a generative system that takes Co-pilot's code suggestions and outputs as its own inputs. Then it refines Co-pilot's suggestions and iterates on them to reduce hallucinations and improve the accuracy of the code. Overall, multi-agent approaches, such as Repilot, are effective simultaneous engagements that add another level of AI generation to an already existing generative process.

\section{Applications} \label{applications}

\begin{figure}[t] \label{fig:applications} 
\centering
\includegraphics[width=.99\linewidth]{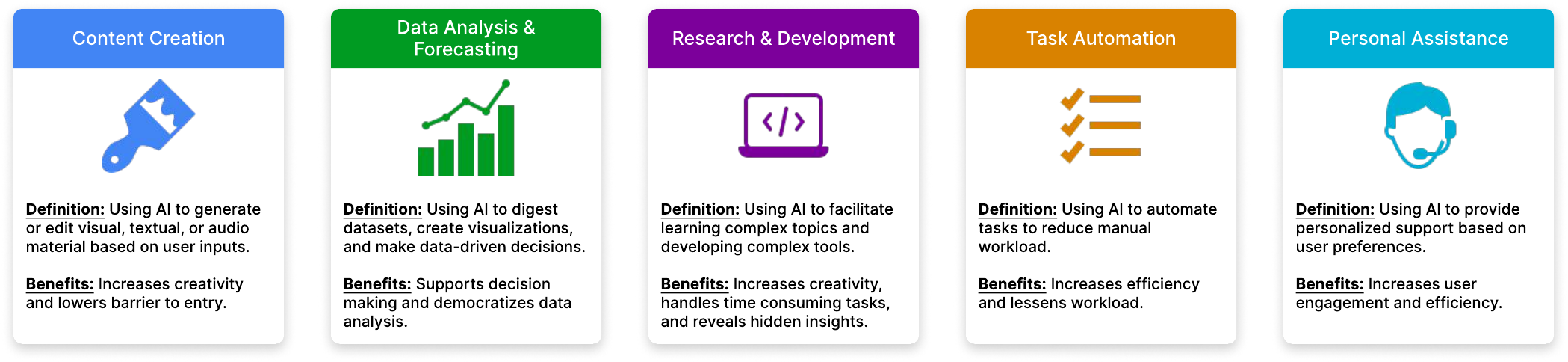}
\caption{%
\textbf{Applications Taxonomy: A high-level visual summary of the different generative AI Applications (Sec.~\ref{applications}).}}
\label{fig:humanAI}
\end{figure}

Generative AI is transformative not only due to its flexibility in engagement but also due to its wide range of applications. As seen in Figure \ref{fig:applications}, the ability to be an effective tool in everything from content creation (Sec. \ref{sec:applications/Content}), to data analysis and forecasting (Sec. \ref{sec:applications/Data}), to research and development (Sec. \ref{sec:applications/development}), to task automation (Sec. \ref{sec:applications/Automation}), and to personal assistance (Sec. \ref{sec:applications/Assist}) makes it important to categorize them in this way. In doing so, we can highlight key applications of generative AI and explicitly explore the specific benefits that every application type provides users. Furthermore, we will focus on the interplay between these applications and the impact it has on the different UI techniques used by the generative system.

\subsection{Content Creation} \label{sec:applications/Content}

\textbf{Overview and Examples} Content creation in the context of generative AI consists of prompting the generative system to create a specific piece of generated material with specific parameters. Content creation consists of anything from creating or editing visual media \citep{liu20233dalleintegratingtexttoimageai, tang2024realfill, jeon2021fashionq, davis2015drawing,wang2024promptcharm, chung2023promptpaint} to creating written content \citep{achiam2023gpt, yuan2022wordcraft, chung2022talebrush, wang2024aesopagent,suh2023structured, wang2024weaver} or audio content \citep{agostinelli2023musiclm,copet2023musicgen,borsos2023audiolm}. The incredible part about generative AI in content creation is that it lowers the barrier to entry for many creatives. Take, for example, \cite{tang2024realfill}'s RealFill, a generative fill system that responsively and accurately "fills" in gaps in existing images or expands them in a way that is in line with the user's requests. This case also exemplifies that content creation generative systems can be used both to edit content and create it from scratch. Furthermore, content creation systems are especially effective at lowering the barrier to entry for many creative domains as they make it easier for the average user to be creative without having to be an expert in the domain. Furthermore, it also speeds up the creative process as the user can spend less time writing, editing, and pushing pixels, and more time creating and brainstorming.

\textbf{Common User Interactions} Content creation platforms can essentially come in all shapes and sizes, but there are some UI layouts (Sec. \ref{uilayouts}) that, through our literature reviews, we have found to be more common. For example, content creation often goes hand-in-hand with conversational user interfaces (Sec. \ref{sec:UIConversational}) and canvas user interfaces (Sec. \ref{sec:UICanvas}). The conversational user interface is especially useful in generating written content but can also be used to generate images. Meanwhile, canvas user interfaces are often best used in generating visual content. Furthermore, content creation systems often use a mix of different user-guided techniques, but there is often a large emphasis on prompting (Sec. \ref{sec:prompting}). All this to say, that while content creation systems use a wide range of user interaction techniques, these were the ones found to be the most common.

\subsection{Data Analysis and Forecasting}\label{sec:applications/Data}

\textbf{Overview and Examples} Data analysis and forecasting is one of the applications where generative AI has the potential to make the largest impact. With data becoming more and more valuable and data sets becoming larger and larger, gleaning insights from them has simultaneously become more important and more time-consuming. Generative AI can help data experts both digest and glean insights from data \citep{goyal2024healai,achiam2023gpt, fu2023challenger} and also help them visualize and present data in an easy-to-understand format \citep{singhfigura11y, 10.1145/2984511.2984588}. In doing so, generative AI makes it easier to make data-backed decisions and increases the efficiency and speed at which those decisions can be made.

\textbf{Common User Interactions} Data Analysis and Forecasting generative platforms take many different shapes, but through our survey, we found that there were many different throughlines. In terms of UI layouts, we found that most data-focused generative systems have mostly are either information visualization canvases, modular user interfaces, or even conversational user interfaces (Sec. \ref{uilayouts}). Modular user interfaces were effective at acting almost as a dashboard and allowing users to adjust many parameters at once, while information visualization canvases were more focused on a single visualization. Conversational UIs were most effective when a user was attempting to ask for specific insights about the data. In terms of common user-guided interactions, data-focused systems often use many system and parameter manipulation techniques to adjust the data's parameters (Sec. \ref{sec:manipulating}). All in all, there is no one correct way to create a data analysis and forecasting generative system, these were just the common through-lines that existed for these types of applications.

\subsection{Research and Development} \label{sec:applications/development}

\textbf{Overview and Examples} Generative AI has begun to revolutionize research and development in much of the same ways that it revolutionizes other fields. In terms of research, generative AI has made it easier to learn complex topics and skills as it allows users to learn in a personalized environment \citep{cao2024scholargpt,malandri2023convxai}. Meanwhile, it also can help develop products and do research-related tasks that normally would be extremely time-consuming \citep{petridis2023anglekindling, chang2023citesee, liang2024storydiffusion, yen2023coladder}. Like many of the other applications, generative AI in the research and development field makes processes more efficient and enables developers and researchers to spend less time on menial tasks.

\textbf{Common User Interactions} Like all applications, research and development systems use a wide range of user interaction techniques. However, we found that there were common design patterns that were used in many of the research and development applications. For one, research and development systems often rely on conversational user interfaces as these interfaces synergize well with the question and answer cadence common in research and learning (Sec. \ref{sec:UIConversational}). Meanwhile, the contextual UI structure can be used in development and research writing settings to give relevant edits and suggestions to the users' development process within the context of what they are working on (Sec. \ref{sec:UIContextual}). 

\subsection{Task Automation} \label{sec:applications/Automation}

\textbf{Overview and Examples} Task automation has become one of the strongest applications of generative AI. Essentially, this application consists of generative AI automating often repetitive tasks that a human would normally do. However, just because the job is repetitive, does not mean it is low-skill, as generative AI is able to automate what are usually considered high-skill tasks \citep{dogheim2023patient,farrokhi2024artificial, jeddi2020remote,kim2019designing,oh2020understanding}. Like many of the other applications, automating tasks increases efficiency and eliminates menial tasks. In doing so, this gives users more time to focus on the larger task they are trying to complete and leaves more time for decision-making.

\textbf{Common User Interactions} Since task automation, by nature, is designed not to have a large level of human interaction, one can imagine that there would not be a large amount of user-guided interactions. For the most part, task automation is mostly done in conversational UIs (Sec. \ref{sec:UIConversational}) as this is where the user initiates the automation. Naturally, in the same way, the most common user-guided interaction is prompting (Sec. \ref{sec:prompting}), as the user will commonly use prompts to initiate the automation. All in all, while many different user interactions can be used to create a task automation system, these two are the most common.

\subsection{Personal Assistance} \label{sec:applications/Assist}

\textbf{Overview and Examples} One of the strongest aspects of generative AI is its ability to give personalized assistance to each of its users based on the user's preferences and individual needs. Most commonly, it can act as a chatbot that can converse with the user and help them with personalized information or advice \citep{achiam2023gpt, fu2023challenger,ang2023socratic} but can also act in place of customer support to increase engagement and streamline interactions \citep{NBERw31161, verma2023generative}. Furthermore, it is often commonly used in line to spell check, provide edits when creating something, or offer personalized, contextual help \citep{fitria2021grammarly,Jakesch_2023} The benefit of this is that users can access personalized and often professional-level assistance at little to no cost. This application is especially impressive as it sometimes serves as a stand-in for low-level medical, legal, or professional advice \citep{li2023chatdoctor, yue2023disc}. All in all, the application of generative AI as a personal assistant changes the way that users can access personalized advice and makes "expert" level guidance more accessible to those who may not normally be able to obtain it.

\textbf{Common User Interactions} 
While personal assistance applications come in many different forms, these applications skew heavily toward having a conversational user interface (Sec. \ref{sec:UIConversational}). This should come as no surprise, as much of the interaction with personal assistance synergizes well with conversational UIs, which are essentially just conversations. Naturally, this also coincides with the fact that the most common user-guided technique is prompting, most specifically text or speech prompts (Sec. \ref{sec:prompting}). However, there are some personal assistant systems that leverage contextual user interfaces (Sec. \ref{sec:UIContextual}), effectively providing in-context personal assistance to the user in real-time. Using contextual user interfaces is effective when creating personal assistants, as they understand the context in which the user is working and can give them specific instructions inline with their task.

\section{Open Problems \& Challenges}\label{sec:open-problems-challenges}
Generative AI is a fast-growing space with many implications for many different fields. For this reason, it is important that research continues in this area, specifically looking at the role that user interfaces and interactions play in generative AI development. In this section, we discuss open problems and highlight important challenges for future work.
\subsection{Accessibility}

\subsubsection{Designing for Disabilities}
Generative artificial intelligence has the opportunity to revolutionize how people with disabilities interact with technology, offering tools that can help disabled users with common difficult tasks like writing or "seeing" who is at a meeting \citep{goodman2022lampost, iyer2023}. Furthermore, \cite{glazko2023} talks about how generative AI can be used to benefit neurodivergent users who have trouble using the correct tone in different messages at work. They can use products like ChatGPT \citep{achiam2023gpt} or Gemini \citep{fu2023challenger} to help adjust the tones of their messages. However, the user experience for disabled users is still lackluster in that many disabled users cannot use it entirely independently. Many times, disabled users need assistance from coworkers to verify that the system understood their original query \citep{glazko2023}.

Future generative artificial intelligence applications need to take more care of users with disabilities when designing interfaces. Many of the works surveyed did not address how their system could be used by users with disabilities, nor did they address how accessibility would play a role in the future. A simple solution is to involve designers or users more in the design process and to center their needs while designing the system's user interfaces. However, extensive research has been done on the role accessibility should play when designing \textit{traditional} user interfaces and experiences \citep{petrie2009evaluation, sauer2020usability, aizpurua2016exploring}, but the field still lacks extensive literature on how designers should specifically design generative AI user interfaces for those with disabilities.

\subsubsection{Designing for Limited Technical Literacy}

Generative artificial intelligence has the opportunity to level the playing field for those traditionally marginalized by technology access. It does this by lowering the barrier to entry for many fields of practice like data science, illustration generation, fictional and nonfictional writing, and so much more. However, many users lack the technical literacy to take full advantage of the aforementioned benefits which leads users to be frustrated with or abandon generative applications altogether. 

The user interfaces of common generative AI applications are easy to use, as many times, the UI elements are limited to chat boxes and conversational history (Sec. \ref{sec:UIConversational}) \citep{achiam2023gpt,fu2023challenger,betker2023improving}. However, many of the applications we surveyed have complicated user interfaces that are most likely targeted at users who already have some understanding of the process. This can be done if the designs place a greater emphasis on educational design elements or even design elements that make discoverability much easier. User interfaces of generative AI have the opportunity to democratize their systems capabilities by designing for users who exhibit a large range of technical literacy, not just ones who are already capable of completing a task without generative AI. Therefore, future research should focus on how user interfaces should be designed to better assist users and give personalized experiences to users with both high and low amounts of technical literacy.

\subsection{The Future of Generative AI}

\subsubsection{Designing for Future User Interfaces}

Just as quickly as it rose to prominence, generative AI will continue to grow and embed itself within new technology. We have already begun to see how generative AI plays roles in virtual \citep{konenkov2024vr,giunchi2024dreamcodevr} and tangible user interfaces \citep{deepscope2019}. This evolution of generative AI technology will necessitate much more research into how user interfaces should be developed for these new mediums. Because these interactions often occur in three dimensions, many of the best practices for designing two-dimensional applications are not applicable. Therefore, additional work needs to be done to outline and survey the user interactions of new technologies as they emerge.

On a similar note, as mentioned, multi-agent approaches, like \cite{10.1145/3611643.3616271}'s Repilot, are becoming more common. Essentially, these multi-agent systems consist of when generative AI agents interact with one another rather than directly with the user. Therefore, as these multi-agent systems emerge, there should be extensive work surveying the common user interactions that exist within them. The user interfaces of multi-agent systems should be transparent and clearly communicate to users what is happening and which agent is acting at what time. In continuing research on emerging generative AI technology, such as this one, we can get a better understanding of the user interactions that are needed to best serve the user. 
\subsubsection{Designing for Growth and Scalability}
As generative AI continues to grow and is forecasted to grow significantly in the coming decade \citep{halal2016forecasts}, it is important the user interfaces grow and evolve with it. In almost every field, generative AI is forecasted to grow, and with that comes a larger need for the application to handle a more diverse set of users and use cases. Furthermore, it is important to retain consistent user interaction patterns as the application grows, preventing users from having to continuously relearn how to use the app with every iteration. One of the major difficulties that comes with scaling up a generative AI application is the increased abilities and complexities that the application must now account for. At the same time, they must account for these complexities while also ensuring that the user interface is straightforward and easy to use. This will address the concern that user interfaces could become overwhelming and cognitively complex with more features being added to the generative AI application. All in all, as generative AI applications grow in size, designers must adapt to the capabilities of the system while also continuing to meet the needs of the user.

\subsection{Ethics}

\subsubsection{Designing for Harmful Bias Mitigation}

Current literature has addressed and surveyed the current state of both bias and harmful bias mitigation in generative AI. Literature like \cite{gallegos2024bias} detail how generative applications inherit biases present within the training data that they were provided. Therefore, generative applications are known to propagate harmful stereotypes, incorrect data, and skewed answers. Given this, there is still a need to explore the role that good design and effective user interaction design techniques can play in mitigating bias in generative artificial intelligence. Most of the user interfaces that were surveyed in this paper do not address that their generative systems may be biased in some harmful way, despite the fact that many of them are biased \citep{gallegos2024bias}.

Future user interface design approaches can potentially address this by designing in features that are transparent about the bias that the system might exhibit. Furthermore, more research can be done to discover if there are ways to leverage user interaction techniques to mitigate harmful biases, such as being able to give explicit feedback on whether or not an answer is explicitly biased. More work can be done to leverage a collaborative approach between users and generative systems to mitigate harmful bias.

\subsubsection{Designing to Prevent Misuse}
Generative AI is very much a two-sided coin: on one hand, it is an amazing invention capable of lowering the barrier to entry for many complex fields. On the other hand, it can be extensively misused. If misused, generative AI can facilitate everything from opinion manipulation and misinformation to plagiarism and much more \citep{marchal2024generative, ferrara2024genai, wach2023dark}. This misuse can have long-lasting and profound negative impacts on society; for this reason, developers can and should design with these considerations in mind when creating generative AI systems.

Given this, it is crucial for future works to consider designing user interfaces that include elements aimed at preventing misuse. Among the generative AI systems we surveyed, there was little to no warning from these systems that a specific query could lead to misuse. For instance, if a user asks a generative system to create a chatbot that responds to social media posts with misinformation, the system should inform the user that this behavior can be considered misuse and explain why. By incorporating such warnings, generative systems would have built-in protections that could prevent misuse before it occurs. This, in turn, could prevent many of the negative societal impacts that generative AI could have moving forward.

\subsection{Data Privacy}
As generative AI grows larger and larger, many systems are integrating into various aspects of daily life. Through this process, users are more likely to surrender sensitive personal data to generative systems, either knowingly or unknowingly. Often, this is for good reason, as generative applications thrive when they can provide personalized experiences to each user. However, there should be guardrails in place to prevent generative AI applications from collecting overly sensitive user data without the user's permission.

That being said, there are few design patterns specifically created for data transparency. For this reason, more research is needed on best practices for communicating to users how their data is being used. There should be user interaction features aimed explicitly at ensuring that users understand what data is being collected and how it is utilized. Furthermore, designers should explore more solutions that allow users to choose which personal data is shared. They can also investigate educational design flows that help users understand the best practices for sharing data with generative systems. By empowering users, designers can ensure that they are well-informed and have control over their data privacy preferences.

\subsection{Open Problems in Designing for Generative AI Interpretation}
With the rise of generative AI, numerous methods have emerged to interpret different types of generative models, such as language models, text-to-image models, and multimodal language models. These methods mainly aim to identify internal components or representations within the generative models that are causally responsible for specific outputs. The ultimate goal is to enhance transparency for end-users while empowering them to control various aspects of the generated content.

In language models, studies like ~\citep{meng2023locatingeditingfactualassociations, meng2023masseditingmemorytransformer, Hartvigsen2022AgingWG} have pinpointed causal components that can be adjusted to alter final outputs. In the context of text-to-image models, research such as ~\citep{basu2024localizing, basu2024on, arad2024refactupdatingtexttoimagemodels, gandikota2023unifiedconcepteditingdiffusion} has identified interpretable subspaces that can influence the generation of specific concepts. Despite significant technical advancements in understanding model components, there remains a gap in the availability of comprehensive, end-to-end systems that allow users to interactively engage with these methods. Recent research has extended these methods to focus on understanding the computational sub-graphs, or ``circuits'' within generative models~\citep{elhage2021mathematical, prakash2024finetuningenhancesexistingmechanisms, hanna2023how, wang2022interpretabilitywildcircuitindirect}. This increased complexity in circuit analysis has highlighted the importance of designing interactive systems to facilitate deeper insights. Recently the community has shifted towards finding controllable steering vectors which when added to the language model can elicit certain behaviours (e.g., improving refusal rate to harmful queries) to an end-user~\citep{arditi2024refusallanguagemodelsmediated, bricken2023monosemanticity, li2024inferencetimeinterventionelicitingtruthful, stolfo2024improvinginstructionfollowinglanguagemodels, cunningham2023sparseautoencodershighlyinterpretable}. There are existing interactive systems for steering vectors in language models, but many are closed-source\footnote{https://goodfire.ai/}. In contrast, Transluce\footnote{https://transluce.org/}, an open and independent technology lab, has initiated efforts to develop interactive tools for understanding language models, aiming to provide users with greater transparency and control. 

With the rise of advanced mechanistic interpretability tools for generative AI, the next logical step is to develop unified interactive systems that allow users to engage with these tools seamlessly. This approach not only enhances user interaction but also paves the way for innovative and robust interactive system development.

\section{Conclusion} \label{sec:conc}
We have presented a comprehensive survey detailing various dimensions of generative AI user interactions and the different design techniques used to facilitate them. We began by expanding on common user interaction concepts, defining a \textit{user-guided interaction} as a form of interaction that is explicitly and deliberately initiated by the user. Furthermore, we detailed the different input mediums that users can utilize when interacting with generative artificial intelligence. Next, we presented four instructional taxonomies that outline how current generative systems are designed for generative AI. These taxonomies include a taxonomy of user-guided interaction techniques for generative AI systems, a taxonomy of common user interface layouts, a taxonomy of human-AI engagement levels, and a taxonomy of applications and use cases for generative AI systems. Our first taxonomy categorizes different types of generative AI user-guided interactions, focusing on interaction patterns explicitly initiated by the user. The second taxonomy addresses key user interface layouts, specifically examining how and when they are employed. Thirdly, the taxonomy on human-AI engagement levels for generative AI systems explores the extent of user involvement in the generative process and the levels of deliberate interaction taking place. Finally, the taxonomy of generative AI applications highlights the various ways generative AIs are utilized and the user interfaces that best align with these use cases. We conclude our survey with several actionable open problems identified during our exploration of generative applications. Through this work, we aim to create a compendium of common generative AI user interactions to lower the barrier to entry for designers and non-designers alike.

\nocite{*}
\bibliography{main}
\bibliographystyle{tmlr}

\end{document}

%% file: preamble.tex
\usepackage{xcolor}
\definecolor{thedarkblue}{RGB}{0,0,120} 
\definecolor{mydarkblue}{rgb}{0,0.08,0.45} 
\definecolor{darkblue}{rgb}{0,0.08,180}
\colorlet{TufteRed}{red!80!black}
\definecolor{theblue}{RGB}{0,0,180}
\colorlet{thered}{TufteRed}
      
\usepackage{microtype}
\usepackage{balance}

\usepackage{enumitem}

\usepackage{booktabs}
\usepackage{tabularx}

\usepackage{amsmath,amssymb,amsthm}

\newcommand{\eat}[1]{\ignorespaces}
\usepackage{comment}

\usepackage{tikz}
\usepackage{verbatim}
\usetikzlibrary{arrows}
\usetikzlibrary{shapes,decorations}
\usetikzlibrary{decorations.pathmorphing} 
\usetikzlibrary{fit}					
\usetikzlibrary{backgrounds}	

\usepackage{ragged2e}
\usepackage{multirow}
\usepackage{microtype}
\usepackage{balance}
\usepackage{setspace}

\graphicspath{{./}{./graphics/}}
\newcolumntype{H}{>{\setbox0=\hbox\bgroup}c<{\egroup}@{}}
\newcolumntype{R}[1]{>{\RaggedLeft\arraybackslash}} 
\newcolumntype{L}[1]{>{\RaggedRight\arraybackslash}} 

\newcommand{\eg}{\emph{e.g.}}
\newcommand{\ie}{\emph{i.e.}}

\newtheorem{Definition}{\bfseries{Definition}}

\AtBeginEnvironment{pmatrix}{\setlength{\arraycolsep}{2pt}}

\renewcommand{\vec}[1]{\boldsymbol{\mathrm{#1}}}

\DeclareMathOperator{\hugeE}{\mbox{\huge\raise-0.3ex\hbox{E}}}
\DeclareMathOperator{\p}{\mathbb{P}}
\DeclareMathOperator{\hugep}{\mbox{\huge\raise-0.3ex\hbox{$\p$}}}

\providecommand{\vx}{\ensuremath{\vec{x}}}


%% file: main.bbl
\begin{thebibliography}{183}
\providecommand{\natexlab}[1]{#1}
\providecommand{\url}[1]{\texttt{#1}}
\expandafter\ifx\csname urlstyle\endcsname\relax
  \providecommand{\doi}[1]{doi: #1}\else
  \providecommand{\doi}{doi: \begingroup \urlstyle{rm}\Url}\fi

\bibitem[Achiam et~al.(2023)Achiam, Adler, Agarwal, Ahmad, Akkaya, Aleman, Almeida, Altenschmidt, Altman, Anadkat, et~al.]{achiam2023gpt}
Josh Achiam, Steven Adler, Sandhini Agarwal, Lama Ahmad, Ilge Akkaya, Florencia~Leoni Aleman, Diogo Almeida, Janko Altenschmidt, Sam Altman, Shyamal Anadkat, et~al.
\newblock Gpt-4 technical report.
\newblock \emph{arXiv preprint arXiv:2303.08774}, 2023.

\bibitem[Agostinelli et~al.(2023)]{agostinelli2023musiclm}
Forest Agostinelli et~al.
\newblock Musiclm: Generating music from text.
\newblock \emph{arXiv preprint arXiv:2301.11325}, 2023.
\newblock URL \url{https://google-research.github.io/seanet/musiclm/examples/}.

\bibitem[Aizpurua et~al.(2016)Aizpurua, Harper, and Vigo]{aizpurua2016exploring}
Amaia Aizpurua, Simon Harper, and Markel Vigo.
\newblock Exploring the relationship between web accessibility and user experience.
\newblock \emph{International Journal of Human-Computer Studies}, 91:\penalty0 13--23, 2016.

\bibitem[Aksan et~al.(2018)Aksan, Pece, and Hilliges]{aksan2018deepwriting}
Emre Aksan, Fabrizio Pece, and Otmar Hilliges.
\newblock Deepwriting: Making digital ink editable via deep generative modeling.
\newblock In \emph{Proceedings of the 2018 CHI Conference on Human Factors in Computing Systems}, pp.\  1--14, 2018.
\newblock \doi{10.1145/3173574.3173708}.

\bibitem[Alayrac et~al.(2022)Alayrac, Donahue, Luc, Miech, Barr, Hasson, Lenc, Mensch, Millican, Reynolds, et~al.]{alayrac2022flamingo}
Jean-Baptiste Alayrac, Jeff Donahue, Pauline Luc, Antoine Miech, Iain Barr, Yana Hasson, Karel Lenc, Arthur Mensch, Katherine Millican, Malcolm Reynolds, et~al.
\newblock Flamingo: a visual language model for few-shot learning.
\newblock \emph{Advances in neural information processing systems}, 35:\penalty0 23716--23736, 2022.

\bibitem[Alonso Gaona~Garc{\'\i}a et~al.(2014)Alonso Gaona~Garc{\'\i}a, Mart{\'\i}n-Moncunill, S{\'a}nchez-Alonso, and Fermoso~Garcia]{alonso2014usability}
Paulo Alonso Gaona~Garc{\'\i}a, David Mart{\'\i}n-Moncunill, Salvador S{\'a}nchez-Alonso, and Ana Fermoso~Garcia.
\newblock A usability study of taxonomy visualisation user interfaces in digital repositories.
\newblock \emph{Online Information Review}, 38\penalty0 (2):\penalty0 284--304, 2014.

\bibitem[Aneja et~al.(2023)Aneja, Thies, Dai, and Nie{\ss}ner]{aneja2023clipface}
Shivangi Aneja, Justus Thies, Angela Dai, and Matthias Nie{\ss}ner.
\newblock Clipface: Text-guided editing of textured 3d morphable models.
\newblock In \emph{ACM SIGGRAPH 2023 Conference Proceedings}, pp.\  1--11, 2023.

\bibitem[Ang et~al.(2023)Ang, Gollapalli, and Ng]{ang2023socratic}
Beng~Heng Ang, Sujatha~Das Gollapalli, and See~Kiong Ng.
\newblock Socratic question generation: A novel dataset, models, and evaluation.
\newblock In \emph{Proceedings of the 17th Conference of the European Chapter of the Association for Computational Linguistics}, pp.\  147--165, 2023.

\bibitem[Antony \& Huang(2024)Antony and Huang]{10.1145/3672277}
Victor~Nikhil Antony and Chien-Ming Huang.
\newblock Id.8: Co-creating visual stories with generative ai.
\newblock 2024.
\newblock ISSN 2160-6455.
\newblock \doi{10.1145/3672277}.
\newblock URL \url{https://doi.org/10.1145/3672277}.

\bibitem[Arad et~al.(2024)Arad, Orgad, and Belinkov]{arad2024refactupdatingtexttoimagemodels}
Dana Arad, Hadas Orgad, and Yonatan Belinkov.
\newblock Refact: Updating text-to-image models by editing the text encoder, 2024.
\newblock URL \url{https://arxiv.org/abs/2306.00738}.

\bibitem[Arakawa et~al.(2022)Arakawa, Yakura, and Kobayashi]{10.1145/3491102.3501839}
Riku Arakawa, Hiromu Yakura, and Sosuke Kobayashi.
\newblock Vocabencounter: Nmt-powered vocabulary learning by presenting computer-generated usages of foreign words into users’ daily lives.
\newblock New York, NY, USA, 2022. Association for Computing Machinery.
\newblock ISBN 9781450391573.
\newblock \doi{10.1145/3491102.3501839}.
\newblock URL \url{https://doi.org/10.1145/3491102.3501839}.

\bibitem[Arakawa et~al.(2023)Arakawa, Yakura, and Goto]{arakawa2023catalyst}
Riku Arakawa, Hiromu Yakura, and Masataka Goto.
\newblock Catalyst: domain-extensible intervention for preventing task procrastination using large generative models.
\newblock In \emph{Proceedings of the 2023 CHI Conference on Human Factors in Computing Systems}, pp.\  1--19, 2023.

\bibitem[Arditi et~al.(2024)Arditi, Obeso, Syed, Paleka, Panickssery, Gurnee, and Nanda]{arditi2024refusallanguagemodelsmediated}
Andy Arditi, Oscar Obeso, Aaquib Syed, Daniel Paleka, Nina Panickssery, Wes Gurnee, and Neel Nanda.
\newblock Refusal in language models is mediated by a single direction, 2024.
\newblock URL \url{https://arxiv.org/abs/2406.11717}.

\bibitem[Bangerl et~al.(2024)Bangerl, Stefan, and Pammer-Schindler]{bangerl2024explorations}
Mia~Magdalena Bangerl, Katharina Stefan, and Viktoria Pammer-Schindler.
\newblock Explorations in human vs. generative ai creative performances: A study on human-ai creative potential.
\newblock 2024.

\bibitem[Barikeri et~al.(2021)Barikeri, Lauscher, Vuli{\'c}, and Glava{\v{s}}]{barikeri2021redditbias}
Soumya Barikeri, Anne Lauscher, Ivan Vuli{\'c}, and Goran Glava{\v{s}}.
\newblock Redditbias: A real-world resource for bias evaluation and debiasing of conversational language models.
\newblock \emph{arXiv preprint arXiv:2106.03521}, 2021.

\bibitem[Barke et~al.(2023{\natexlab{a}})Barke, James, and Polikarpova]{10.1145/3586030}
Shraddha Barke, Michael~B. James, and Nadia Polikarpova.
\newblock Grounded copilot: How programmers interact with code-generating models.
\newblock 7\penalty0 (OOPSLA1), 2023{\natexlab{a}}.
\newblock \doi{10.1145/3586030}.
\newblock URL \url{https://doi.org/10.1145/3586030}.

\bibitem[Barke et~al.(2023{\natexlab{b}})Barke, James, and Polikarpova]{barke2023grounded}
Shraddha Barke, Michael~B James, and Nadia Polikarpova.
\newblock Grounded copilot: How programmers interact with code-generating models.
\newblock \emph{Proceedings of the ACM on Programming Languages}, 7\penalty0 (OOPSLA1):\penalty0 85--111, 2023{\natexlab{b}}.

\bibitem[Basu et~al.(2024{\natexlab{a}})Basu, Rezaei, Kattakinda, Morariu, Zhao, Rossi, Manjunatha, and Feizi]{basu2024on}
Samyadeep Basu, Keivan Rezaei, Priyatham Kattakinda, Vlad~I Morariu, Nanxuan Zhao, Ryan~A. Rossi, Varun Manjunatha, and Soheil Feizi.
\newblock On mechanistic knowledge localization in text-to-image generative models.
\newblock In \emph{Forty-first International Conference on Machine Learning}, 2024{\natexlab{a}}.
\newblock URL \url{https://openreview.net/forum?id=fsVBsxjRER}.

\bibitem[Basu et~al.(2024{\natexlab{b}})Basu, Zhao, Morariu, Feizi, and Manjunatha]{basu2024localizing}
Samyadeep Basu, Nanxuan Zhao, Vlad~I Morariu, Soheil Feizi, and Varun Manjunatha.
\newblock Localizing and editing knowledge in text-to-image generative models.
\newblock In \emph{The Twelfth International Conference on Learning Representations}, 2024{\natexlab{b}}.
\newblock URL \url{https://openreview.net/forum?id=Qmw9ne6SOQ}.

\bibitem[Betker et~al.(2023)Betker, Goh, Jing, Brooks, Wang, Li, Ouyang, Zhuang, Lee, Guo, et~al.]{betker2023improving}
James Betker, Gabriel Goh, Li~Jing, Tim Brooks, Jianfeng Wang, Linjie Li, Long Ouyang, Juntang Zhuang, Joyce Lee, Yufei Guo, et~al.
\newblock Improving image generation with better captions.
\newblock \emph{Computer Science. https://cdn. openai. com/papers/dall-e-3. pdf}, 2\penalty0 (3):\penalty0 8, 2023.

\bibitem[Birgit Endrass (Augsburg~University \& Matthias Rehm (Aalborg~University(2011)Birgit Endrass (Augsburg~University and Matthias Rehm (Aalborg~University]{towardsculturally}
Elisabeth André (Augsburg University~Germany) Birgit Endrass (Augsburg~University, Germany) and Denmark) Matthias Rehm (Aalborg~University.
\newblock Towards culturally-aware virtual agent systems.
\newblock \emph{Handbook of Research on Culturally-Aware Information Technology: Perspectives and Models}, 2011.

\bibitem[Borsos et~al.(2023)Borsos, Marinier, Vincent, Kharitonov, Pietquin, Sharifi, Roblek, Teboul, Grangier, Tagliasacchi, et~al.]{borsos2023audiolm}
Zal{\'a}n Borsos, Rapha{\"e}l Marinier, Damien Vincent, Eugene Kharitonov, Olivier Pietquin, Matt Sharifi, Dominik Roblek, Olivier Teboul, David Grangier, Marco Tagliasacchi, et~al.
\newblock Audiolm: a language modeling approach to audio generation.
\newblock \emph{IEEE/ACM Transactions on Audio, Speech, and Language Processing}, 2023.

\bibitem[Bricken et~al.(2023)Bricken, Templeton, Batson, Chen, Jermyn, Conerly, Turner, Anil, Denison, Askell, Lasenby, Wu, Kravec, Schiefer, Maxwell, Joseph, Hatfield-Dodds, Tamkin, Nguyen, McLean, Burke, Hume, Carter, Henighan, and Olah]{bricken2023monosemanticity}
Trenton Bricken, Adly Templeton, Joshua Batson, Brian Chen, Adam Jermyn, Tom Conerly, Nick Turner, Cem Anil, Carson Denison, Amanda Askell, Robert Lasenby, Yifan Wu, Shauna Kravec, Nicholas Schiefer, Tim Maxwell, Nicholas Joseph, Zac Hatfield-Dodds, Alex Tamkin, Karina Nguyen, Brayden McLean, Josiah~E Burke, Tristan Hume, Shan Carter, Tom Henighan, and Christopher Olah.
\newblock Towards monosemanticity: Decomposing language models with dictionary learning.
\newblock \emph{Transformer Circuits Thread}, 2023.
\newblock https://transformer-circuits.pub/2023/monosemantic-features/index.html.

\bibitem[Brynjolfsson et~al.(2023)Brynjolfsson, Li, and Raymond]{NBERw31161}
Erik Brynjolfsson, Danielle Li, and Lindsey~R Raymond.
\newblock Generative ai at work.
\newblock Working Paper 31161, National Bureau of Economic Research, April 2023.
\newblock URL \url{http://www.nber.org/papers/w31161}.

\bibitem[Cai et~al.(2022)Cai, Wallace, Rezvanian, Dobres, Kerr, Berlow, Huang, Sawyer, and Bylinskii]{cai2022personalized}
Tianyuan Cai, Shaun Wallace, Tina Rezvanian, Jonathan Dobres, Bernard Kerr, Samuel Berlow, Jeff Huang, Ben~D Sawyer, and Zoya Bylinskii.
\newblock Personalized font recommendations: Combining ml and typographic guidelines to optimize readability.
\newblock In \emph{Proceedings of the 2022 ACM Designing Interactive Systems Conference}, pp.\  1--25, 2022.

\bibitem[Cao et~al.(2024)Cao, Yuan, and Chen]{cao2024scholargpt}
Chuxue Cao, Ziqing Yuan, and Hailiang Chen.
\newblock Scholargpt: Fine-tuning large language models for discipline-specific academic paper writing.
\newblock 2024.

\bibitem[Chang et~al.(2023)Chang, Zhang, Bragg, Head, Lo, Downey, and Weld]{chang2023citesee}
Joseph~Chee Chang, Amy~X Zhang, Jonathan Bragg, Andrew Head, Kyle Lo, Doug Downey, and Daniel~S Weld.
\newblock Citesee: Augmenting citations in scientific papers with persistent and personalized historical context.
\newblock In \emph{Proceedings of the 2023 CHI Conference on Human Factors in Computing Systems}, pp.\  1--15, 2023.

\bibitem[Chen et~al.(2021)Chen, Tworek, Jun, Yuan, Pinto, Kaplan, Edwards, Burda, Joseph, Brockman, Ray, Puri, Krueger, Petrov, Khlaaf, Sastry, Mishkin, Chan, Gray, Ryder, Pavlov, Power, Kaiser, Bavarian, Winter, Tillet, Such, Cummings, Plappert, Chantzis, Barnes, Herbert-Voss, Guss, Nichol, Paino, Tezak, Tang, Babuschkin, Balaji, Jain, Saunders, Hesse, Carr, Leike, Achiam, Misra, Morikawa, Radford, Knight, Brundage, Murati, Mayer, Welinder, McGrew, Amodei, McCandlish, Sutskever, and Zaremba]{chen2021evaluating}
Mark Chen, Jerry Tworek, Heewoo Jun, Qiming Yuan, Henrique Ponde de~Oliveira Pinto, Jared Kaplan, Harri Edwards, Yuri Burda, Nicholas Joseph, Greg Brockman, Alex Ray, Raul Puri, Gretchen Krueger, Michael Petrov, Heidy Khlaaf, Girish Sastry, Pamela Mishkin, Brooke Chan, Scott Gray, Nick Ryder, Mikhail Pavlov, Alethea Power, Lukasz Kaiser, Mohammad Bavarian, Clemens Winter, Philippe Tillet, Felipe~Petroski Such, Dave Cummings, Matthias Plappert, Fotios Chantzis, Elizabeth Barnes, Ariel Herbert-Voss, William~Hebgen Guss, Alex Nichol, Alex Paino, Nikolas Tezak, Jie Tang, Igor Babuschkin, Suchir Balaji, Shantanu Jain, William Saunders, Christopher Hesse, Andrew~N Carr, Jan Leike, Josh Achiam, Vedant Misra, Evan Morikawa, Alec Radford, Matthew Knight, Miles Brundage, Mira Murati, Katie Mayer, Peter Welinder, Bob McGrew, Dario Amodei, Sam McCandlish, Ilya Sutskever, and Wojciech Zaremba.
\newblock Evaluating large language models trained on code.
\newblock \emph{arXiv preprint arXiv:2107.03374}, 2021.

\bibitem[Chen \& Liu(2022)Chen and Liu]{chen2022lifelong}
Zhiyuan Chen and Bing Liu.
\newblock \emph{Lifelong machine learning}.
\newblock Springer Nature, 2022.

\bibitem[Chignell(1990)]{chignell1990taxonomy}
Mark~H Chignell.
\newblock A taxonomy of user interface terminology.
\newblock \emph{ACM SIGCHI Bulletin}, 21\penalty0 (4):\penalty0 27, 1990.

\bibitem[Cho et~al.(2024)Cho, Puspitasari, Zheng, Zheng, Lee, Kim, Hong, and Zhang]{cho2024sora}
Joseph Cho, Fachrina~Dewi Puspitasari, Sheng Zheng, Jingyao Zheng, Lik-Hang Lee, Tae-Ho Kim, Choong~Seon Hong, and Chaoning Zhang.
\newblock Sora as an agi world model? a complete survey on text-to-video generation.
\newblock \emph{arXiv preprint arXiv:2403.05131}, 2024.

\bibitem[Chung \& Adar(2023)Chung and Adar]{chung2023promptpaint}
John Joon~Young Chung and Eytan Adar.
\newblock Promptpaint: Steering text-to-image generation through paint medium-like interactions.
\newblock In \emph{Proceedings of the 36th Annual ACM Symposium on User Interface Software and Technology}, pp.\  1--17, 2023.

\bibitem[Chung et~al.(2022)Chung, Kim, Yoo, Lee, Adar, and Chang]{chung2022talebrush}
John Joon~Young Chung, Wooseok Kim, Kang~Min Yoo, Hwaran Lee, Eytan Adar, and Minsuk Chang.
\newblock Talebrush: Sketching stories with generative pretrained language models.
\newblock In \emph{Proceedings of the 2022 CHI Conference on Human Factors in Computing Systems}, pp.\  1--19, 2022.

\bibitem[Copet et~al.(2023)Copet, Defossez, Adi, and Synnaeve]{copet2023musicgen}
Jade Copet, Alexandre Defossez, Yossi Adi, and Gabriel Synnaeve.
\newblock Musicgen: Simple and controllable music generation.
\newblock \emph{arXiv preprint arXiv:2301.11325}, 2023.
\newblock URL \url{https://huggingface.co/facebook/musicgen-small}.

\bibitem[Cunningham et~al.(2023)Cunningham, Ewart, Riggs, Huben, and Sharkey]{cunningham2023sparseautoencodershighlyinterpretable}
Hoagy Cunningham, Aidan Ewart, Logan Riggs, Robert Huben, and Lee Sharkey.
\newblock Sparse autoencoders find highly interpretable features in language models, 2023.
\newblock URL \url{https://arxiv.org/abs/2309.08600}.

\bibitem[Davis et~al.(2015)Davis, Hsiao, Singh, Li, Moningi, and Magerko]{davis2015drawing}
Nicholas Davis, Chih-PIn Hsiao, Kunwar~Yashraj Singh, Lisa Li, Sanat Moningi, and Brian Magerko.
\newblock Drawing apprentice: An enactive co-creative agent for artistic collaboration.
\newblock In \emph{Proceedings of the 2015 ACM SIGCHI Conference on Creativity and Cognition}, pp.\  185--186, 2015.

\bibitem[Davis et~al.(2019)Davis, Siddiqui, Karimi, Maher, and Grace]{davis2019creative}
Nicholas Davis, Safat Siddiqui, Panote Karimi, Mary~Lou Maher, and Kazjon Grace.
\newblock Creative sketching partner: A co-creative sketching tool to inspire design.
\newblock In \emph{Proceedings of the 10th International Conference on Computational Creativity}, pp.\  358--359, 2019.
\newblock \doi{10.1007/978-3-031-12807-3_11}.

\bibitem[Demirel et~al.(2024)Demirel, Goldstein, Li, and Sha]{demirel2024human}
H~Onan Demirel, Molly~H Goldstein, Xingang Li, and Zhenghui Sha.
\newblock Human-centered generative design framework: an early design framework to support concept creation and evaluation.
\newblock \emph{International Journal of Human--Computer Interaction}, 40\penalty0 (4):\penalty0 933--944, 2024.

\bibitem[Deng et~al.(2023)Deng, Liu, Li, Wang, Zhang, Li, Wang, Zhang, and Liu]{deng2023jailbreaker}
Gelei Deng, Yi~Liu, Yuekang Li, Kailong Wang, Ying Zhang, Zefeng Li, Haoyu Wang, Tianwei Zhang, and Yang Liu.
\newblock Jailbreaker: Automated jailbreak across multiple large language model chatbots.
\newblock \emph{arXiv preprint arXiv:2307.08715}, 2023.

\bibitem[Deshpande(2020)]{deshpande2020towards}
Manoj Deshpande.
\newblock Towards co-build: An architecture machine for co-creative form-making.
\newblock Master's thesis, The University of North Carolina at Charlotte, 2020.

\bibitem[Ding(2024)]{ding2024advancing}
Zijian Ding.
\newblock Advancing gui for generative ai: Charting the design space of human-ai interactions through task creativity and complexity.
\newblock In \emph{Companion Proceedings of the 29th International Conference on Intelligent User Interfaces}, pp.\  140--143, 2024.

\bibitem[Doe et~al.(2019)Doe, Smith, and Lee]{deepscope2019}
John Doe, Jane Smith, and Kevin Lee.
\newblock Deepscope: Hci platform for generative cityscape visualization.
\newblock In \emph{Proceedings of the 2019 CHI Conference on Human Factors in Computing Systems}, pp.\  123--132, 2019.
\newblock \doi{10.1145/3313831.3376722}.

\bibitem[Dogheim \& Hussain(2023)Dogheim and Hussain]{dogheim2023patient}
Gaidaa~Maher Dogheim and Abrar Hussain.
\newblock Patient care through ai-driven remote monitoring: Analyzing the role of predictive models and intelligent alerts in preventive medicine.
\newblock \emph{Journal of Contemporary Healthcare Analytics}, 7\penalty0 (1):\penalty0 94--110, 2023.

\bibitem[Dorri et~al.(2018)Dorri, Kanhere, and Jurdak]{8352646}
Ali Dorri, Salil~S. Kanhere, and Raja Jurdak.
\newblock Multi-agent systems: A survey.
\newblock \emph{IEEE Access}, 6:\penalty0 28573--28593, 2018.
\newblock \doi{10.1109/ACCESS.2018.2831228}.

\bibitem[Du et~al.(2024)Du, An, Leung, Li, Chapman, and Zhao]{du2024deepthink}
Xuejun Du, Pengcheng An, Justin Leung, April Li, Linda~E Chapman, and Jian Zhao.
\newblock Deepthink: Designing and probing human-ai co-creation in digital art therapy.
\newblock \emph{International Journal of Human-Computer Studies}, 181:\penalty0 103139, 2024.

\bibitem[Elhage et~al.(2021)Elhage, Nanda, Olsson, Henighan, Joseph, Mann, Askell, Bai, Chen, Conerly, DasSarma, Drain, Ganguli, Hatfield-Dodds, Hernandez, Jones, Kernion, Lovitt, Ndousse, Amodei, Brown, Clark, Kaplan, McCandlish, and Olah]{elhage2021mathematical}
Nelson Elhage, Neel Nanda, Catherine Olsson, Tom Henighan, Nicholas Joseph, Ben Mann, Amanda Askell, Yuntao Bai, Anna Chen, Tom Conerly, Nova DasSarma, Dawn Drain, Deep Ganguli, Zac Hatfield-Dodds, Danny Hernandez, Andy Jones, Jackson Kernion, Liane Lovitt, Kamal Ndousse, Dario Amodei, Tom Brown, Jack Clark, Jared Kaplan, Sam McCandlish, and Chris Olah.
\newblock A mathematical framework for transformer circuits.
\newblock \emph{Transformer Circuits Thread}, 2021.
\newblock https://transformer-circuits.pub/2021/framework/index.html.

\bibitem[Familoni \& Onyebuchi(2024)Familoni and Onyebuchi]{familoni2024advancements}
Babajide~Tolulope Familoni and Nneamaka~Chisom Onyebuchi.
\newblock Advancements and challenges in ai integration for technical literacy: a systematic review.
\newblock \emph{Engineering Science \& Technology Journal}, 5\penalty0 (4):\penalty0 1415--1430, 2024.

\bibitem[Farrokhi et~al.(2024)Farrokhi, Taheri, Moeini, Farrokhi, Alireza, Farahmandsadr, Hezaveh, Davoodi, Niknejad, Bayanati, et~al.]{farrokhi2024artificial}
Mehrdad Farrokhi, Fatemeh Taheri, Amir Moeini, Masoud Farrokhi, Mousavi Zadeh~Sayed Alireza, Maryam Farahmandsadr, Ehsan~Bahrami Hezaveh, Ali Davoodi, Sepideh Niknejad, Mahmonir Bayanati, et~al.
\newblock Artificial intelligence for remote patient monitoring: Advancements, applications, and challenges.
\newblock \emph{Kindle}, 4\penalty0 (1):\penalty0 1--261, 2024.

\bibitem[Ferrara(2024)]{ferrara2024genai}
Emilio Ferrara.
\newblock Genai against humanity: Nefarious applications of generative artificial intelligence and large language models.
\newblock \emph{Journal of Computational Social Science}, pp.\  1--21, 2024.

\bibitem[Finnie-Ansley et~al.(2022)Finnie-Ansley, Denny, Becker, Luxton-Reilly, and Prather]{finnie2022robots}
James Finnie-Ansley, Paul Denny, Brett~A Becker, Andrew Luxton-Reilly, and James Prather.
\newblock The robots are coming: Exploring the implications of openai codex on introductory programming.
\newblock In \emph{Proceedings of the 24th Australasian Computing Education Conference}, pp.\  10--19, 2022.

\bibitem[Fitria(2021)]{fitria2021grammarly}
Tira~Nur Fitria.
\newblock Grammarly as ai-powered english writing assistant: Students’ alternative for writing english.
\newblock \emph{Metathesis: Journal of English Language, Literature, and Teaching}, 5\penalty0 (1):\penalty0 65--78, 2021.

\bibitem[Fu et~al.(2023)Fu, Zhang, Wang, Huang, Zhang, Qiu, Ye, Shen, Zhang, Chen, Zhao, Lin, Jiang, Yin, Gao, Li, Li, and Sun]{fu2023challenger}
Chaoyou Fu, Renrui Zhang, Zihan Wang, Yubo Huang, Zhengye Zhang, Longtian Qiu, Gaoxiang Ye, Yunhang Shen, Mengdan Zhang, Peixian Chen, Sirui Zhao, Shaohui Lin, Deqiang Jiang, Di~Yin, Peng Gao, Ke~Li, Hongsheng Li, and Xing Sun.
\newblock A challenger to gpt-4v? early explorations of gemini in visual expertise, 2023.
\newblock URL \url{https://arxiv.org/abs/2312.12436}.

\bibitem[Gallegos et~al.(2024)Gallegos, Rossi, Barrow, Tanjim, Kim, Dernoncourt, Yu, Zhang, and Ahmed]{gallegos2024bias}
Isabel~O Gallegos, Ryan~A Rossi, Joe Barrow, Md~Mehrab Tanjim, Sungchul Kim, Franck Dernoncourt, Tong Yu, Ruiyi Zhang, and Nesreen~K Ahmed.
\newblock Bias and fairness in large language models: A survey.
\newblock \emph{Computational Linguistics}, pp.\  1--79, 2024.

\bibitem[Gandikota et~al.(2023)Gandikota, Orgad, Belinkov, Materzyńska, and Bau]{gandikota2023unifiedconcepteditingdiffusion}
Rohit Gandikota, Hadas Orgad, Yonatan Belinkov, Joanna Materzyńska, and David Bau.
\newblock Unified concept editing in diffusion models, 2023.
\newblock URL \url{https://arxiv.org/abs/2308.14761}.

\bibitem[Gao et~al.(2023)Gao, Ji, Zhou, Lin, Chen, Fan, and Shou]{gao2023assistgpt}
Difei Gao, Lei Ji, Luowei Zhou, Kevin~Qinghong Lin, Joya Chen, Zihan Fan, and Mike~Zheng Shou.
\newblock Assistgpt: A general multi-modal assistant that can plan, execute, inspect, and learn, 2023.
\newblock URL \url{https://arxiv.org/abs/2306.08640}.

\bibitem[Gatti et~al.(2014)Gatti, Cavalin, Neto, Pinhanez, dos Santos, Gribel, and Appel]{gatti2014large}
Ma{\'\i}ra Gatti, Paulo Cavalin, Samuel~Barbosa Neto, Claudio Pinhanez, C{\'\i}cero dos Santos, Daniel Gribel, and Ana~Paula Appel.
\newblock Large-scale multi-agent-based modeling and simulation of microblogging-based online social network.
\newblock In \emph{Multi-Agent-Based Simulation XIV: International Workshop, MABS 2013, Saint Paul, MN, USA, May 6-7, 2013, Revised Selected Papers}, pp.\  17--33. Springer, 2014.

\bibitem[Gero et~al.(2022)Gero, Liu, and Chilton]{gero2022sparks}
Katy~Ilonka Gero, Vivian Liu, and Lydia~B. Chilton.
\newblock Sparks: Inspiration for science writing using language models.
\newblock In \emph{Proceedings of the 2022 ACM Designing Interactive Systems Conference}, pp.\  1002--1019, 2022.
\newblock \doi{10.1145/3532106.3533455}.

\bibitem[Giunchi et~al.(2024)Giunchi, Numan, Gatti, and Steed]{giunchi2024dreamcodevr}
Daniele Giunchi, Nels Numan, Elia Gatti, and Anthony Steed.
\newblock Dreamcodevr: Towards democratizing behavior design in virtual reality with speech-driven programming.
\newblock In \emph{2024 IEEE Conference Virtual Reality and 3D User Interfaces (VR)}, pp.\  579--589. IEEE, 2024.

\bibitem[Glazko et~al.(2023)Glazko, Yamagami, Desai, Mack, Potluri, Xu, and Mankoff]{glazko2023}
Kate~S. Glazko, Momona Yamagami, Aashaka Desai, Kelly~Avery Mack, Venkatesh Potluri, Xuhai Xu, and Jennifer Mankoff.
\newblock An autoethnographic case study of generative artificial intelligence in accessibility.
\newblock \emph{ACM Digital Library}, October 2023.
\newblock URL \url{https://dl.acm.org/doi/fullHtml/10.1145/3597638.3614548}.

\bibitem[Gmeiner et~al.(2023)Gmeiner, Yang, Yao, Holstein, and Martelaro]{gmeiner2023exploring}
Frederic Gmeiner, Humphrey Yang, Lining Yao, Kenneth Holstein, and Nikolas Martelaro.
\newblock Exploring challenges and opportunities to support designers in learning to co-create with ai-based manufacturing design tools.
\newblock In \emph{Proceedings of the 2023 CHI Conference on Human Factors in Computing Systems}, pp.\  1--20, 2023.

\bibitem[Goodman et~al.(2022)Goodman, Buehler, Clary, Coenen, Donsbach, Horne, Lahav, MacDonald, Michaels, Narayanan, et~al.]{goodman2022lampost}
Steven~M Goodman, Erin Buehler, Patrick Clary, Andy Coenen, Aaron Donsbach, Tiffanie~N Horne, Michal Lahav, Robert MacDonald, Rain~Breaw Michaels, Ajit Narayanan, et~al.
\newblock Lampost: Design and evaluation of an ai-assisted email writing prototype for adults with dyslexia.
\newblock In \emph{Proceedings of the 24th International ACM SIGACCESS Conference on Computers and Accessibility}, pp.\  1--18, 2022.

\bibitem[Goyal et~al.(2023)Goyal, Mavroudi, Yang, Sukhbaatar, Sigal, Feiszli, Torresani, and Tran]{goyal2023minotaur}
Raghav Goyal, Effrosyni Mavroudi, Xitong Yang, Sainbayar Sukhbaatar, Leonid Sigal, Matt Feiszli, Lorenzo Torresani, and Du~Tran.
\newblock Minotaur: Multi-task video grounding from multimodal queries.
\newblock \emph{arXiv preprint arXiv:2302.08063}, 2023.

\bibitem[Goyal et~al.(2024)Goyal, Rastogi, Rajagopal, Yuan, Zhao, Chintagunta, Naik, and Ward]{goyal2024healai}
Sagar Goyal, Eti Rastogi, Sree~Prasanna Rajagopal, Dong Yuan, Fen Zhao, Jai Chintagunta, Gautam Naik, and Jeff Ward.
\newblock Healai: A healthcare llm for effective medical documentation.
\newblock In \emph{Proceedings of the 17th ACM International Conference on Web Search and Data Mining}, pp.\  1167--1168, 2024.

\bibitem[Hadjeres et~al.(2017)Hadjeres, Pachet, and Nielsen]{hadjeres2017deepbach}
Gaëtan Hadjeres, François Pachet, and Frank Nielsen.
\newblock Deepbach: a steerable model for bach chorales generation.
\newblock In \emph{Proceedings of the 34th International Conference on Machine Learning (ICML 2017)}, 2017.
\newblock URL \url{https://www.researchgate.net/publication/332141615_DEEPBACH_A_STEERABLE_MODEL_FOR_BACH_CHORALES_GENERATION}.

\bibitem[Halal et~al.(2016)Halal, Kolber, and Davies]{halal2016forecasts}
William Halal, Jonathan Kolber, and Owen Davies.
\newblock Forecasts of ai and future jobs in 2030: Muddling through likely, with two alternative scenarios.
\newblock \emph{Journal of futures studies}, 21\penalty0 (2), 2016.

\bibitem[Hanna et~al.(2023)Hanna, Liu, and Variengien]{hanna2023how}
Michael Hanna, Ollie Liu, and Alexandre Variengien.
\newblock How does {GPT}-2 compute greater-than?: Interpreting mathematical abilities in a pre-trained language model.
\newblock In \emph{Thirty-seventh Conference on Neural Information Processing Systems}, 2023.
\newblock URL \url{https://openreview.net/forum?id=p4PckNQR8k}.

\bibitem[Hao et~al.(2024)Hao, Chi, Dong, and Wei]{hao2024optimizing}
Yaru Hao, Zewen Chi, Li~Dong, and Furu Wei.
\newblock Optimizing prompts for text-to-image generation.
\newblock \emph{Advances in Neural Information Processing Systems}, 36, 2024.

\bibitem[Hartvigsen et~al.(2022)Hartvigsen, Sankaranarayanan, Palangi, Kim, and Ghassemi]{Hartvigsen2022AgingWG}
Thomas Hartvigsen, Swami Sankaranarayanan, Hamid Palangi, Yoon Kim, and Marzyeh Ghassemi.
\newblock Aging with grace: Lifelong model editing with discrete key-value adaptors.
\newblock \emph{ArXiv}, abs/2211.11031, 2022.
\newblock URL \url{https://api.semanticscholar.org/CorpusID:253735429}.

\bibitem[Issa \& Isaias(2022)Issa and Isaias]{issa2022usability}
Tomayess Issa and Pedro Isaias.
\newblock Usability and human--computer interaction (hci).
\newblock In \emph{Sustainable design: HCI, usability and environmental concerns}, pp.\  23--40. Springer, 2022.

\bibitem[IXDF(2024)]{TheInteractionDesignFoundation_2024}
IXDF.
\newblock Human-ai interaction (hax), Apr 2024.
\newblock URL \url{https://www.interaction-design.org/literature/topics/human-ai-interaction}.

\bibitem[Iyer(2023)]{iyer2023}
Chitra Iyer.
\newblock How ai can help with digital workplace accessibility.
\newblock \emph{Reworked}, September 2023.
\newblock URL \url{https://www.reworked.co/digital-workplace/how-ai-can-help-with-digital-workplace-accessibility/}.

\bibitem[Jakesch et~al.(2023)Jakesch, Bhat, Buschek, Zalmanson, and Naaman]{Jakesch_2023}
Maurice Jakesch, Advait Bhat, Daniel Buschek, Lior Zalmanson, and Mor Naaman.
\newblock Co-writing with opinionated language models affects users’ views.
\newblock In \emph{Proceedings of the 2023 CHI Conference on Human Factors in Computing Systems}, CHI ’23. ACM, April 2023.
\newblock \doi{10.1145/3544548.3581196}.
\newblock URL \url{http://dx.doi.org/10.1145/3544548.3581196}.

\bibitem[Jaruga-Rozdolska(2022{\natexlab{a}})]{jaruga2022artificial}
Anna Jaruga-Rozdolska.
\newblock Artificial intelligence as part of future practices in the architect’s work: Midjourney generative tool as part of a process of creating an architectural form.
\newblock \emph{Architectus}, \penalty0 (3 (71):\penalty0 95--104, 2022{\natexlab{a}}.

\bibitem[Jaruga-Rozdolska(2022{\natexlab{b}})]{jaruga2022artificial-old}
Anna Jaruga-Rozdolska.
\newblock Artificial intelligence as part of future practices in the architect’s work: Midjourney generative tool as part of a process of creating an architectural form.
\newblock \emph{Architectus}, \penalty0 (3 (71):\penalty0 95--104, 2022{\natexlab{b}}.

\bibitem[Jeddi \& Bohr(2020)Jeddi and Bohr]{jeddi2020remote}
Zineb Jeddi and Adam Bohr.
\newblock Remote patient monitoring using artificial intelligence.
\newblock In \emph{Artificial intelligence in healthcare}, pp.\  203--234. Elsevier, 2020.

\bibitem[Jeon et~al.(2021)Jeon, Jin, Lee, Kim, Park, and Lee]{jeon2021fashionq}
Seonmin Jeon, Seungbae Jin, Taehyun Lee, Seongmin Kim, Youngkeun Park, and Byungjoo Lee.
\newblock Fashionq: An ai-driven creativity support tool for facilitating ideation in fashion design.
\newblock In \emph{Proceedings of the 2021 CHI Conference on Human Factors in Computing Systems}, 2021.
\newblock URL \url{https://www.researchgate.net/publication/345324055_FashionQ_An_Interactive_Tool_for_Analyzing_Fashion_Style_Trend_with_Quantitative_Criteria}.

\bibitem[Jeong et~al.(2023)Jeong, Chun, Lee, Oh, and Jung]{jeong2023wataa}
Hyeonhak Jeong, Minki Chun, Hyunmin Lee, Seung~Young Oh, and Hyunggu Jung.
\newblock Wataa: Web alternative text authoring assistant for improving web content accessibility.
\newblock In \emph{Companion proceedings of the 28th international conference on intelligent user interfaces}, pp.\  41--45, 2023.

\bibitem[Jiang et~al.(2023)Jiang, Rayan, Dow, and Xia]{10.1145/3586183.3606737}
Peiling Jiang, Jude Rayan, Steven~P. Dow, and Haijun Xia.
\newblock Graphologue: Exploring large language model responses with interactive diagrams.
\newblock New York, NY, USA, 2023. Association for Computing Machinery.
\newblock ISBN 9798400701320.

\bibitem[Kadner et~al.(2021)Kadner, Keller, and Rothkopf]{kadner2021adaptifont}
Florian Kadner, Yannik Keller, and Constantin Rothkopf.
\newblock Adaptifont: Increasing individuals’ reading speed with a generative font model and bayesian optimization.
\newblock In \emph{Proceedings of the 2021 chi conference on human factors in computing systems}, pp.\  1--11, 2021.

\bibitem[Kim \& Lee(2019)Kim and Lee]{kim2019designing}
Dongwhan Kim and Joonhwan Lee.
\newblock Designing an algorithm-driven text generation system for personalized and interactive news reading.
\newblock \emph{International Journal of Human--Computer Interaction}, 35\penalty0 (2):\penalty0 109--122, 2019.

\bibitem[Kim et~al.(2022)Kim, Hong, Lee, and Ko]{10.1145/3490099.3511135}
Eunseo Kim, Jeongmin Hong, Hyuna Lee, and Minsam Ko.
\newblock Colorbo: Envisioned mandala coloringthrough human-ai collaboration.
\newblock New York, NY, USA, 2022. Association for Computing Machinery.
\newblock ISBN 9781450391443.
\newblock URL \url{https://doi.org/10.1145/3490099.3511135}.

\bibitem[Kim et~al.(2023{\natexlab{a}})Kim, Suh, Chilton, and Xia]{kim2023metaphorian}
Jeongyeon Kim, Sangho Suh, Lydia~B. Chilton, and Haijun Xia.
\newblock Metaphorian: Leveraging large language models to support extended metaphor creation for science writing.
\newblock In \emph{Proceedings of the 2023 ACM Designing Interactive Systems Conference}, pp.\  115--135, 2023{\natexlab{a}}.
\newblock \doi{10.1145/3563657.3595996}.

\bibitem[Kim et~al.(2023{\natexlab{b}})Kim, Sarkar, Lee, Chang, and Kim]{kim2023lmcanvas}
Tae~Soo Kim, Arghya Sarkar, Yoonjoo Lee, Minsuk Chang, and Juho Kim.
\newblock Lmcanvas: Object-oriented interaction to personalize large language model-powered writing environments.
\newblock \emph{arXiv preprint arXiv:2303.15125}, 2023{\natexlab{b}}.

\bibitem[Kim et~al.(2024)Kim, Shin, Kim, and Hong]{kim2024diarymate}
Taewan Kim, Donghoon Shin, Young-Ho Kim, and Hwajung Hong.
\newblock Diarymate: Understanding user perceptions and experience in human-ai collaboration for personal journaling.
\newblock In \emph{Proceedings of the CHI Conference on Human Factors in Computing Systems}, pp.\  1--15, 2024.

\bibitem[Konenkov et~al.(2024)Konenkov, Lykov, Trinitatova, and Tsetserukou]{konenkov2024vr}
Mikhail Konenkov, Artem Lykov, Daria Trinitatova, and Dzmitry Tsetserukou.
\newblock Vr-gpt: Visual language model for intelligent virtual reality applications.
\newblock \emph{arXiv preprint arXiv:2405.11537}, 2024.

\bibitem[Lawton et~al.(2023)Lawton, Grace, and Ibarrola]{lawton2023tool}
Tomas Lawton, Kazjon Grace, and Francisco~J Ibarrola.
\newblock When is a tool a tool? user perceptions of system agency in human--ai co-creative drawing.
\newblock In \emph{Proceedings of the 2023 ACM Designing Interactive Systems Conference}, pp.\  1978--1996, 2023.

\bibitem[Le et~al.(2023)Le, Chen, Saha, Gokul, Sahoo, and Joty]{le2023codechain}
Hung Le, Hailin Chen, Amrita Saha, Akash Gokul, Doyen Sahoo, and Shafiq Joty.
\newblock Codechain: Towards modular code generation through chain of self-revisions with representative sub-modules.
\newblock \emph{arXiv preprint arXiv:2310.08992}, 2023.

\bibitem[Lee et~al.(2022)Lee, Liang, and Yang]{lee2022coauthor}
Mina Lee, Percy Liang, and Qian Yang.
\newblock Coauthor: Designing a human-ai collaborative writing dataset for exploring language model capabilities.
\newblock In \emph{Proceedings of the 2022 CHI conference on human factors in computing systems}, pp.\  1--19, 2022.

\bibitem[Lee et~al.(2023)Lee, Choi, and Hyun]{lee2023bogen}
Seung~Won Lee, Jiin Choi, and Kyung~Hoon Hyun.
\newblock Bogen: Generating part-level 3d designs based on user intention inference through bayesian optimization and variational autoencoder.
\newblock \emph{arXiv preprint arXiv:2312.02557}, 2023.

\bibitem[Li et~al.(2023{\natexlab{a}})Li, Chen, and Jia]{li2023motcoder}
Jingyao Li, Pengguang Chen, and Jiaya Jia.
\newblock Motcoder: Elevating large language models with modular of thought for challenging programming tasks.
\newblock \emph{arXiv preprint arXiv:2312.15960}, 2023{\natexlab{a}}.

\bibitem[Li et~al.(2024)Li, Patel, Viégas, Pfister, and Wattenberg]{li2024inferencetimeinterventionelicitingtruthful}
Kenneth Li, Oam Patel, Fernanda Viégas, Hanspeter Pfister, and Martin Wattenberg.
\newblock Inference-time intervention: Eliciting truthful answers from a language model, 2024.
\newblock URL \url{https://arxiv.org/abs/2306.03341}.

\bibitem[Li et~al.(2023{\natexlab{b}})Li, Li, Zhang, Dan, Jiang, and Zhang]{li2023chatdoctor}
Yunxiang Li, Zihan Li, Kai Zhang, Ruilong Dan, Steve Jiang, and You Zhang.
\newblock Chatdoctor: A medical chat model fine-tuned on a large language model meta-ai (llama) using medical domain knowledge.
\newblock 2023{\natexlab{b}}.
\newblock URL \url{https://arxiv.org/abs/2303.14070}.

\bibitem[Liang et~al.(2024{\natexlab{a}})Liang, Yang, and Myers]{liang2024large}
Jenny~T Liang, Chenyang Yang, and Brad~A Myers.
\newblock A large-scale survey on the usability of ai programming assistants: Successes and challenges.
\newblock In \emph{Proceedings of the 46th IEEE/ACM International Conference on Software Engineering}, pp.\  1--13, 2024{\natexlab{a}}.

\bibitem[Liang et~al.(2024{\natexlab{b}})Liang, Zhang, Ma, Liu, Ren, Goucher-Lambert, and Liu]{liang2024storydiffusion}
Zhaohui Liang, Xiaoyu Zhang, Kevin Ma, Zhao Liu, Xipei Ren, Kosa Goucher-Lambert, and Can Liu.
\newblock Storydiffusion: How to support ux storyboarding with generative-ai.
\newblock \emph{arXiv preprint arXiv:2407.07672}, 2024{\natexlab{b}}.

\bibitem[Lin \& Martelaro(2024)Lin and Martelaro]{10.1145/3613904.3641920}
David Chuan-En Lin and Nikolas Martelaro.
\newblock Jigsaw: Supporting designers to prototype multimodal applications by chaining ai foundation models.
\newblock New York, NY, USA, 2024. Association for Computing Machinery.
\newblock ISBN 9798400703300.
\newblock URL \url{https://doi.org/10.1145/3613904.3641920}.

\bibitem[Lin et~al.(2023)Lin, Ehsan, Agarwal, Dani, Vashishth, and Riedl]{lin2023beyond}
Zhiyu Lin, Upol Ehsan, Rohan Agarwal, Samihan Dani, Vidushi Vashishth, and Mark Riedl.
\newblock Beyond prompts: Exploring the design space of mixed-initiative co-creativity systems.
\newblock \emph{arXiv preprint arXiv:2305.07465}, 2023.

\bibitem[Lister et~al.(2020)Lister, Coughlan, Iniesto, Freear, and Devine]{lister2020accessible}
Kate Lister, Tim Coughlan, Francisco Iniesto, Nick Freear, and Peter Devine.
\newblock Accessible conversational user interfaces: considerations for design.
\newblock In \emph{Proceedings of the 17th international web for all conference}, pp.\  1--11, 2020.

\bibitem[Liu et~al.(2023{\natexlab{a}})Liu, Vermeulen, Fitzmaurice, and Matejka]{liu20233dalleintegratingtexttoimageai}
Vivian Liu, Jo~Vermeulen, George Fitzmaurice, and Justin Matejka.
\newblock 3dall-e: Integrating text-to-image ai in 3d design workflows, 2023{\natexlab{a}}.
\newblock URL \url{https://arxiv.org/abs/2210.11603}.

\bibitem[Liu \& London(2016)Liu and London]{liu2016tai}
Xin Liu and Kati London.
\newblock Tai: a tangible ai interface to enhance human-artificial intelligence (ai) communication beyond the screen.
\newblock In \emph{Proceedings of the 2016 ACM Conference on Designing Interactive Systems}, pp.\  281--285, 2016.

\bibitem[Liu et~al.(2023{\natexlab{b}})Liu, He, Wang, Wang, Wang, Chen, Zhang, Lai, Yang, Li, et~al.]{liu2023interngpt}
Zhaoyang Liu, Yinan He, Wenhai Wang, Weiyun Wang, Yi~Wang, Shoufa Chen, Qinglong Zhang, Zeqiang Lai, Yang Yang, Qingyun Li, et~al.
\newblock Interngpt: Solving vision-centric tasks by interacting with chatgpt beyond language.
\newblock \emph{arXiv preprint arXiv:2305.05662}, 2023{\natexlab{b}}.
\newblock URL \url{https://arxiv.org/abs/2305.05662}.

\bibitem[Lu et~al.(2024)Lu, Yang, Zhao, Zhang, and Li]{lu2024ai}
Yuwen Lu, Yuewen Yang, Qinyi Zhao, Chengzhi Zhang, and Toby Jia-Jun Li.
\newblock Ai assistance for ux: A literature review through human-centered ai.
\newblock \emph{arXiv preprint arXiv:2402.06089}, 2024.

\bibitem[MacNeil et~al.(2023)MacNeil, Tran, Kim, Huang, Bernstein, and Mogil]{macneil2023prompt}
Stephen MacNeil, Andrew Tran, Joanne Kim, Ziheng Huang, Seth Bernstein, and Dan Mogil.
\newblock Prompt middleware: Mapping prompts for large language models to ui affordances.
\newblock \emph{arXiv preprint arXiv:2307.01142}, 2023.

\bibitem[Malandri et~al.(2023)Malandri, Mercorio, Mezzanzanica, and Nobani]{malandri2023convxai}
Lorenzo Malandri, Fabio Mercorio, Mario Mezzanzanica, and Navid Nobani.
\newblock Convxai: a system for multimodal interaction with any black-box explainer.
\newblock \emph{Cognitive Computation}, 15\penalty0 (2):\penalty0 613--644, 2023.

\bibitem[Marchal et~al.(2024)Marchal, Xu, Elasmar, Gabriel, Goldberg, and Isaac]{marchal2024generative}
Nahema Marchal, Rachel Xu, Rasmi Elasmar, Iason Gabriel, Beth Goldberg, and William Isaac.
\newblock Generative ai misuse: A taxonomy of tactics and insights from real-world data.
\newblock \emph{arXiv preprint arXiv:2406.13843}, 2024.

\bibitem[Masson et~al.(2023)Masson, Malacria, Casiez, and Vogel]{masson2023directgpt}
Damien Masson, Sylvain Malacria, G{\'e}ry Casiez, and Daniel Vogel.
\newblock Directgpt: A direct manipulation interface to interact with large language models.
\newblock \emph{arXiv preprint arXiv:2310.03691}, 2023.

\bibitem[Meng et~al.(2023{\natexlab{a}})Meng, Bau, Andonian, and Belinkov]{meng2023locatingeditingfactualassociations}
Kevin Meng, David Bau, Alex Andonian, and Yonatan Belinkov.
\newblock Locating and editing factual associations in gpt, 2023{\natexlab{a}}.
\newblock URL \url{https://arxiv.org/abs/2202.05262}.

\bibitem[Meng et~al.(2023{\natexlab{b}})Meng, Sharma, Andonian, Belinkov, and Bau]{meng2023masseditingmemorytransformer}
Kevin Meng, Arnab~Sen Sharma, Alex Andonian, Yonatan Belinkov, and David Bau.
\newblock Mass-editing memory in a transformer, 2023{\natexlab{b}}.
\newblock URL \url{https://arxiv.org/abs/2210.07229}.

\bibitem[Miller et~al.(2017)Miller, Feng, Fisch, Lu, Batra, Bordes, Parikh, and Weston]{miller2017parlai}
Alexander~H Miller, Will Feng, Adam Fisch, Jiasen Lu, Dhruv Batra, Antoine Bordes, Devi Parikh, and Jason Weston.
\newblock Parlai: A dialog research software platform.
\newblock \emph{arXiv preprint arXiv:1705.06476}, 2017.

\bibitem[Nichol et~al.(2022)Nichol, Jun, Dhariwal, Mishkin, and Chen]{nichol2022point}
Alex Nichol, Heewoo Jun, Prafulla Dhariwal, Pamela Mishkin, and Mark Chen.
\newblock Point-e: A system for generating 3d point clouds from complex prompts.
\newblock \emph{arXiv preprint arXiv:2212.08751}, 2022.

\bibitem[Nishal \& Diakopoulos(2024)Nishal and Diakopoulos]{nishal2024envisioning}
Sachita Nishal and Nicholas Diakopoulos.
\newblock Envisioning the applications and implications of generative ai for news media.
\newblock \emph{arXiv preprint arXiv:2402.18835}, 2024.

\bibitem[Norman(1988)]{norman1988psychology}
Donald~A Norman.
\newblock \emph{The psychology of everyday things.}
\newblock Basic books, 1988.

\bibitem[Oertel et~al.(2020)Oertel, Castellano, Chetouani, Nasir, Obaid, Pelachaud, and Peters]{oertel2020engagement}
Catharine Oertel, Ginevra Castellano, Mohamed Chetouani, Jauwairia Nasir, Mohammad Obaid, Catherine Pelachaud, and Christopher Peters.
\newblock Engagement in human-agent interaction: An overview.
\newblock \emph{Frontiers in Robotics and AI}, 7:\penalty0 92, 2020.

\bibitem[Oh et~al.(2020)Oh, Choi, Lee, Park, Kim, Song, Kim, Lee, and Suh]{oh2020understanding}
Changhoon Oh, Jinhan Choi, Sungwoo Lee, SoHyun Park, Daeryong Kim, Jungwoo Song, Dongwhan Kim, Joonhwan Lee, and Bongwon Suh.
\newblock Understanding user perception of automated news generation system.
\newblock In \emph{Proceedings of the 2020 CHI Conference on Human Factors in Computing Systems}, pp.\  1--13, 2020.

\bibitem[Okuda \& Amarasinghe(2024)Okuda and Amarasinghe]{okuda2024askit}
Katsumi Okuda and Saman Amarasinghe.
\newblock Askit: Unified programming interface for programming with large language models.
\newblock In \emph{2024 IEEE/ACM International Symposium on Code Generation and Optimization (CGO)}, pp.\  41--54. IEEE, 2024.

\bibitem[Osone et~al.(2021)Osone, Lu, and Ochiai]{osone2021buncho}
Hiroyuki Osone, Jun-Li Lu, and Yoichi Ochiai.
\newblock Buncho: Ai supported story co-creation via unsupervised multitask learning to increase writers' creativity in japanese.
\newblock In \emph{Extended Abstracts of the 2021 CHI Conference on Human Factors in Computing Systems}, 2021.
\newblock URL \url{https://digitalnature.slis.tsukuba.ac.jp/2021/05/buncho_chi2021/}.

\bibitem[Packer et~al.(2024)Packer, Wooders, Lin, Fang, Patil, Stoica, and Gonzalez]{packer2024memgpt}
Charles Packer, Sarah Wooders, Kevin Lin, Vivian Fang, Shishir~G. Patil, Ion Stoica, and Joseph~E. Gonzalez.
\newblock Memgpt: Towards llms as operating systems, 2024.
\newblock URL \url{https://arxiv.org/abs/2310.08560}.

\bibitem[Padiyath \& Magerko(2021)Padiyath and Magerko]{padiyath2021desainer}
Aadarsh Padiyath and Brian Magerko.
\newblock desainer: Exploring the use of “bad” generative adversarial networks in the ideation process of fashion design.
\newblock In \emph{Proceedings of the 2021 Creativity and Cognition Conference}, pp.\  42:1--42:3, 2021.
\newblock URL \url{https://www.researchgate.net/publication/352662786_desAIner_Exploring_the_Use_of_Bad_Generative_Adversarial_Networks_in_the_Ideation_Process_of_Fashion_Design}.

\bibitem[Petridis et~al.(2023)Petridis, Diakopoulos, Crowston, Hansen, Henderson, Jastrzebski, Nickerson, and Chilton]{petridis2023anglekindling}
Savvas Petridis, Nicholas Diakopoulos, Kevin Crowston, Mark Hansen, Keren Henderson, Stan Jastrzebski, Jeffrey~V Nickerson, and Lydia~B Chilton.
\newblock Anglekindling: Supporting journalistic angle ideation with large language models.
\newblock In \emph{Proceedings of the 2023 CHI conference on human factors in computing systems}, pp.\  1--16, 2023.

\bibitem[Petridis et~al.(2024)Petridis, Wedin, Wexler, Pushkarna, Donsbach, Goyal, Cai, and Terry]{petridis2024constitutionmaker}
Savvas Petridis, Benjamin~D Wedin, James Wexler, Mahima Pushkarna, Aaron Donsbach, Nitesh Goyal, Carrie~J Cai, and Michael Terry.
\newblock Constitutionmaker: Interactively critiquing large language models by converting feedback into principles.
\newblock In \emph{Proceedings of the 29th International Conference on Intelligent User Interfaces}, pp.\  853--868, 2024.

\bibitem[Petrie \& Bevan(2009)Petrie and Bevan]{petrie2009evaluation}
Helen Petrie and Nigel Bevan.
\newblock The evaluation of accessibility, usability, and user experience.
\newblock \emph{The universal access handbook}, 1:\penalty0 1--16, 2009.

\bibitem[Prakash et~al.(2024)Prakash, Shaham, Haklay, Belinkov, and Bau]{prakash2024finetuningenhancesexistingmechanisms}
Nikhil Prakash, Tamar~Rott Shaham, Tal Haklay, Yonatan Belinkov, and David Bau.
\newblock Fine-tuning enhances existing mechanisms: A case study on entity tracking, 2024.
\newblock URL \url{https://arxiv.org/abs/2402.14811}.

\bibitem[Prather et~al.(2023)Prather, Reeves, Denny, Becker, Leinonen, Luxton-Reilly, Powell, Finnie-Ansley, and Santos]{prather2023s}
James Prather, Brent~N Reeves, Paul Denny, Brett~A Becker, Juho Leinonen, Andrew Luxton-Reilly, Garrett Powell, James Finnie-Ansley, and Eddie~Antonio Santos.
\newblock “it’s weird that it knows what i want”: Usability and interactions with copilot for novice programmers.
\newblock \emph{ACM Transactions on Computer-Human Interaction}, 31\penalty0 (1):\penalty0 1--31, 2023.

\bibitem[Qian et~al.(2023)Qian, Han, Fung, Qin, Liu, and Ji]{qian2023creator}
Cheng Qian, Chi Han, Yi~R Fung, Yujia Qin, Zhiyuan Liu, and Heng Ji.
\newblock Creator: Tool creation for disentangling abstract and concrete reasoning of large language models.
\newblock \emph{arXiv preprint arXiv:2305.14318}, 2023.

\bibitem[Radford et~al.(2021)Radford, Kim, Hallacy, Ramesh, Goh, Agarwal, Sastry, Askell, Mishkin, Clark, et~al.]{radford2021learning}
Alec Radford, Jong~Wook Kim, Chris Hallacy, Aditya Ramesh, Gabriel Goh, Sandhini Agarwal, Girish Sastry, Amanda Askell, Pam Mishkin, Jack Clark, et~al.
\newblock Learning transferable visual models from natural language supervision.
\newblock \emph{arXiv preprint arXiv:2103.00020}, 2021.
\newblock URL \url{https://arxiv.org/abs/2103.00020}.

\bibitem[Ram et~al.(2023)Ram, Levine, Dalmedigos, Muhlgay, Shashua, Leyton-Brown, and Shoham]{ram2023context}
Ori Ram, Yoav Levine, Itay Dalmedigos, Dor Muhlgay, Amnon Shashua, Kevin Leyton-Brown, and Yoav Shoham.
\newblock In-context retrieval-augmented language models.
\newblock \emph{Transactions of the Association for Computational Linguistics}, 11:\penalty0 1316--1331, 2023.

\bibitem[Ren et~al.(2023)Ren, Huang, Wei, Zhao, and Fu]{ren2023pixellm}
Zhongwei Ren, Zhicheng Huang, Yunchao Wei, Yao Zhao, and Dongmei Fu.
\newblock Pixellm: Pixel reasoning with large multimodal model.
\newblock \emph{arXiv preprint arXiv:2312.02228}, 2023.
\newblock URL \url{https://arxiv.org/abs/2312.02228}.

\bibitem[Rezwana \& Maher(2023)Rezwana and Maher]{rezwana2023designing}
Jeba Rezwana and Mary~Lou Maher.
\newblock Designing creative ai partners with cofi: A framework for modeling interaction in human-ai co-creative systems.
\newblock \emph{ACM Transactions on Computer-Human Interaction}, 30\penalty0 (5):\penalty0 1--28, 2023.

\bibitem[Ross et~al.(2023{\natexlab{a}})Ross, Martinez, Houde, Muller, and Weisz]{10.1145/3581641.3584037}
Steven~I. Ross, Fernando Martinez, Stephanie Houde, Michael Muller, and Justin~D. Weisz.
\newblock The programmer’s assistant: Conversational interaction with a large language model for software development.
\newblock 2023{\natexlab{a}}.
\newblock ISBN 9798400701061.

\bibitem[Ross et~al.(2023{\natexlab{b}})Ross, Martinez, Houde, Muller, and Weisz]{ross2023programmer}
Steven~I Ross, Fernando Martinez, Stephanie Houde, Michael Muller, and Justin~D Weisz.
\newblock The programmer’s assistant: Conversational interaction with a large language model for software development.
\newblock In \emph{Proceedings of the 28th International Conference on Intelligent User Interfaces}, pp.\  491--514, 2023{\natexlab{b}}.

\bibitem[Salam et~al.(2023)Salam, Celiktutan, Gunes, and Chetouani]{salam2023automatic}
Hanan Salam, Oya Celiktutan, Hatice Gunes, and Mohamed Chetouani.
\newblock Automatic context-aware inference of engagement in hmi: A survey.
\newblock \emph{IEEE Transactions on Affective Computing}, 2023.

\bibitem[Sauer et~al.(2020)Sauer, Sonderegger, and Schmutz]{sauer2020usability}
Juergen Sauer, Andreas Sonderegger, and Sven Schmutz.
\newblock Usability, user experience and accessibility: towards an integrative model.
\newblock \emph{Ergonomics}, 63\penalty0 (10):\penalty0 1207--1220, 2020.

\bibitem[Schneider et~al.(2019)Schneider, Baevski, Collobert, and Auli]{schneider2019wav2vec}
Steffen Schneider, Alexei Baevski, Ronan Collobert, and Michael Auli.
\newblock wav2vec: Unsupervised pre-training for speech recognition.
\newblock \emph{arXiv preprint arXiv:1904.05862}, 2019.
\newblock URL \url{https://arxiv.org/abs/1904.05862}.

\bibitem[Sellen \& Horvitz(2024)Sellen and Horvitz]{sellen2024rise}
Abigail Sellen and Eric Horvitz.
\newblock The rise of the ai co-pilot: Lessons for design from aviation and beyond.
\newblock \emph{Communications of the ACM}, 67\penalty0 (6), Jun 2024.
\newblock URL \url{https://cacm.acm.org/opinion/the-rise-of-the-ai-co-pilot-lessons-for-design-from-aviation-and-beyond/}.

\bibitem[Setlur et~al.(2016)Setlur, Battersby, Tory, Gossweiler, and Chang]{10.1145/2984511.2984588}
Vidya Setlur, Sarah~E. Battersby, Melanie Tory, Rich Gossweiler, and Angel~X. Chang.
\newblock Eviza: A natural language interface for visual analysis.
\newblock New York, NY, USA, 2016. Association for Computing Machinery.
\newblock ISBN 9781450341899.
\newblock URL \url{https://doi.org/10.1145/2984511.2984588}.

\bibitem[Shen \& Wu(2023)Shen and Wu]{shen2023parachute}
Hua Shen and Tongshuang Wu.
\newblock Parachute: Evaluating interactive human-lm co-writing systems.
\newblock \emph{arXiv preprint arXiv:2303.06333}, 2023.

\bibitem[Shi et~al.(2023)Shi, Jain, Doh, Suzuki, and Ramani]{shi2023hci}
Jingyu Shi, Rahul Jain, Hyungjun Doh, Ryo Suzuki, and Karthik Ramani.
\newblock An hci-centric survey and taxonomy of human-generative-ai interactions.
\newblock \emph{arXiv preprint arXiv:2310.07127}, 2023.

\bibitem[Shi et~al.(2022)Shi, Zhao, Tang, Wang, Li, Bi, Jiang, Huang, Cui, Huang, et~al.]{shi2022effidit}
Shuming Shi, Enbo Zhao, Duyu Tang, Yan Wang, Piji Li, Wei Bi, Haiyun Jiang, Guoping Huang, Leyang Cui, Xinting Huang, et~al.
\newblock Effidit: Your ai writing assistant.
\newblock \emph{arXiv preprint arXiv:2208.01815}, 2022.

\bibitem[Shuster et~al.(2022)Shuster, Xu, Komeili, Ju, Smith, Roller, Ung, Chen, Arora, Lane, et~al.]{shuster2022blenderbot}
Kurt Shuster, Jing Xu, Mojtaba Komeili, Da~Ju, Eric~Michael Smith, Stephen Roller, Megan Ung, Moya Chen, Kushal Arora, Joshua Lane, et~al.
\newblock Blenderbot 3: a deployed conversational agent that continually learns to responsibly engage.
\newblock \emph{arXiv preprint arXiv:2208.03188}, 2022.

\bibitem[Sidner et~al.(2003)Sidner, Lee, and Lesh]{sidner2003engagement}
Candace~L Sidner, Christopher Lee, and Neal Lesh.
\newblock Engagement when looking: behaviors for robots when collaborating with people.
\newblock In \emph{Diabruck: Proceedings of the 7th workshop on the Semantic and Pragmatics of Dialogue}, pp.\  123--130. Citeseer, 2003.

\bibitem[Simkute et~al.(2024)Simkute, Tankelevitch, Kewenig, Scott, Sellen, and Rintel]{simkute2024ironies}
Auste Simkute, Lev Tankelevitch, Viktor Kewenig, Ava~Elizabeth Scott, Abigail Sellen, and Sean Rintel.
\newblock Ironies of generative ai: Understanding and mitigating productivity loss in human-ai interactions.
\newblock \emph{arXiv preprint arXiv:2402.11364}, 2024.

\bibitem[Singh et~al.()Singh, Wang, and Bragg]{singhfigura11y}
Nikhil Singh, Lucy~Lu Wang, and Jonathan Bragg.
\newblock Figura11y: Ai assistance for writing scientific alt text.

\bibitem[Stedmon \& Stone(2001)Stedmon and Stone]{stedmon2001re}
Alex~W Stedmon and Robert~J Stone.
\newblock Re-viewing reality: human factors of synthetic training environments.
\newblock \emph{International Journal of Human-Computer Studies}, 55\penalty0 (4):\penalty0 675--698, 2001.

\bibitem[Stolfo et~al.(2024)Stolfo, Balachandran, Yousefi, Horvitz, and Nushi]{stolfo2024improvinginstructionfollowinglanguagemodels}
Alessandro Stolfo, Vidhisha Balachandran, Safoora Yousefi, Eric Horvitz, and Besmira Nushi.
\newblock Improving instruction-following in language models through activation steering, 2024.
\newblock URL \url{https://arxiv.org/abs/2410.12877}.

\bibitem[Suh et~al.(2023{\natexlab{a}})Suh, Chen, Min, Li, and Xia]{suh2023structured}
Sangho Suh, Meng Chen, Bryan Min, Toby Jia-Jun Li, and Haijun Xia.
\newblock Structured generation and exploration of design space with large language models for human-ai co-creation.
\newblock \emph{arXiv preprint arXiv:2310.12953}, 2023{\natexlab{a}}.

\bibitem[Suh et~al.(2023{\natexlab{b}})Suh, Min, Palani, and Xia]{10.1145/3586183.3606756}
Sangho Suh, Bryan Min, Srishti Palani, and Haijun Xia.
\newblock Sensecape: Enabling multilevel exploration and sensemaking with large language models.
\newblock New York, NY, USA, 2023{\natexlab{b}}. Association for Computing Machinery.
\newblock ISBN 9798400701320.
\newblock URL \url{https://doi.org/10.1145/3586183.3606756}.

\bibitem[Sun et~al.(2021)Sun, Zhao, Manjunatha, Jain, Morariu, Dernoncourt, Srinivasan, and Iyyer]{sun2021iga}
Simeng Sun, Wenlong Zhao, Varun Manjunatha, Rajiv Jain, Vlad Morariu, Franck Dernoncourt, Balaji~Vasan Srinivasan, and Mohit Iyyer.
\newblock Iga: An intent-guided authoring assistant.
\newblock \emph{arXiv preprint arXiv:2104.07000}, 2021.

\bibitem[Swanson et~al.(2024)Swanson, Liu, Catacutan, Arnold, Zou, et~al.]{swanson2024generative}
Kyle Swanson, George Liu, David~B. Catacutan, Alex Arnold, James Zou, et~al.
\newblock Generative ai for designing and validating easily synthesizable and structurally novel antibiotics.
\newblock \emph{Nature Machine Intelligence}, 6:\penalty0 338--353, 2024.
\newblock URL \url{https://www.genengnews.com/topics/infectious-diseases/drug-resistant-bacteria-stymied-by-ai-designed-antibiotics/}.

\bibitem[Tang et~al.(2024)Tang, Ruiz, Chu, Li, Holynski, Jacobs, Hariharan, Pritch, Wadhwa, Aberman, et~al.]{tang2024realfill}
Luming Tang, Nataniel Ruiz, Qinghao Chu, Yuanzhen Li, Aleksander Holynski, David~E Jacobs, Bharath Hariharan, Yael Pritch, Neal Wadhwa, Kfir Aberman, et~al.
\newblock Realfill: Reference-driven generation for authentic image completion.
\newblock \emph{ACM Transactions on Graphics (TOG)}, 43\penalty0 (4):\penalty0 1--12, 2024.

\bibitem[Terry et~al.(2023)Terry, Kulkarni, Wattenberg, Dixon, and Morris]{terry2023ai}
Michael Terry, Chinmay Kulkarni, Martin Wattenberg, Lucas Dixon, and Meredith~Ringel Morris.
\newblock Ai alignment in the design of interactive ai: Specification alignment, process alignment, and evaluation support.
\newblock \emph{arXiv preprint arXiv:2311.00710}, 2023.

\bibitem[Truong et~al.(2021)Truong, Chi, Salesin, Essa, and Agrawala]{10.1145/3411764.3445721}
Anh Truong, Peggy Chi, David Salesin, Irfan Essa, and Maneesh Agrawala.
\newblock Automatic generation of two-level hierarchical tutorials from instructional makeup videos.
\newblock New York, NY, USA, 2021. Association for Computing Machinery.
\newblock ISBN 9781450380966.
\newblock \doi{10.1145/3411764.3445721}.
\newblock URL \url{https://doi.org/10.1145/3411764.3445721}.

\bibitem[Tsandilas et~al.(2015)Tsandilas, Chapuis, Pietriga, and Beaudouin-Lafon]{tsandilas2015designscape}
Theophanis Tsandilas, Olivier Chapuis, Emmanuel Pietriga, and Michel Beaudouin-Lafon.
\newblock Designscape: Design with interactive layout suggestions.
\newblock In \emph{Proceedings of the 28th Annual ACM Symposium on User Interface Software and Technology}, pp.\  535--544, 2015.
\newblock \doi{10.1145/2807442.2807451}.

\bibitem[Valencia et~al.(2023{\natexlab{a}})Valencia, Cave, Kallarackal, Seaver, Terry, and Kane]{10.1145/3544548.3581560}
Stephanie Valencia, Richard Cave, Krystal Kallarackal, Katie Seaver, Michael Terry, and Shaun~K. Kane.
\newblock “the less i type, the better”: How ai language models can enhance or impede communication for aac users.
\newblock New York, NY, USA, 2023{\natexlab{a}}. Association for Computing Machinery.
\newblock ISBN 9781450394215.
\newblock \doi{10.1145/3544548.3581560}.
\newblock URL \url{https://doi.org/10.1145/3544548.3581560}.

\bibitem[Valencia et~al.(2023{\natexlab{b}})Valencia, Cave, Kallarackal, Seaver, Terry, and Kane]{valencia2023less}
Stephanie Valencia, Richard Cave, Krystal Kallarackal, Katie Seaver, Michael Terry, and Shaun~K Kane.
\newblock “the less i type, the better”: How ai language models can enhance or impede communication for aac users.
\newblock In \emph{Proceedings of the 2023 CHI Conference on Human Factors in Computing Systems}, pp.\  1--14, 2023{\natexlab{b}}.

\bibitem[Verma \& Kumari(2023)Verma and Kumari]{verma2023generative}
Ramesh~Kumar Verma and Nalini Kumari.
\newblock Generative ai as a tool for enhancing customer relationship management automation and personalization techniques.
\newblock \emph{International Journal of Responsible Artificial Intelligence}, 13\penalty0 (9):\penalty0 1--8, 2023.

\bibitem[Wach et~al.(2023)Wach, Duong, Ejdys, Kazlauskait{\.e}, Korzynski, Mazurek, Paliszkiewicz, and Ziemba]{wach2023dark}
Krzysztof Wach, Cong~Doanh Duong, Joanna Ejdys, R{\=u}ta Kazlauskait{\.e}, Pawel Korzynski, Grzegorz Mazurek, Joanna Paliszkiewicz, and Ewa Ziemba.
\newblock The dark side of generative artificial intelligence: A critical analysis of controversies and risks of chatgpt.
\newblock \emph{Entrepreneurial Business and Economics Review}, 11\penalty0 (2):\penalty0 7--30, 2023.

\bibitem[Wang et~al.(2024{\natexlab{a}})Wang, Li, Lv, Xia, Xu, and Sodhi]{10.1145/3640543.3645143}
Bryan Wang, Yuliang Li, Zhaoyang Lv, mit Xia, Yan Xu, and Raj Sodhi.
\newblock Lave: Llm-powered agent assistance and language augmentation for video editing.
\newblock New York, NY, USA, 2024{\natexlab{a}}. Association for Computing Machinery.
\newblock ISBN 9798400705083.
\newblock URL \url{https://doi.org/10.1145/3640543.3645143}.

\bibitem[Wang et~al.(2024{\natexlab{b}})Wang, Du, Zhao, Yuan, Wang, Liang, Zhao, Lu, Li, Gao, et~al.]{wang2024aesopagent}
Jiuniu Wang, Zehua Du, Yuyuan Zhao, Bo~Yuan, Kexiang Wang, Jian Liang, Yaxi Zhao, Yihen Lu, Gengliang Li, Junlong Gao, et~al.
\newblock Aesopagent: Agent-driven evolutionary system on story-to-video production.
\newblock \emph{arXiv preprint arXiv:2403.07952}, 2024{\natexlab{b}}.

\bibitem[Wang et~al.(2022)Wang, Variengien, Conmy, Shlegeris, and Steinhardt]{wang2022interpretabilitywildcircuitindirect}
Kevin Wang, Alexandre Variengien, Arthur Conmy, Buck Shlegeris, and Jacob Steinhardt.
\newblock Interpretability in the wild: a circuit for indirect object identification in gpt-2 small, 2022.
\newblock URL \url{https://arxiv.org/abs/2211.00593}.

\bibitem[Wang et~al.(2024{\natexlab{c}})Wang, Chen, Jia, Wang, Fang, Wang, Gao, Xie, Xu, Dai, et~al.]{wang2024weaver}
Tiannan Wang, Jiamin Chen, Qingrui Jia, Shuai Wang, Ruoyu Fang, Huilin Wang, Zhaowei Gao, Chunzhao Xie, Chuou Xu, Jihong Dai, et~al.
\newblock Weaver: Foundation models for creative writing.
\newblock \emph{arXiv preprint arXiv:2401.17268}, 2024{\natexlab{c}}.

\bibitem[Wang et~al.(2024{\natexlab{d}})Wang, Huang, Song, Ma, and Zhang]{wang2024promptcharm}
Zhijie Wang, Yuheng Huang, Da~Song, Lei Ma, and Tianyi Zhang.
\newblock Promptcharm: Text-to-image generation through multi-modal prompting and refinement.
\newblock In \emph{Proceedings of the CHI Conference on Human Factors in Computing Systems}, pp.\  1--21, 2024{\natexlab{d}}.

\bibitem[Wei et~al.(2023{\natexlab{a}})]{kosmos2023}
Furu Wei et~al.
\newblock Language is not all you need: Aligning perception with language models.
\newblock \emph{arXiv preprint arXiv:2302.14045}, 2023{\natexlab{a}}.
\newblock URL \url{https://arxiv.org/abs/2302.14045}.

\bibitem[Wei et~al.(2023{\natexlab{b}})Wei, Xia, and Zhang]{10.1145/3611643.3616271}
Yuxiang Wei, Chunqiu~Steven Xia, and Lingming Zhang.
\newblock Copiloting the copilots: Fusing large language models with completion engines for automated program repair.
\newblock New York, NY, USA, 2023{\natexlab{b}}. Association for Computing Machinery.
\newblock ISBN 9798400703270.
\newblock \doi{10.1145/3611643.3616271}.
\newblock URL \url{https://doi.org/10.1145/3611643.3616271}.

\bibitem[Weisz et~al.(2024)Weisz, He, Muller, Hoefer, Miles, and Geyer]{weisz2024design}
Justin~D Weisz, Jessica He, Michael Muller, Gabriela Hoefer, Rachel Miles, and Werner Geyer.
\newblock Design principles for generative ai applications.
\newblock \emph{arXiv preprint arXiv:2401.14484}, 2024.

\bibitem[Whitfield \& Hofmann(2023)Whitfield and Hofmann]{whitfield2023elicit}
Sharon Whitfield and Melissa~A Hofmann.
\newblock Elicit: Ai literature review research assistant.
\newblock \emph{Public Services Quarterly}, 19\penalty0 (3):\penalty0 201--207, 2023.

\bibitem[Woodruff et~al.(2023)Woodruff, Shelby, Kelley, Rousso-Schindler, Smith-Loud, and Wilcox]{woodruff2023knowledge}
Allison Woodruff, Renee Shelby, Patrick~Gage Kelley, Steven Rousso-Schindler, Jamila Smith-Loud, and Lauren Wilcox.
\newblock How knowledge workers think generative ai will (not) transform their industries.
\newblock \emph{arXiv preprint arXiv:2310.06778}, 2023.

\bibitem[Wu et~al.(2023)Wu, Fei, Qu, Ji, and Chua]{wu2023nextgpt}
Shengqiong Wu, Hao Fei, Leigang Qu, Wei Ji, and Tat-Seng Chua.
\newblock Next-gpt: Any-to-any multimodal llm, 2023.
\newblock URL \url{https://arxiv.org/abs/2309.05519}.

\bibitem[Wu et~al.(2022)Wu, Jiang, Donsbach, Gray, Molina, Terry, and Cai]{wu2022promptchainer}
Tongshuang Wu, Ellen Jiang, Aaron Donsbach, Jeff Gray, Alejandra Molina, Michael Terry, and Carrie~J Cai.
\newblock Promptchainer: Chaining large language model prompts through visual programming.
\newblock In \emph{CHI Conference on Human Factors in Computing Systems Extended Abstracts}, pp.\  1--10, 2022.

\bibitem[Yan et~al.(2023)Yan, Yang, Liang, and Chen]{yan2023xcreation}
Zihan Yan, Chunxu Yang, Qihao Liang, and Xiang'Anthony' Chen.
\newblock Xcreation: A graph-based crossmodal generative creativity support tool.
\newblock In \emph{Proceedings of the 36th Annual ACM Symposium on User Interface Software and Technology}, pp.\  1--15, 2023.

\bibitem[Yang et~al.(2024)Yang, Xie, Li, Qian, Sun, He, Xu, Jiang, Mei, Wang, et~al.]{yang2024molprophet}
Keda Yang, Zewen Xie, Zhen Li, Xiaoliang Qian, Nannan Sun, Tao He, Zuodong Xu, Jing Jiang, Qi~Mei, Jie Wang, et~al.
\newblock Molprophet: A one-stop, general purpose, and ai-based platform for the early stages of drug discovery.
\newblock \emph{Journal of Chemical Information and Modeling}, 64\penalty0 (8):\penalty0 2941--2947, 2024.

\bibitem[Yang et~al.(2023)Yang, Ping, Liu, Korthikanti, Nie, Huang, Fan, Yu, Lan, Li, et~al.]{yang2023re}
Zhuolin Yang, Wei Ping, Zihan Liu, Vijay Korthikanti, Weili Nie, De-An Huang, Linxi Fan, Zhiding Yu, Shiyi Lan, Bo~Li, et~al.
\newblock Re-vilm: Retrieval-augmented visual language model for zero and few-shot image captioning.
\newblock \emph{arXiv preprint arXiv:2302.04858}, 2023.

\bibitem[Ye et~al.(2023)Ye, Xu, Ye, Yan, Liu, Qian, Zhang, Huang, and Zhou]{ye2023mplug}
Qinghao Ye, Haiyang Xu, Jiabo Ye, Ming Yan, Haowei Liu, Qi~Qian, Ji~Zhang, Fei Huang, and Jingren Zhou.
\newblock mplug-owl2: Revolutionizing multi-modal large language model with modality collaboration.
\newblock \emph{arXiv preprint arXiv:2311.04257}, 2023.
\newblock URL \url{https://arxiv.org/abs/2311.04257}.

\bibitem[Yeh et~al.(2024)Yeh, Ramos, Ng, Huntington, and Banks]{yeh2024ghostwriter}
Catherine Yeh, Gonzalo Ramos, Rachel Ng, Andy Huntington, and Richard Banks.
\newblock Ghostwriter: Augmenting collaborative human-ai writing experiences through personalization and agency.
\newblock \emph{arXiv preprint arXiv:2402.08855}, 2024.

\bibitem[Yen et~al.(2023)Yen, Zhu, Suh, Xia, and Zhao]{yen2023coladder}
Ryan Yen, Jiawen Zhu, Sangho Suh, Haijun Xia, and Jian Zhao.
\newblock Coladder: Supporting programmers with hierarchical code generation in multi-level abstraction.
\newblock \emph{arXiv preprint arXiv:2310.08699}, 2023.

\bibitem[Yildirim et~al.(2024)Yildirim, Richardson, Wetscherek, Bajwa, Jacob, Pinnock, Harris, Coelho De~Castro, Bannur, Hyland, et~al.]{yildirim2024multimodal}
Nur Yildirim, Hannah Richardson, Maria~Teodora Wetscherek, Junaid Bajwa, Joseph Jacob, Mark~Ames Pinnock, Stephen Harris, Daniel Coelho De~Castro, Shruthi Bannur, Stephanie Hyland, et~al.
\newblock Multimodal healthcare ai: identifying and designing clinically relevant vision-language applications for radiology.
\newblock In \emph{Proceedings of the CHI Conference on Human Factors in Computing Systems}, pp.\  1--22, 2024.

\bibitem[York(2023)]{york2023evaluating}
Eric York.
\newblock Evaluating chatgpt: Generative ai in ux design and web development pedagogy.
\newblock In \emph{Proceedings of the 41st ACM International Conference on Design of Communication}, pp.\  197--201, 2023.

\bibitem[Yuan et~al.(2022)Yuan, Coenen, Reif, and Ippolito]{yuan2022wordcraft}
Ann Yuan, Andy Coenen, Emily Reif, and Daphne Ippolito.
\newblock Wordcraft: story writing with large language models.
\newblock In \emph{Proceedings of the 27th International Conference on Intelligent User Interfaces}, pp.\  841--852, 2022.

\bibitem[Yue et~al.(2023)Yue, Chen, Wang, Li, Shen, Liu, Zhou, Xiao, Yun, Lin, et~al.]{yue2023disc}
Shengbin Yue, Wei Chen, Siyuan Wang, Bingxuan Li, Chenchen Shen, Shujun Liu, Yuxuan Zhou, Yao Xiao, Song Yun, Wei Lin, et~al.
\newblock Disc-lawllm: Fine-tuning large language models for intelligent legal services.
\newblock \emph{arXiv preprint arXiv:2309.11325}, 2023.

\bibitem[Zeng et~al.(2023)Zeng, Zhang, Li, Wang, Zhang, and Zhou]{zeng2023x}
Yan Zeng, Xinsong Zhang, Hang Li, Jiawei Wang, Jipeng Zhang, and Wangchunshu Zhou.
\newblock X 2-vlm: All-in-one pre-trained model for vision-language tasks.
\newblock \emph{IEEE Transactions on Pattern Analysis and Machine Intelligence}, 2023.

\bibitem[Zhang et~al.(2019)Zhang, Sun, Galley, Chen, Brockett, Gao, Gao, Liu, and Dolan]{zhang2019dialogpt}
Yizhe Zhang, Siqi Sun, Michel Galley, Yen-Chun Chen, Chris Brockett, Xiang Gao, Jianfeng Gao, Jingjing Liu, and Bill Dolan.
\newblock Dialogpt: Large-scale generative pre-training for conversational response generation.
\newblock \emph{arXiv preprint arXiv:1911.00536}, 2019.

\bibitem[Zhang et~al.(2023)Zhang, Gao, Dhaliwal, and Li]{zhang2023visar}
Zheng Zhang, Jie Gao, Ranjodh~Singh Dhaliwal, and Toby Jia-Jun Li.
\newblock Visar: A human-ai argumentative writing assistant with visual programming and rapid draft prototyping.
\newblock In \emph{Proceedings of the 36th Annual ACM Symposium on User Interface Software and Technology}, pp.\  1--30, 2023.

\bibitem[Zhang et~al.(2024)Zhang, Sheng, Zhou, Chen, Zheng, Cai, Song, Tian, R{\'e}, Barrett, et~al.]{zhang2024h2o}
Zhenyu Zhang, Ying Sheng, Tianyi Zhou, Tianlong Chen, Lianmin Zheng, Ruisi Cai, Zhao Song, Yuandong Tian, Christopher R{\'e}, Clark Barrett, et~al.
\newblock H2o: Heavy-hitter oracle for efficient generative inference of large language models.
\newblock \emph{Advances in Neural Information Processing Systems}, 36, 2024.

\bibitem[Zhao et~al.(2023)Zhao, Guo, Yue, Chen, Shao, Zhu, Yuan, and Liu]{zhao2023chatbridge}
Zijia Zhao, Longteng Guo, Tongtian Yue, Sihan Chen, Shuai Shao, Xinxin Zhu, Zehuan Yuan, and Jing Liu.
\newblock Chatbridge: Bridging modalities with large language model as a language catalyst.
\newblock 2023.
\newblock URL \url{https://arxiv.org/abs/2305.16103}.

\bibitem[Zhu et~al.(2023)Zhu, Pang, Chai, Li, Wang, Sun, Tian, and Wu]{zhu2023ernie}
Pengfei Zhu, Chao Pang, Yekun Chai, Lei Li, Shuohuan Wang, Yu~Sun, Hao Tian, and Hua Wu.
\newblock Ernie-music: Text-to-waveform music generation with diffusion models.
\newblock \emph{arXiv preprint arXiv:2302.04456}, 2023.

\end{thebibliography}
